\documentclass{comnet}

\usepackage{amsmath} 
\usepackage{amsthm} 
\usepackage{graphicx}
\usepackage{xcolor} 
\usepackage{algorithmic} 
\usepackage{subfig} 
\usepackage{url} 
\usepackage{enumitem} 
\usepackage{hyperref} 
\usepackage{placeins}

\setcitestyle{numbers,square}

\usepackage[dvipsnames]{xcolor}

\begin{document}

\title{Multilayer Analysis of the Global Trade Network}

\shorttitle{Multilayer Analysis of the Global Trade Network} 
\shortauthorlist{Multilayer Analysis of the Global Trade Network}

\author{
\name{Chenyang Li}
\address{School of Mathematical Sciences, Queen Mary University of London, London, UK}
\name{Leonardo Brogi}
\address{
  Department of Economics and Statistics, Università di Siena, Italy}
\name{Andrea Civilini}
\address{
Sorbonne Universit\'e, Paris Brain Institute (ICM), CNRS UMR7225, INRIA Paris, 
INSERM U1127, H\^opital de la Piti\'e-Salp\^etri\`ere, AP-HP, Paris 75013, France}
\name{Piero Mazzarisi}
\address{Department of Economics and Statistics, Università di Siena, Italy}
\name{Nicola Perra}
\address{School of Mathematical Sciences, Queen Mary University of London, London, UK}
\and
\name{Vito Latora$^*$}
    \address{School of Mathematical Sciences, Queen Mary University of London, London, UK \\ Dipartimento di Fisica ed Astronomia, Universit\`a di Catania, and INFN Catania, Italy\\
  Complexity Science Hub Vienna, Vienna, Austria}
\email{$^*$Corresponding author:  v.latora@qmul.ac.uk}}
\maketitle

\begin{abstract}
{Global trade is more than a single network of aggregate flows. Beneath the observable exchange of products among economies lies a complex multilayer structure, formed by thousands of product-specific trade relationships that differ in their similarity,  interdependence, and temporal evolution. 
Using the CEPII's BACI database, which records bilateral product-level trade flows between economies, we represent the global trade network from 1995 to 2024 as a temporal multilayer network, with economies as nodes and directed weighted trade flows as edges. To investigate product-level organisation and cross-layer similarity, temporal structural change, and the structural role of individual economies, we introduce a random-walk-based similarity measure 
that provides a unified framework for comparing weighted and directed trade layers. 
Our results show that the global trade network remains relatively stable over short periods but undergoes gradual structural change over longer timescales. We also find that similarity-based product communities only partially align with the official product taxonomy, indicating that products 
assigned to the same official category do not necessarily exhibit similar trade-network structures. Finally, we show that an economy's structural influence is not always determined by its trade volume. These results highlight the value of multilayer network analysis for revealing patterns in global trade that remain hidden at the aggregate level.}

{Complex Systems; Multilayer Networks; Global Trade Network; Measures of Graph Similarity.} 
\end{abstract}

\section{Introduction}
\label{sec:intro}

Global trade can be represented as a directed weighted network in which economies are nodes, trade relations are directed edges, and edge weights measure the value of trade flows. This network perspective is useful because it captures structural relations that are not visible from aggregate trade volumes alone. A large empirical literature has used network methods to study the International Trade Network (also known as the World Trade Web) characterising its topology, centrality structure, reciprocity, clustering, core--periphery organisation, and temporal evolution \citep{garlaschelli_fitness_2004,garlaschelli_structure_2005,fagiolo_worldtrade_2009,fagiolo_worldtrade_2010,debenedictis_world_2011}. These studies show that trade networks are highly heterogeneous and evolve through structural changes that are not fully captured by aggregate trade volumes. BACI data have also been used explicitly for network analysis of world trade, providing an empirical basis for studying international trade as a weighted network of economy-to-economy relations \citep{debenedictis_network_2013}. More recent work has further examined how trade-network structure changes around crises, policy shifts, and global disruptions \citep{hoang_reshaping_2023,kosztyan_trade_2024}.

A related strand of research moves beyond aggregate trade networks by considering sectoral, commodity-specific, or multilayer representations. Commodity-specific trade networks have been shown to display heterogeneous structures that can differ substantially from the aggregate international trade network \citep{barigozzi_multinetwork_2010,barigozzi_identifying_2011}. Community structure in the World Trade Web has also been studied as a way to understand the organisation of trade relations and the propagation of economic shocks \citep{piccardi_communities_2012}. More recent multilayer approaches to world trade have examined nestedness and economic complexity, showing that product-level trade layers contain hierarchical information about countries' competitiveness and productive structure \citep{ren_bridging_2020}.

In parallel, economic complexity research has shown that product-level export baskets contain information about productive capabilities and the organisation of production. The product-space literature links related products to countries' diversification paths, while fitness--complexity approaches infer country capabilities and product complexity from country--product export matrices \citep{hidalgo_product_2007,hidalgo_building_2009,tacchella_new_2012}. These studies are closely related to our motivation because they treat products as informative units of analysis. However, their main object is usually the country--product export matrix, revealed comparative advantage, or export-basket structure, rather than the bilateral economy-to-economy trade-flow network induced by each individual product.

Taken together, these strands of literature leave an important gap. Aggregate trade-network studies describe the overall position of economies in the world trade system, but they may hide substantial product-level heterogeneity. Commodity- and sector-specific studies show that different product domains can have distinct network structures, yet they usually focus on selected sectors or on the positions of economies within product-specific networks, rather than comparing thousands of product layers with one another. Economic-complexity approaches capture product relatedness and country capabilities, but they do not directly ask whether two products are traded through similar bilateral flow structures. This gap is particularly relevant for product-level trade data. For a given product and year, economies can be represented as nodes, directed edges as exports from one economy to another, and edge weights as the value of those exports. In this sense, each product induces one layer of a multilayer trade network. An economy may therefore be central in one product layer but peripheral in another. Furthermore, two products may have similar total trade volumes but very different exporter--importer structures. Treating products as layers makes it possible to compare these structures systematically across products, aggregation levels, and years.

In this context, we study the Global Trade Network (GTN) as a time-resolved multilayer directed weighted network. The multilayer representation follows the general idea that complex systems can be described through multiple layers of interactions among the same set of nodes \citep{battiston_structural_2014}. It is particularly suitable for international trade, where thousands of products generate distinct but comparable patterns of exchange among economies. We build the GTN from BACI trade data covering 234 economies and 5,018 products over 1995--2024 \citep{CEPII:2010-23}. Product layers are also aggregated into chapter and section layers according to the Harmonized System, an international product classification used in customs statistics \citep{oliver_harmonised_1987}. This gives three levels of analysis: product, chapter, and section.

On this basis, the paper addresses three intertwined research questions:

\begin{enumerate}[itemsep=2pt, topsep=4pt, parsep=0pt]
    \item How stable is the structure of trade layers over time, and how does this stability vary across product, chapter, and section levels?
    \item To what extent does the administrative HS taxonomy reflect the structure of trade flows between products?
    \item Which economies exert the greatest  
    influence on the GTN, and how does this influence shift when their trade patterns change?
\end{enumerate}

To address these questions, we introduce JS-Walk, a novel graph similarity measure that compares node-level random-walk transition profiles through Jensen--Shannon divergence. Its construction draws on random-walk representations of network structure and on the symmetric and bounded properties of Jensen--Shannon divergence for comparing probability distributions~\citep{lovasz_random_1993,lin1991divergence,endres2003new}. We use JS-Walk to quantify temporal structural change and to identify product communities from similarities in trade-flow structure. We then extend the analysis to the economy level with a leave-one-out PageRank perturbation measure of network-structural influence \cite{page1999pagerank,gleich2015pagerank}, which quantifies how strongly the removal of an economy perturbs the centrality structure of the network. Finally, node-level JS-Walk distances are incorporated into this framework to distinguish influence associated with relatively stable trade patterns from influence located in structurally changing layers.

Interestingly, we find that the GTN is highly stable over short time horizons but changes gradually over longer periods. Product communities inferred from trade-flow similarity only partially align with the official HS taxonomy, indicating that realised trade relationships contain structural information not fully captured by administrative product categories. Finally, the economy influence analysis shows that network-structural influence is not equivalent to trade volume: economies can matter because of their structural position across product layers, and the same economy may play different roles on the export and import sides. For example, the USA remains highly influential on both sides, while China shows a more pronounced influence on the import side than on the export side. Overall, our results highlight how a multilayer network analysis can reveal temporal, product-level, and economy-level structures that remain hidden in aggregate trade statistics.

The remainder of the paper is organised as follows. Section~\ref{sec:data} describes the BACI data and the construction of the multilayer trade network. Section~\ref{sec:methods} introduces the methodology, covering graph similarity via JS-Walk, temporal analysis, product similarity analysis, and the economy influence framework. Section~\ref{sec:results} presents the empirical results on temporal similarity, trade-based product taxonomy, and economy influence. Section~\ref{sec:discussion} discusses the implications of these results, and summarises the main conclusions. Additional robustness checks, benchmark comparisons, detailed ARI matrices, threshold sensitivity, and extended product-community and economy-influence results are provided in the Supplementary Material.

\section{Data and Network Construction}
\label{sec:data}

\begin{figure}[!h]
    \centering\includegraphics[width=5.5in]{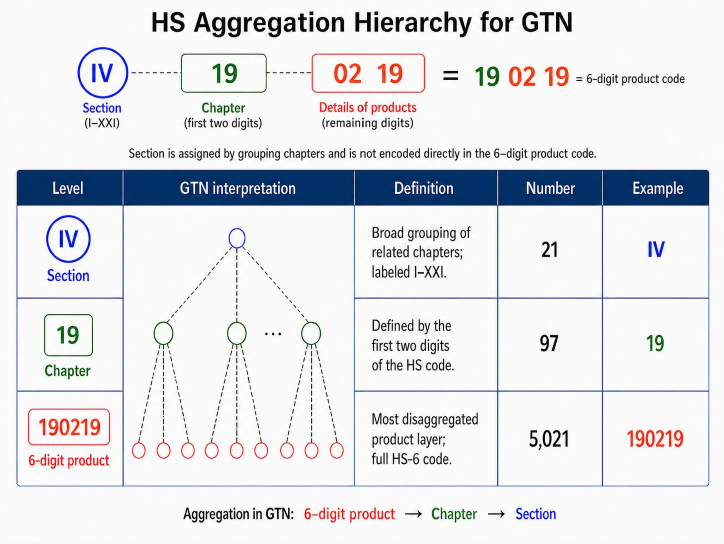}
    \caption{\textbf{Product aggregation hierarchy} In the HS, each product is classified by a six-digit code. The first two digits of a product identify the corresponding {\em chapter}. Chapters are then grouped into broader groups, named  {\em sections}, labelled by Roman numerals. The example shows HS code 190219, the ``\texttt{Pasta}'' product: its first two digits assign it to Chapter 19, which belongs to Section IV. }
    \label{fig:hs}
\end{figure}

We use CEPII's \footnote{Centre d'Études Prospectives et d'Informations Internationales, generally referred to by its acronym CEPII, is a French institute for research in international economics}  BACI \footnote{Base pour l’Analyse du Commerce Internationa--Database for International Trade Analysis.} trade data covering $N = 234$ economies, representing world countries/regions, and $L = 5{,}018$ distinct products, representing the different goods imported and exported by the different economies in the period 1995--2024 \citep{CEPII:2010-23}. BACI provides harmonised bilateral trade flows at the product level, making it suitable for constructing product-specific trade networks as a function of time.
Products are classified according to the Harmonized System (HS), an international nomenclature developed for the classification of traded goods in customs and trade statistics \citep{wco_hs_faq, wco_hs_convention}. The HS provides a logical and hierarchical structure, as illustrated in Figure~\ref{fig:hs}. In this hierarchy, each {\em product} is identified by a six-digit HS code (hereafter, HS6). Products are grouped into classes called {\em chapters}, while chapters are, in turn, arranged within broader classes, named {\em sections} \citep{wco_hs_faq}. In our data, the HS nomenclature contains 96 active chapters and 21 sections (chapter numbering runs from 01 to 97, with Chapter 77 reserved for possible future use \citep{wco_chapter77}). Therefore, we use the HS hierarchy to construct three aggregation levels in the Global Trade Network: product layers, chapter layers and section layers. For example, the HS code 190219 denotes the ``\texttt{Pasta}'' product (more precisely ``Food preparations; pasta, uncooked (excluding that containing eggs), not stuffed or otherwise prepared.''). The first two digits assign this product to Chapter 19, which covers ``Preparations of cereals, flour, starch or milk, and pastrycooks' products, and this chapter belongs to Section IV, which covers ``Prepared foodstuffs; Beverages, Spirits and Vinegar; Tobacco and Manufactured tobacco substitutes.''

For each product, the dataset provides information about the directed import-export flows among each pair of economies in terms of trade values measured in thousand USD. By using the data, we can describe the global trade network (GTN) in the form of a directed weighted multilayer network which is changing in time. The $N$ nodes of such a network represent countries/regions, while each of the $L$ layers codes for the trade of a different product, at a given year $t$. The presence of a directed edge from one economy to another indicates the existence of a trade from an exporter to an importer, and is associated with a weight equal to the corresponding value of the trade volume in that year. Thus, different products correspond to the different layers of the multilayer network, while the set of economies remains fixed across layers. More precisely, this construction follows the general idea of a multiplex network, a special case of a multilayer network, where the same set of nodes can be connected through different types of relations, each one forming a layer \citep{battiston_structural_2014,battiston_multiplex_2026}. Formally, for each year $t$, at its finest scale of product level, the GTN is represented as a multiplex network: 
\begin{equation}
    \mathcal{M}(t) = \bigl(\mathbf{W}^\ell(t)\bigr)_{\ell \in \mathcal{L}}
    \label{eq:multiplex-gtn}
\end{equation}
\noindent where $\mathcal{L}$ denotes the set of products, with $|\mathcal{L}| \equiv L = 5,021$. For each product layer $\ell \in \mathcal{L}$,
\begin{equation}
    \mathbf{W}^\ell(t) = [w_{ij}^\ell(t)] \in \mathbb{R}_{\geq 0}^{N \times N},
\label{eq:product-layer-matrix}
\end{equation}
is the $N \times N$ weighted adjacency matrix of layer $\ell$,with $N=234$ and $\ell=1,2,\ldots,L$, which we refer to as the {\em exporter-importer matrix} of product $\ell$. The weight $w_{ij}^\ell(t)$ denotes the trading volume of the exported product $\ell$ from economy $i$ to economy $j$ in year $t$.

To study the trade structure at coarser aggregation levels, we aggregate product layers within a Chapter (or a Section). If $\mathcal{C}_\alpha \subseteq \mathcal{L}$ denotes the set of products belonging to Chapter (or Section) $\alpha$, the aggregated layer is 
\begin{equation}
    \mathbf{W}^\alpha(t) = \sum_{\ell \in \mathcal{C}_\alpha} \mathbf{W}^\ell(t).
    \label{eq:layer-aggregation}
\end{equation}
This allows the same framework to be applied consistently at the product, chapter, and section levels. Therefore, in what follows, $\alpha$ is used to indicate a generic layer index either at the product, at the chapter, or section level. Finally, to analyse GTN at full aggregation, we combine all product layers into a single annual trade network,
\begin{equation}
    \mathbf{W}^{\mathrm{all}}(t) = \sum_{\ell \in \mathcal{L}} \mathbf{W}^{\ell}(t),
    \label{eq:all-products-network}
\end{equation}
with $\mathbf{W}^{\mathrm{all}}(t)$ representing the total bilateral trade flows across all products in year $t$. 

Because the GTN is directed, each trade layer admits two complementary representations. The matrix $\mathbf{W}^{\alpha}(t)$ is defined by considering flows from exporters to the importers, with $w_{ij}^{\alpha}(t)$ denoting the trade value exported from economy $i$ to economy $j$. We define the \emph{export-side representation} as
\begin{equation}
\label{eq:export-side}
\mathbf{W}^{\alpha,\mathrm{export}}(t) = \mathbf{W}^{\alpha}(t),
\end{equation} 
and the \emph{import-side representation} as
\begin{equation} 
\label{eq:import-side}
\mathbf{W}^{\alpha,\mathrm{import}}(t) = \left(\mathbf{W}^{\alpha}(t)\right)^\top.
\end{equation} 

Whenever the same expression applies to both representations, we use $r \in \{\mathrm{export},\mathrm{import}\}$ as the side index and write $\mathbf{W}^{\alpha,r}(t)$ for the corresponding matrix. As discussed in detail in Section~\ref{sec:jswalk}, row~$i$ of the export-side representation describes the distribution of economy $i$'s exports across destination economies, whereas row~$i$ of the import-side representation describes the distribution of its imports across origin economies. The two representations, therefore, contain the same bilateral trade flows and differ only in whether rows are indexed by exporters or importers.

\begin{figure}[!h]
\centering\includegraphics[width=5.5in]{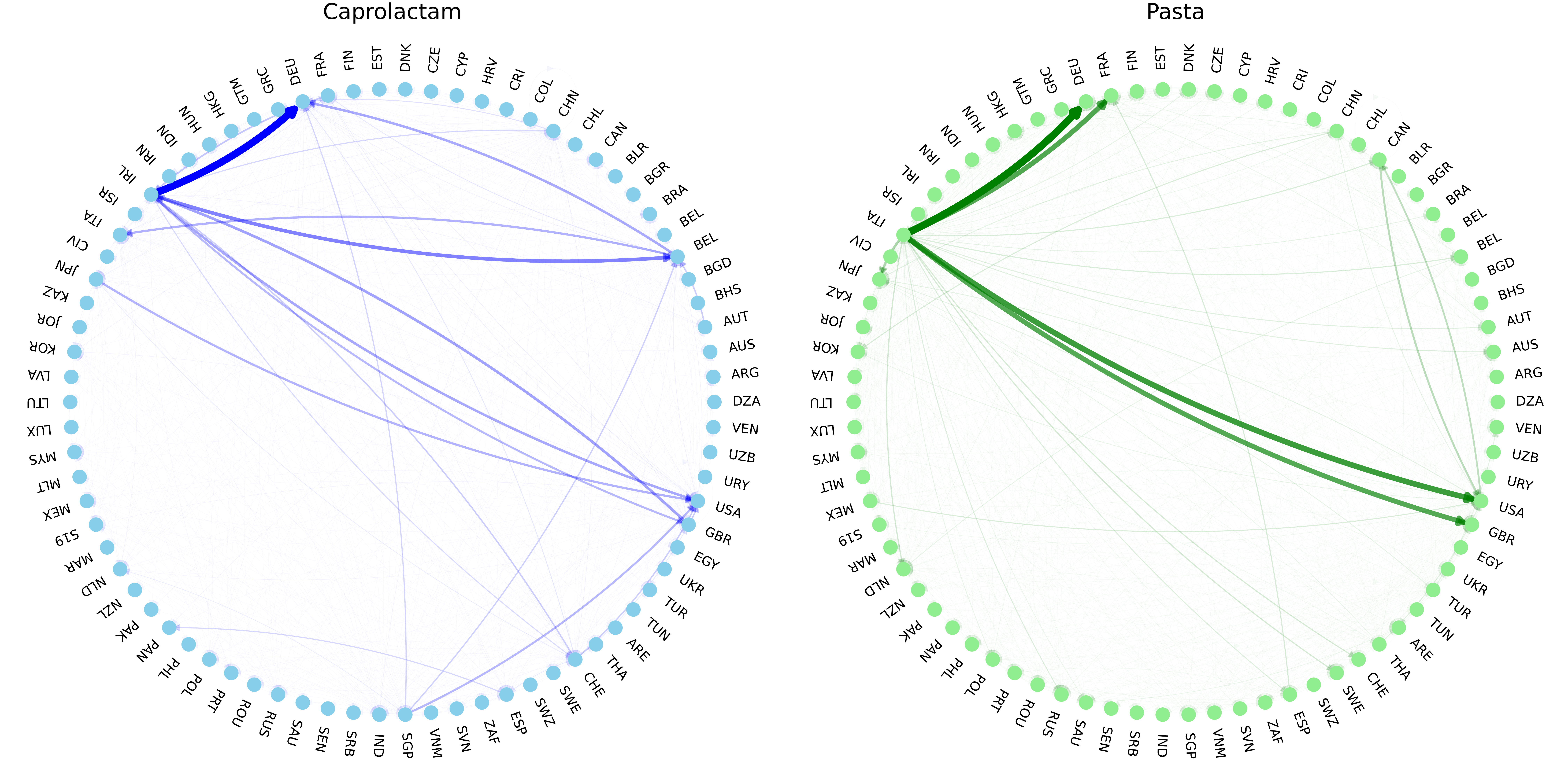}
    \caption{\textbf{Illustrative example of two product-level layers of the GTN.} We show two layers of the global trade network, corresponding respectively to the two products ``\texttt{Caprolactam}'' and ``\texttt{Pasta}'' in 2010. Nodes describe world economies, while directed edges represent trade flows from one economy to another. To make the two layers directly comparable, both panels are first restricted to the common set of economies present in the two product layers. For visual clarity, only the top $15\%$ of edges by trade value are then retained.Economies disconnected after this thresholding step are omitted from the visualisation. Edge thickness is proportional to the trade value $w_{ij}$. The contrast between the two layers illustrates that products defined over a shared economy set can still exhibit markedly different trade-flow structures.}
    \label{fig:layer_examples_products}
\end{figure}

To illustrate the constructed global trade network,  Figure~\ref{fig:layer_examples_products} shows two representative product layers in 2010. The two layers are defined on the same set of economies and use the same exporter--importer edge definition as above, but their flow structures differ substantially. For visual clarity, the diagrams display only the top 15$\%$ of trade links ranked by trade value, while the quantitative analysis in the following sections uses the full weighted adjacency matrices. 

Finally, throughout the paper, $\mathcal{T}=\{1,\dots,T\}$ denotes the set of annual time indices, with $T=30$.
The corresponding calendar year is denoted by $Y_t$, with $Y_t = Y_1 + t - 1$, and $Y_1 = 1995$ and $Y_T=2024$. 
We use $t$ in equations as the serial time index, while $Y_t$ is used when referring to the actual calendar year in the text and figures.
\FloatBarrier

\section{Methods}
\label{sec:methods}

\subsection{Graph similarity through random walk}
\label{sec:jswalk}

Comparing trade layers across products, aggregation levels, and different years requires a definition of a graph similarity that applies uniformly to the full range of graphs that arise in our framework. The Global Trade Network is intrinsically directed (exports and imports are asymmetric), weighted (the monetary value of trade links varies greatly across economy pairs and products), and of widely varying density across products, chapters, and sections. Therefore, the choice of a proper measure of graph similarity is key and can strongly influence the results of the analysis. 
This problem is part of a broader literature on graphs and networks comparison, where different similarity measures emphasise different aspects of network structure \citep{koutra_deltacon_2013,bagrow_information-theoretic_2019,piccardi_metrics_2023}.

\paragraph{Limitations of existing measures.} 

We surveyed a range of established graph-similarity measures and found each of them to be only partially adequate for our purposes. Methods based on local features and ego-networks compare two graphs by summarising the neighbourhood structure around nodes into graph-level descriptors. Examples include NetSimile, which constructs size-independent graph signatures from local and neighbourhood features \citep{berlingerio2012netsimile}, and more recent Egonet which constructs distribution distances based on degree, clustering coefficient, and persistence \citep{piccardi_metrics_2023}. While useful for broad network comparison, these descriptors cannot be used to directly compare the full economy-by-economy trade-transition profiles that define each product layer in the GTN. DeltaCon \citep{koutra_deltacon_2013} relies on fast belief propagation through an attenuation parameter whose value must be chosen heuristically and materially affects the resulting score. Information-theoretic graph similarity measures provide principled ways to compare graphs \citep{bagrow_information-theoretic_2019,felippe_network_2024}. Their suitability depends, however, on how the graph is constructed in each specific case. In the GTN setting, our aim is not only to compare graphs at a global level, but to compare, for each economy, the full distribution of trades with other economies, and across the different types of products. This motivates an information-theoretic formulation based directly on node-level transition profiles in directed weighted product layers. 
A further class of simple baselines is provided by set- and vector-based metrics, such as binary and weighted Jaccard similarity, Euclidean distance, and Manhattan distance \citep{hand_principles_2001}. Applied to the GTN, however, these measures are not well suited to our objective. Binary Jaccard compares only edge existence and therefore discards trade values. Weighted Jaccard retains edge weights, but treats each exporter--importer edge as an independent coordinate rather than comparing the full destination or origin profile of each economy. Euclidean and Manhattan distances can be dominated by a small number of very large trade flows. 

We require a similarity measure that is: (i) \emph{parameter-free}, so that results are reproducible and not depending on the choices of tuning parameters; (ii) \emph{versatile}, in the sense that the same expression can be applied to weighted or unweighted, directed or undirected graphs by changing only the underlying adjacency matrix; (iii) \emph{consistent across aggregation levels}, so that product-, chapter-, and section-level layers, as well as export- and import-side representations, are compared using the same definition of similarity; (iv) \emph{profile-based}, in the sense that it compares the full neighbourhood profile of each node rather than treating individual edges as independent matching units; (v) \emph{bounded}, so that similarity values are directly comparable across years and aggregation levels; and (vi) \emph{node-resolved}, so that the contribution of individual nodes to overall dissimilarity can be recovered. 

To the best of our knowledge, no off-the-shelf graph similarity measure combines all these properties in a form directly suited to the GTN setting. This motivates the JS-Walk similarity introduced below. JS-Walk compares each node's transition profile and then aggregates the node-level differences into a graph-level similarity that takes values in $[0,1]$, thus satisfying all the above requirements.

\paragraph{JS-Walk similarity.}

Our graph similarity measure is based on random-walk transition profiles. For a graph with non-negative edge weights, row-normalising the adjacency matrix gives a stochastic transition matrix whose $i$-th row represents the distribution of a one-step random walk leaving node $i$ \citep{lovasz_random_1993}.
JS-Walk compares two graphs by applying Jensen--Shannon divergence to the corresponding node-level transition distributions, and then averaging the resulting distances over all nodes. More specifically, Jensen--Shannon divergence is an information-theoretic divergence derived from Shannon entropy and provides a symmetric and bounded comparison between probability distributions \citep{cover_elements_2006}. Moreover, the square root of Jensen--Shannon divergence defines a metric \citep{endres2003new}. This yields a bounded, parameter-free similarity score. 

Let $G$ be a graph on $N$ nodes with non-negative ${N \times N}$ weight matrix $W = [w_{ij}]$, where $w_{ij}\ge 0$. Let $P = [p_{ij}]$ be the row-normalised transition-profile matrix of a standard random walk on $G$, defined as:

\begin{equation}
    p_{ij} =
    \begin{cases}
    \frac{w_{ij}}{\sum_{k} w_{ik}},&\text{if} \quad \sum_k w_{ik}>0, \\
    0, &\text{if} \quad \sum_k w_{ik}=0.
    \end{cases}
    \label{eq:rw}
\end{equation}
Notice that rows with zero outgoing weights are retained as rows of zero so that $\sum_{j} p_{ij} = 1$ for each non-zero row, while $\sum_j p_{ij}=0$ for rows of zero. For this reason, matrix $P$ is not strictly speaking row-stochastic. We denote the row-normalisation operation in Eq.~\eqref{eq:rw} by $\mathcal{R}$, so that we can write $P=\mathcal{R}(W)$.
Let 
$P_i = (p_{i1},p_{i2}, \ldots, p_{iN})$ denote the $i$-th row of matrix $P$. We refer to $P_i$ as the transition profile of node $i$, since it gives the probability of reaching the different possible destinations in one random-walk step from node $i$. We can then define the Shannon entropy $H(P_i)$ associated with the transition profile $P_i$ as: 
\begin{equation}
    H(P_i) = -\sum_{j} p_{ij}\,\log_2 p_{ij}, \qquad \forall i,
    \label{eq:entropy}
\end{equation}
with the standard convention $0\log_2 0=0$  \citep{cover_elements_2006}. The entropy $H(P_i)$ quantifies how concentrated or dispersed the transition distribution of node $i$ is across its neighbours, reflecting both the number of outgoing links and the relative distribution of their weights. For the GTN, applying the normalisation in Eq.~\eqref{eq:rw} to the export- and import-side representations defined in Eqs.~\eqref{eq:export-side} and \eqref{eq:import-side} yields the corresponding transition matrices \(P^{\alpha,\mathrm{export}}(t)\) and \(P^{\alpha,\mathrm{import}}(t)\), collectively denoted by $P^{\alpha,r}(t)$, when the same statement applies to either side. Row \(i\) of \(P^{\alpha,\mathrm{export}}(t)\) captures the distribution of economy \(i\)'s exports across destination economies, whereas row \(i\) of \(P^{\alpha,\mathrm{import}}(t)\) gives the distribution of its imports across origin economies.

Now, given two graphs $G$ and $G'$ on the same node set, with transition matrices $P^{G}$ and $P^{G'}$, the node-level Jensen--Shannon divergence between their outgoing distributions at node $i$ is

\begin{equation}
    \mathrm{JSD}_i = H \!\left(\frac{P_i^{G} + P_i^{G'}}{2}\right) - \frac{1}{2}\bigl[H(P_i^{G}) + H(P_i^{G'})\bigr].
    \label{eq:jsd}
\end{equation}
By construction, $\mathrm{JSD}_i \in [0,1]$. If both rows are zero, then $\mathrm{JSD}_i=0$. If one row is zero and the other is non-zero, then $\mathrm{JSD}_i>0$ recording the appearance or disappearance of an outgoing profile. For two non-zero rows, $\mathrm{JSD}_i = 0$ if and only if $P^{G}_i = P^{G'}_i$, namely the two distributions are identical, while $\mathrm{JSD}_i=1$ occurs when the two distributions (namely the two rows) have disjoint support.
For two non-zero transition profiles, the square root of Jensen--Shannon divergence defines a metric \citep{endres2003new}. Here, we extend the Jensen--Shannon divergence to the case involving rows of zero with the following convention: when both rows are zero, their distance is zero; when exactly one row is zero, Eq.~\eqref{eq:jsd} gives $\mathrm{JSD}_i=\frac{1}{2}$. This extension preserves the metric properties of row-level distance. Averaging these distances over the common node set defines the JS-Walk distance.

\begin{equation}
    D(P^G, P^{G'}) = \frac{1}{N} \sum_{i} \sqrt{\mathrm{JSD}_i}.
    \label{eq:jswalk-dist}
\end{equation}
The corresponding similarity of the two graphs $G$ and $G'$ can be written as: 
\begin{equation}
S(G, G') \equiv
    S(P^G, P^{G'}) = 1 - D(P^G, P^{G'}) \;\in\; [0,1].
    \label{eq:jswalk-sim}
\end{equation}
JS-Walk provides a unified algebraic formulation that naturally applies to directed and undirected, weighted and unweighted graphs. Moreover, being based on graph transition matrices, the comparison of two graphs contains no free tuning parameter. Finally, the resulting similarity is bounded between $0$ and $1$, making values directly comparable across products, aggregation levels and years. Notice also that the distance decomposes additively over economies through the terms $\sqrt{\mathrm{JSD}_i}$, allowing the node-level contributions underlying the resulting layer similarity to be identified from the same computation. We use $S(\cdot,\cdot)$ as the key metric in all subsequent analyses of layer evolution (Section~\ref{sec:temporal}), product communities (Section~\ref{sec:communities}), and economy-level structural roles (Section~\ref{sec:economy_results}).
\FloatBarrier

\subsection{Temporal analysis framework}
\label{sec:temporal_method}

We first use JS-Walk to quantify how the structure of trade layers evolves over time. Since the data are observed annually, we treat the GTN as a sequence of network snapshots, a common representation for time-resolved network data observed at discrete intervals \citep{holme_temporal_2012}. As defined in Section~\ref{sec:data}, $t=1$ corresponds to year 1995 and $t=T$ corresponds to year 2024. For a fixed aggregation level (i.e., product, chapter or section), or the all-products aggregate, and for $r\in \{\mathrm{export}, \mathrm{import}\}$, where $r$ indexes the export or import side, we write\(\{P^r_t\}_{t\in\mathcal{T}}\) for the corresponding sequence of side-specific annual transition matrices.  

For each sequence of annual networks, we have evaluated graph similarity using three temporal references.
First, the year-on-year similarity is defined as
\begin{equation}
    S_{\mathrm{yoy}}^r(t) = S(P^r_t, P^r_{t-1}),\qquad t=2,\ldots,T.
    \label{eq:temporal-yoy}
\end{equation} which compares each annual network with that of the preceding year capturing, in this way, short-run structural persistence. The similarity with respect to the first year in the dataset is defined as 
\begin{equation}
    S_{\mathrm{ini}}^r(t) = S(P^r_t, P^r_1), \qquad t=2,\ldots,T.
    \label{eq:temporal-initial}
\end{equation}
measures cumulative drift away from the beginning of the sample, where $P^r_1$ is the transition matrix for 1995. Finally, we compare each annual transition matrix with the transition matrix associated with the trade flows aggregated over the full observation period. We first define:

\begin{equation}
    \overline{\mathbf{W}}^{\,r}=\sum_{t \in \mathcal{T}}\mathbf{W}_t^r
\end{equation}
and then obtain:
\begin{equation}
    \overline{P}^{\,r} = \mathcal{R}\!\left(\overline{\mathbf{W}}^{\,r}\right),
\label{eq:temporal-average}
\end{equation}
and compute: 
\begin{equation}
    S_{\mathrm{avg}}^r(t) = S(P^r_t,\overline{P}^r).
    \label{eq:temporal-average-similarity}
\end{equation}
which measures deviations from the product-, chapter-, section-, or aggregate-specific long-run profile.

As a supplementary chapter-level summary of short-run persistence, we also compute the year-on-year similarity series for each official HS chapter and each side $r \in \{\mathrm{export},\mathrm{import}\}$. Products are first aggregated into the chapter-level trade layer $\mathbf{W}^\alpha(t)$ as defined in Section~\ref{sec:data}, and the same year-on-year measure $S_{\mathrm{yoy}}^r(t)$ is then evaluated for each chapter. In the Supplementary Material, we report the minimum, maximum, interquartile range, median, and mean of these consecutive-year comparisons. This provides a complete analysis of how short-run persistence varies across the HS hierarchy.

For the aggregate export and import networks, we also compare the observed temporal similarities with a year-label permutation null model \citep{good_permutation_2005}. This null model preserves the observed annual networks but randomly reshuffles their chronological order. It therefore provides a baseline for the similarity patterns expected if the same set of annual network states were observed, but without their actual temporal sequence. Observed similarities are reported together with the null mean and the corresponding 95\% confidence band in Supplementary Material.
\FloatBarrier

\subsection{Product similarity analysis}
\label{sec:community}

The HS taxonomy provides an administrative classification of traded goods, but this classification does not necessarily align with the trade-flow structure revealed by actual trade patterns. A natural question is therefore whether the communities induced by actual trade flows are related to this official taxonomy, and, where they diverge, how. Building on the product layers defined in Section~\ref{sec:data}, JS-Walk provides a way to address this question quantitatively: the similarity between two layers measures how similar the two products are in the way they organise trade across economies. Clustering on top of this similarity recovers product communities endogenously, from trade flows alone.
Applying JS-Walk to every pair of products yields the product-level similarity matrix
\begin{equation}
    S_{a,b}^{r}(t) = S\!\left(P^{a,r}(t), P^{b,r}(t)\right),
    \qquad a,b \in \mathcal{L}.
    \label{eq:Sprod}
\end{equation}
By construction, {$S^r(t)=[S^r_{a,b}(t)]$} is symmetric and bounded in $[0,1]$. The same procedure is applied to the import-side transition matrices when constructing import-side product similarities. This matrix is then used as the affinity matrix for the clustering procedures described below.

\begin{figure}[!h]
\centering
\includegraphics[width=5.5in]{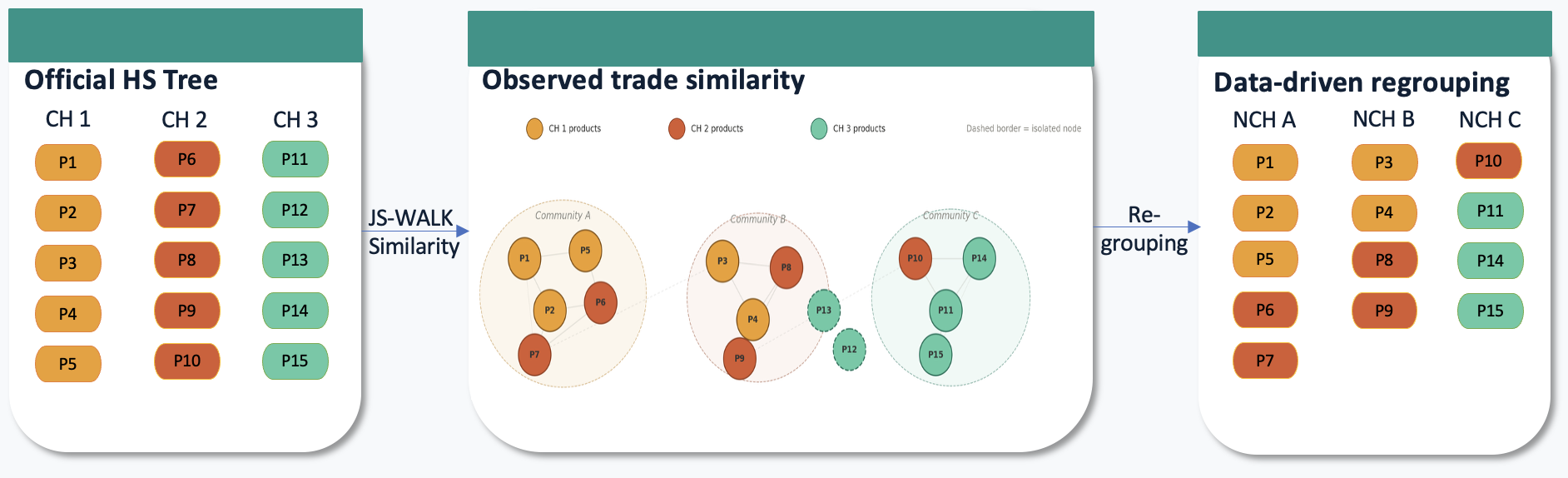}
\caption{
    \textbf{Schematic of the JS-Walk--based reconstruction of the HS taxonomy.}
    \textit{Left:} the official HS tree, in which products $P_1,\dots,P_{15}$ are administratively assigned to three chapters $\mathrm{CH}\,1$--$\mathrm{CH}\,3$.
    \textit{Centre:} pairwise JS-Walk similarities computed from product-level transition matrices place the same products into three communities recovered by spectral clustering. The recovered communities do not necessarily coincide with the official chapters and may mix products from different HS groups.
    \textit{Right:} reading off the community labels yields a data-driven regrouping into new chapters; the same procedure applied at the next level yields new sections. Note how the diagram is illustrative; the empirical application uses 5{,}021 products, 96 new chapters and 21 new sections.}
\label{fig:regrouping_schematic}
\end{figure}

We recover product communities at two levels of resolution through a constrained bottom-up procedure that mirrors the nesting of the HS taxonomy. The logic of the procedure is summarised in Fig.~\ref{fig:regrouping_schematic}: products are first compared through their pairwise JS-Walk similarities, clustered into new chapters, and then aggregated again to obtain new sections.

At the lower level, spectral clustering \citep{ng2002spectral} is applied directly to $S$ as a precomputed affinity matrix with a fixed number of clusters $k = 96$ (matching the number of active HS chapters in the data), yielding the \emph{new-chapter} labels for the 5{,}021 products. Spectral clustering is appropriate here because it accepts a precomputed similarity matrix, allows the cluster count to be fixed exactly, and rests on the same random-walk foundation as JS-Walk through the normalised graph Laplacian~\citep{ng2002spectral}. 

At the upper level, to obtain a chapter-level transition matrix, products within each new chapter are aggregated into a single composite layer by summing their adjacency matrices and then row-normalising the resulting aggregate matrix. Pairwise JS-Walk between these new-chapter transition matrices yields a chapter-level similarity matrix on which spectral clustering with $k = 21$ produces the \emph{new-section} labels. Aggregating at the adjacency level rather than averaging already-stochastic transition matrices preserves the interpretation of treating a new chapter as a single composite trade layer. By construction, every new chapter is contained in exactly one new section, so the recovered hierarchy is strictly nested in the same sense as the official HS.

To put the constrained reconstruction in context, we benchmark it against the Leiden community detection method~\citep{traag2019leiden} applied to the weighted similarity graph derived from $S$. Leiden is a community-detection algorithm that optimizes a generalised modularity and lets the number of communities emerge from the data. It searches for groups of nodes that are more strongly connected to each other than expected under a null model, and it improves the partition through local node moves, refinement, and aggregation steps. We sweep its resolution parameter $\gamma \in \{0.5, 1.0, 1.5, 2.0, 3.0, 5.0, 10.0\}$ and retain the partition with the highest modularity. 
The two methods address related but distinct questions. Spectral clustering finds a partition into a pre-specified number of groups, while Leiden returns a data-driven number of communities. Since our aim is to provide a trade-flow-based classification alternative to the official hierarchy, we chose to use Leiden as the benchmark to ensure that our results were not random or resulted from data noise. At each level of the hierarchy we report three Adjusted Rand Indices (ARI) \citep{hubert1985comparing}: ours versus HS, Leiden versus HS, and ours versus Leiden. 
These comparisons allow to investigate two different things: not only the agreement with the administrative taxonomy, but also the agreement between the two clustering algorithms. 

Beyond the ARIs comparisons, we further characterise how the reconstructed taxonomy differs from the official HS hierarchy. In the main text, we focus on the section-level correspondence between the two classifications, which provides an interpretable overview of where the administrative and trade-flow-based taxonomies align or diverge. Finer chapter-level fragmentation and the temporal evolution of this divergence are reported in the Supplementary Material.

First, a $21 \times 21$ section--community heatmap is computed from the reconstructed taxonomy obtained from the 1995--2024 time-averaged product similarity matrix. Each row corresponds to an official HS section and each column to a reconstructed new section; cell $(o,s)$ counts the number of official HS chapters in section $o$ whose products are predominantly assigned to new section $s$. A row-concentrated pattern indicates that the trade-flow taxonomy largely preserves the administrative grouping, whereas dispersed rows indicate where an official HS section is split across several reconstructed sections.

As a finer supplementary analysis, we also examine chapter fragmentation. For each official HS chapter, we record the number of reconstructed new chapters to which its products are assigned, together with the share of its products assigned to the dominant reconstructed chapter. Chapters with products concentrated in one reconstructed chapter are interpreted as relatively coherent, whereas chapters whose products are spread across many reconstructed chapters are interpreted as fragmented. This analysis identifies broad or heterogeneous HS chapters whose products are reorganised by trade-flow similarity.

As an additional analysis reported in the Supplementary material, we compute, for each new section $s$, a yearly trade-value divergence from the official taxonomy,
\begin{equation}
    \Delta_s(t) = 1 - \max_{o \in \mathcal{O}}
        \frac{V_{s,o}(t)}{V_s(t)},
    \label{eq:divergence}
\end{equation}
where $V_s(t)$ is the total trade value in new section $s$ in year $t$, $V_{s,o}(t)$ is the trade value that falls inside official HS section $o$, and $\mathcal{O}$ ranges over the 21 HS sections. Tracking $\Delta_s(t)$ across 1995--2024 allows us to investigate whether the alignment between flow-based and administrative taxonomies is stable, drifting, or punctuated by structural breaks.
\FloatBarrier

\subsection{Economy influence indexes}
\label{sec:influence}

The previous sections compare trade layers and recover product communities from trade-flow similarity. We now move from layer-level structure to economy-level influence. The objective is not to rank economies by trade volume, but to quantify how strongly the removal of an economy perturbs the centrality structure of the Global Trade Network. Hence, an economy is influential if, when its participation in a given product or chapter layer is removed, the PageRank centralities of other economies change substantially. 

\paragraph{Layer-level perturbation.}
For a fixed aggregation level, let $\mathcal{A}$ denote the set of layers used in the influence calculation. In the main product-level analysis, $\mathcal{A}=\mathcal{L}$ is the set of product layers; in chapter- or section-level analyses, $\mathcal{A}$ denotes the corresponding set of aggregated layers defined in Section~\ref{sec:data}. We index a generic layer in this set by $\alpha \in \mathcal{A}$. 
For each layer $\alpha$, year $t$, and side $r \in \{\mathrm{export},\mathrm{import}\}$, the influence analysis uses the side-specific trade-flow matrix $\mathbf{W}^{\alpha,r}(t)$ defined in Eqs.~\eqref{eq:export-side} and \eqref{eq:import-side}. Thus, the export side uses the exporter--importer matrix, whereas the import side uses its transpose.
Let $c^{\alpha,r}_{j,t}$ denote the PageRank centrality of economy $j$ computed from $\mathbf{W}^{\alpha,r}(t)$ \citep{page1999pagerank,gleich2015pagerank}. PageRank is used here only as the centrality measure in the leave-one-out influence calculation. We use the standard damping factor $d=0.85$ and a uniform teleportation distribution over all economies. Nodes with zero outgoing strength are treated using the standard PageRank convention, with their probability mass redistributed uniformly across the node set. This differs from JS-Walk, which compares the original row-normalised trade-partner profiles directly, without teleportation or dangling-node redistribution.
The layer-level influence of economy $i$ is defined by a leave-one-out perturbation:
\begin{equation}
    I^{\alpha,r}_{i,t}
    =
    \frac{1}{N-1}
    \sum_{j\neq i}
    \left|
        c^{\alpha,r}_{j,t}
        -
        c^{\alpha,r(-i)}_{j,t}
    \right|,
    \label{eq:loo-pr-impact}
\end{equation}
where 
$c^{\alpha,r(-i)}_{j,t}$ is the PageRank centrality of economy $j$ after economy $i$ is removed from that layer. Operationally, removal is implemented by setting both the row and the column of economy $i$ to zero while keeping it in the node set as an isolated node. This corresponds to an embargo-like counterfactual: the economy remains part of the world system but ceases to participate in the relevant trade layer.
The quantity in Eq.~\eqref{eq:loo-pr-impact} is a network-perturbation measure. It does not quantify the trade volume of economy $i$, nor the PageRank centrality of $i$ itself. Rather, it measures how much the PageRank centralities of all other economies change when $i$ is removed from the corresponding export or import side.

\paragraph{Aggregating influence across layers.}
Economies participate unevenly across products and chapters. To aggregate layer-level impacts, we weight each layer by its importance in the economy's own trade portfolio. On the export side, the layer weight is the share of layer $\alpha$ in economy $i$'s total exports:
\begin{equation}
    \omega^{\alpha,\mathrm{export}}_{i,t} = \frac{\sum_j w^{\alpha}_{ij}(t)}{\sum_{\beta \in \mathcal{A}}\sum_j w^{\beta}_{ij}(t)}.
    \label{eq:export-share-weight}
\end{equation}
On the import side, the layer weight is the share of layer $\alpha$ in economy $i$'s total imports:
\begin{equation}
    \omega^{\alpha,\mathrm{import}}_{i,t} = \frac{\sum_j w^{\alpha}_{ji}(t)}{\sum_{\beta \in \mathcal{A}}\sum_j w^{\beta}_{ji}(t)}.
    \label{eq:import-share-weight}
\end{equation}
If economy $i$ has zero total trade on side $r$ in year $t$, we set $\omega^{\alpha,r}_{i,t}=0$ for all $\alpha \in \mathcal{A}$.
These weights ensure that a layer contributes strongly to an economy's aggregate influence only when that layer is economically relevant for that economy.
The annual baseline influence of economy $i$ in side $r$ is obtained by summing its weighted PageRank impact across all layers:
\begin{equation}
    I^{r}_{i,t}
    =
    \sum_{\alpha \in \mathcal{A}}
    \omega^{\alpha,r}_{i,t}
    I^{\alpha,r}_{i,t},
    \qquad r \in \{\mathrm{export},\mathrm{import}\}.
    \label{eq:annual-baseline-influence}
\end{equation}
The export version combines export-share weights with PageRank impact computed on the exporter--importer graph. The import version combines import-share weights with PageRank impact computed on the transposed importer--exporter graph. Both are view-consistent: each influence measure uses the same orientation as the corresponding JS-Walk trade profile.
For a full-period summary, we define total baseline influence over the sample period $\mathcal{T}$
\begin{equation}
    I^{r,\mathrm{tot}}_{i}
    =
    \sum_{t=1}^{T}
    \sum_{\alpha \in \mathcal{A}}
    \omega^{\alpha,r}_{i,t}
    I^{\alpha,r}_{i,t}.
    \label{eq:total-baseline-influence}
\end{equation}
In the empirical analysis, annual values are retained whenever temporal dynamics are of interest, because cumulative sums obscure year-to-year variations. The baseline measures introduced in Eq.~\eqref{eq:annual-baseline-influence} and Eq.~\eqref{eq:total-baseline-influence} therefore combine  two components: the structural effect of removing economy $i$ from each layer, measured by $I^{\alpha,r}_{i,t}$, and the economic relevance of that layer to the economy, measured by $\omega^{\alpha,r}_{i,t}$. They do not, however, distinguish whether this structural influence is associated with trade-partner configurations that remain stable over time or with configurations that are themselves undergoing substantial change. For this reason, we also consider the following measure.

\paragraph{Shock-modulated influence.}
To add this temporal dimension to the baseline measures, we exploit the node-resolved structure of JS-Walk introduced in Section~\ref{sec:jswalk}. For each economy and layer, the row-level JS-Walk distance between consecutive years measures how much that economy's distribution of trade partners has changed. We use this quantity to modulate the corresponding layer-level removal impact. In this way, a layer contributes strongly to the shock-modulated influence only when economy $i$ is both structurally influential in that layer and undergoing a substantial change in its own trade-partner profile.

For each economy $i$, layer $\alpha$, side $r$, and year $t>1$, we define:
\begin{equation}
    s^{\alpha,r}_{i,t}
    =
    \sqrt{
    \mathrm{JSD}
    \!\left(
        P^{\alpha,r}_{i,t},
        P^{\alpha,r}_{i,t-1}
    \right)
    },
    \label{eq:node-shock}
\end{equation}
where $P^{\alpha,r}_{i,t}$ denotes economy $i$'s transition profile in layer $\alpha$, side $r$, and year $t$, as defined in Section~\ref{sec:jswalk}. Accordingly, $s^{\alpha,r}_{i,t}$ is the row-level JS-Walk distance between the economy's trade-partner profiles in two consecutive years.
The annual shock-modulated influence can be defined as:
\begin{equation}
    I^{r,\mathrm{shock}}_{i,t}
    =
    \sum_{\alpha \in \mathcal{A}}
    \omega^{\alpha,r}_{i,t}
    I^{\alpha,r}_{i,t}s^{\alpha,r}_{i,t},
    \qquad t>1.
    \label{eq:annual-shock-influence}
\end{equation}
The corresponding influence over the whole period is: 
\begin{equation}
    I^{r,\mathrm{shock,tot}}_{i}
    =
    \sum_{t=2}^{T}
    \sum_{\alpha \in \mathcal{A}}
    \omega^{\alpha,r}_{i,t}
    I^{\alpha,r}_{i,t}s^{\alpha,r}_{i,t}.
    \label{eq:total-shock-influence}
\end{equation}
where the time summation starts at $t=2$ (corresponding to year 1996), because the shock term compares year $t$ with year $t-1$ and is undefined for the first year of the sample.

The baseline and shock-modulated measures therefore answer complementary questions. The baseline score identifies economies whose removal strongly perturbs the network across economically relevant layers, irrespective of whether those layers are temporally stable. The shock-modulated score gives greater weight to the subset of this influence that coincides with changes in the economy's own trade-partner configuration. It should therefore be interpreted as a dynamic modulation of baseline influence, rather than as an independent measure of influence.

The side-specific formulation above is the main specification used in this paper: export-side influence combines PageRank impact on the exporter--importer graph with export-share weights and export-side JS-Walk shocks, while import-side influence combines PageRank impact on the transposed importer--exporter graph with import-share weights and import-side shocks.
The same two graph orientations can also be read from a role-based perspective. The exporter--importer graph follows trade flows toward destination markets and therefore emphasises perturbations in the downstream flow structure. The transposed importer--exporter graph follows origin profiles from importers back to exporters and therefore emphasises perturbations in the upstream source structure. Consequently, if the research question is reframed from export--import sides to supplier--demand roles, the same leave-one-out PageRank logic can be used to construct alternative role-based variants. A supplier-side version would evaluate PageRank impact on the transposed graph while retaining export-side weights and shocks, whereas a demand-side version would evaluate PageRank impact on the original graph while retaining import-side weights and shocks. We treat this as a possible sensitivity extension rather than the main specification, because the main framework is designed to remain consistent with the export and import sides used in the JS-Walk analysis.


For each aggregation level, we compute annual scores and annual ranks for baseline and shock-modulated influence on both export and import sides, with rank one denoting the highest influence in a given year. The main empirical presentation focuses on these annual product-level rankings, together with a cross-sectional comparison of baseline influence in the export and import sides in the final year of the sample. When a single full-period summary is required, we use the cumulative quantities defined in Eqs.~\eqref{eq:total-baseline-influence} and~\eqref{eq:total-shock-influence}, rather than mean annual scores.

Additional results are reported in the Supplementary Material. These include chapter-level robustness checks and comparisons with the fitness index introduced in Ref.~\citep{tacchella_new_2012} computed from the yearly RCA matrix. The fitness comparison is used only as an external reference for productive capability and product complexity, not as a benchmark method for network influence.



\section{Results}
\label{sec:results}

\subsection{Temporal dynamics in the global-wide trading}
\label{sec:temporal}

\begin{figure}[!h]
\centering\includegraphics[width=6.9in]{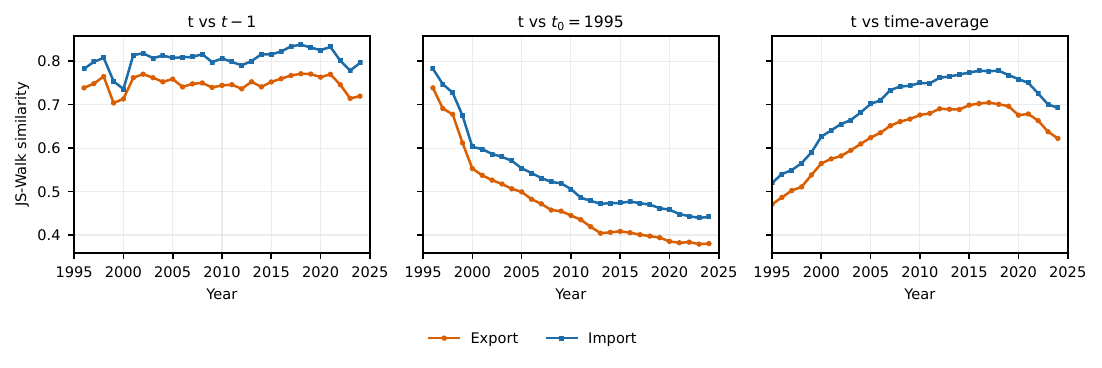}
\caption{\textbf{Temporal evolution of the GTN on the export and import sides.} Each panel reports JS-Walk similarity $S$ between annual networks aggregated across all products over 1995--2024, evaluated under three temporal references: consecutive years (\emph{left}, $t$ vs.\ $t-1$), the first year of the sample (\emph{centre}, $t$ vs.\ $t_0 = 1995$), and the full-period aggregate transition matrix (\emph{right}, $t$ vs.\ $\overline{P}$). The export-side series (orange) and import-side series (blue) are overlaid in each panel. The import side is constructed from the row-normalised transposed adjacency matrix, so that rows represent importing economies and columns represent import origins. The year-on-year panel captures short-run persistence, the fixed-initial-year panel captures cumulative drift from the beginning of the sample, and the full-period aggregate panel shows deviations from the long-run aggregate trade profile.}
\label{fig:temp_aggregate}
\end{figure}

We first examine how the aggregate Global Trade Network evolves over time when all products are combined into a single annual export/import network. Figure~\ref{fig:temp_aggregate} reports JS-Walk similarity for the aggregate export and import networks under three temporal references: year-on-year comparison, comparison with the initial network, and comparison with the full-period aggregate transition matrix. 

The year-on-year similarity remains persistently high throughout the sample on both export and import sides. This indicates that, at the annual scale, the aggregate GTN is structurally persistent: consecutive years preserve similar trade-partner transition profiles. The import side lies systematically above the export side, showing that import-origin profiles are more stable than export-destination profiles. In other words, the set of suppliers from which economies import changes more slowly than the set of destination markets to which exporters sell.

The fixed-initial-year comparison reveals a different pattern. Similarity to 1995 declines gradually over the sample, indicating cumulative structural drift away from the initial configuration. This decline is faster in the early years and more gradual thereafter, suggesting that the aggregate GTN evolves through the accumulation of small year-to-year changes rather than through frequent discontinuous rewiring. The time-average comparison complements this interpretation. Similarity to the long-run average rises from the early years, remains relatively high around the middle of the sample, and declines toward the end. This pattern indicates that the beginning and end of the sample are both relatively far from the long-run average profile, consistent with a gradual transition in the aggregate trade structure over the 1995--2024 period.

The corresponding year-label permutation null model is reported in Supplementary Material. The null model preserves the set of annual networks but destroys their chronological order. The observed year-on-year similarities lie above the null mean, confirming that consecutive years are more similar than would be expected from randomly ordered annual network states. The fixed-initial-year trajectories also display a systematic decline that is not reproduced by the nearly flat null mean, indicating that the long-run drift is tied to the chronological ordering of the observed networks.

\begin{figure}[!h]
\centering\includegraphics[width=6.5in]{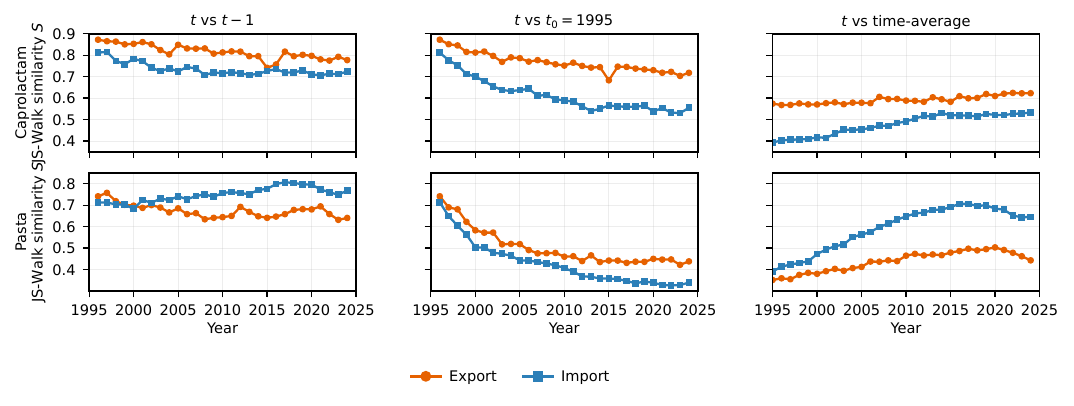}
\caption{\textbf{Examples of temporal evolution of the GTN on the export and import sides at product-level.}  JS-Walk similarity is shown, as example, for two selected products, namely ``\texttt{Caprolactam}'' (HS 293379) and ``\texttt{Pasta}'' (HS 190219). Columns report the three temporal references ($t$ vs.\ $t-1$, $t$ vs.\ $t_0$, and $t$ vs.\ the full-period aggregated transition matrix), while rows correspond to the two products. Export (orange) and import (blue) sides are overlaid in each panel. The y-axis range differs across product rows to make within-product temporal variation visible.
}
\label{fig:temp_case_product}
\end{figure}

To assess whether this aggregate pattern hides product-level heterogeneity, Figure~\ref{fig:temp_case_product} compares two product layers as examples: ``\texttt{Caprolactam}'' and ``\texttt{Pasta}''. Both products show long-run drift away from 1995, but their temporal profiles differ substantially from the aggregate and from each other. ``\texttt{Caprolactam}'' displays higher export-side than import-side similarity across the three references, reversing the aggregate pattern in which the import side is more stable. ``\texttt{Pasta}'', by contrast, shows a lower import-side similarity to the initial configuration than the aggregate, together with a different pattern of year-on-year fluctuations. These contrasts show that the aggregate trajectory might not be representative of all product layers. Instead, the aggregate GTN should be understood as the superposition of many product-specific layers whose temporal dynamics can differ markedly in stability, asymmetry, and volatility.

In the Supplementary Material we show that this heterogeneity is not limited to the two selected product examples. A chapter-level persistence summary reports the distribution of year-on-year JS-Walk similarity for each official HS chapter after aggregating all products within that chapter. The variation across chapters in the range, mean and median similarity confirms that the high persistence observed in the aggregate GTN masks substantial heterogeneity across lower-level trade layers.

The additional chapter- and section-level case studies shown in the Supplementary Information reinforce this interpretation. ``\texttt{Organic Chemicals}'' and ``\texttt{Cereals}'', as well as ``\texttt{Chemicals}'' and ``\texttt{Food, drink \& tobacco}'', display the same broad distinction between short-run persistence and long-run drift, but differ in the magnitude of their year-on-year stability and in the separation between export and import sides. These supplementary analyses show that aggregation smooths product-specific and section-specific dynamics, making the multilayer representation essential for understanding how different parts of the trade system evolve.

\subsection{Emergent trade-based product taxonomy}

\label{sec:communities}
\begin{figure}[p]
\centering
\includegraphics[width=6.5in]{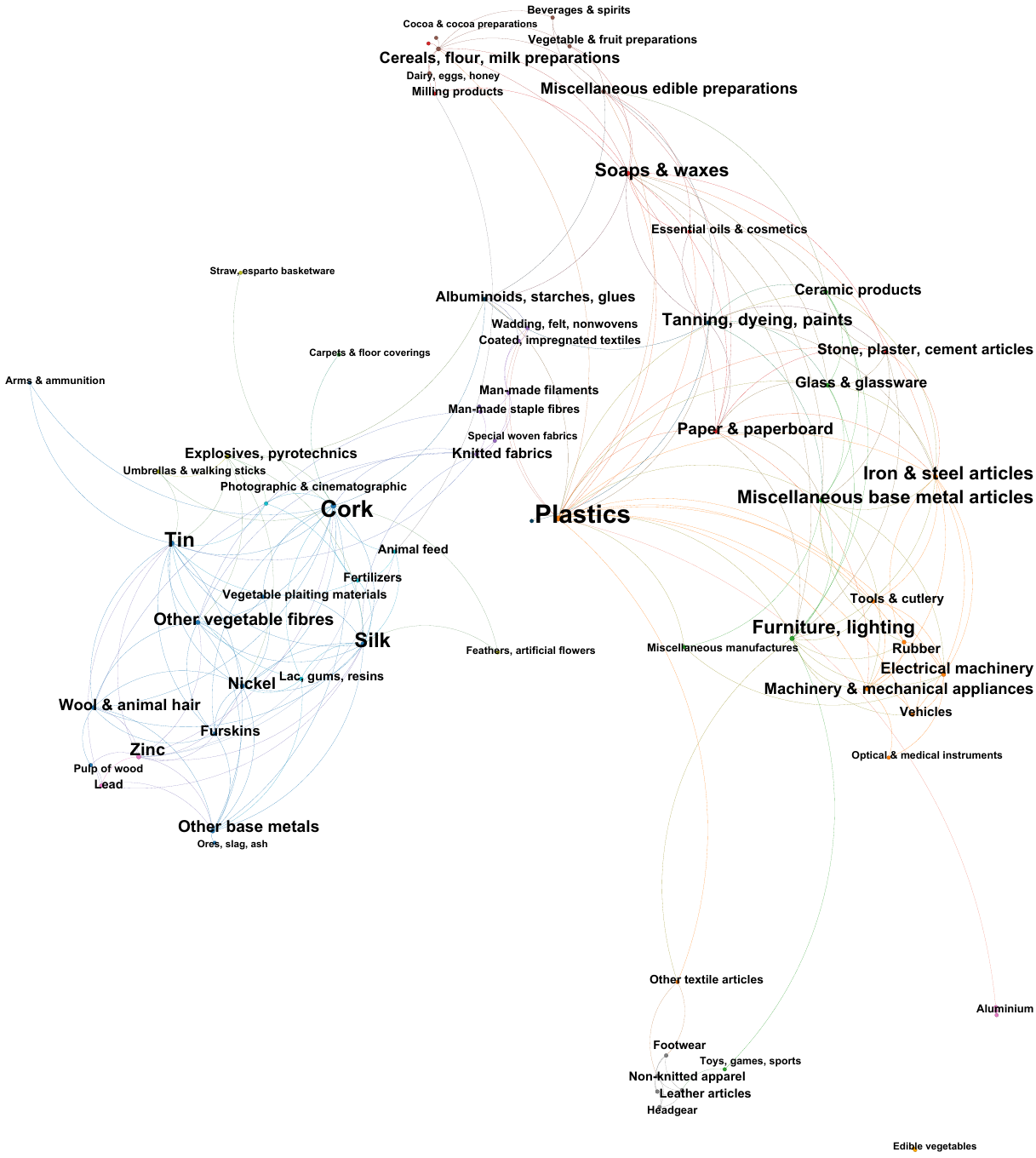}
\caption{\textbf{Community structure in the trade-flow similarity network.} Network nodes represent the 96 official HS product chapters present in the data, and are sized by total chapter-level trade volume in year 2010. Edges indicate pairwise JS-Walk similarity between chapter-level export-flow networks. For visual clarity, only the edges with the $5\%$ largest similarity values are shown. Node colours denote the communities obtained by spectral clustering, where the number of clusters was fixed to $k=21$, equal to the number of official HS sections. 
(Layout: ForceAtlas2 in Gephi).}
\label{fig:motivation_chapter_export}
\end{figure}

We now examine whether products that are close in trade-flow structure are also close in the official HS taxonomy. As a coarse-grained preview of the product-level analysis, Fig.~\ref{fig:motivation_chapter_export} shows an export-side similarity network between official HS chapters in 2010. Nodes represent official HS chapters, while edges retain only the strongest (i.e., top $5\%$) pairwise JS-Walk similarities for visual clarity. If the HS taxonomy fully captured trade-flow organisation, the similarity network would be expected to align closely with official HS sections. Instead, the network already shows cross-category structure: some chapters from different HS sections are connected by strong trade-flow similarity, suggesting that products can be administratively distant but structurally close in the GTN. For example, ``\texttt{Machinery and mechanical appliances}'' and ``\texttt{Electrical machinery}'' cluster with ``\texttt{Optical and medical instruments}'', while ``\texttt{Cork}'' groups with several base-metal and natural-fibre chapters. This motivates the systematic HS6-level reconstruction and evaluation reported below.

We then reconstruct product groups directly from trade-flow similarity. The HS taxonomy groups products according to material, use, and customs definitions, whereas our approach groups products according to the similarity of their economy-to-economy trade profiles. Agreement between the two classifications would indicate that the administrative taxonomy captures the main organisation of trade flows; disagreement would indicate that products from different HS categories are traded through similar global structures.
%
Table~\ref{tab:clustering_evaluation} quantifies this relationship using Adjusted Rand Index (ARI) scores. The ARI values are positive and far above the random baseline, indicating that the HS taxonomy and the flow-based taxonomy are not unrelated. However, the values remain moderate rather than high. This is an important result: while the HS classification contains some information about trade-flow structure, it does not fully determine it. Products that are similar in customs terms are not always traded through similar networks, and products from different HS categories may share common trade-flow patterns.

\begin{table}[!h]
\caption{\textbf{Comparing clusters of products with the official HS taxonomy.} \emph{Spectral} uses $k$ matching the number of active HS groups at each level: 96 active chapters and 21 sections. Chapter 77 is reserved and absent from the active product chapters. \emph{Leiden} finds the value of $k$ by modularity optimization on a $k$-NN sparsified similarity graph ($k_{\text{NN}}=30$); the obtained $k$ is shown in parentheses. \emph{Spectral@Leiden's $k$} applies spectral clustering,  with a value of $k$ fixed to be equal to the one of the Leiden algorithm, for a head-to-head comparison free of cluster-count effects. \emph{Random p95} is the 95th-percentile ARI from 100 random clusterings at HS-matched $k$, marking the expected noise levels. All non-random methods exceed the random expectations by more than two orders of magnitude. Bold entries indicate the head-to-head winner.}
\label{tab:clustering_evaluation}
\centering
\begin{tabular}{|l|l|l|l|l|l|}
\hline
\textbf{side} & \textbf{level} & \textbf{Spectral} & \textbf{Leiden} & \textbf{Spectral@Leiden's k} & \textbf{Random p95}\\
\hline
Export & Chapter & 0.154 & 0.039 (k=8) & \textbf{0.083} & 0.0004\\
\hline
Export & Section & 0.106 & 0.062 (k=8) & \textbf{0.148} & 0.0005\\
\hline
Import & Chapter & 0.164 & \textbf{0.135} (k=9) & 0.107 & 0.0004\\
\hline
Import & Section & 0.107 & \textbf{0.183} (k=9) & 0.163 & 0.0005\\
\hline
\end{tabular}
\end{table}

\begin{figure}[!h]
\centering
\includegraphics[width=5.5in]{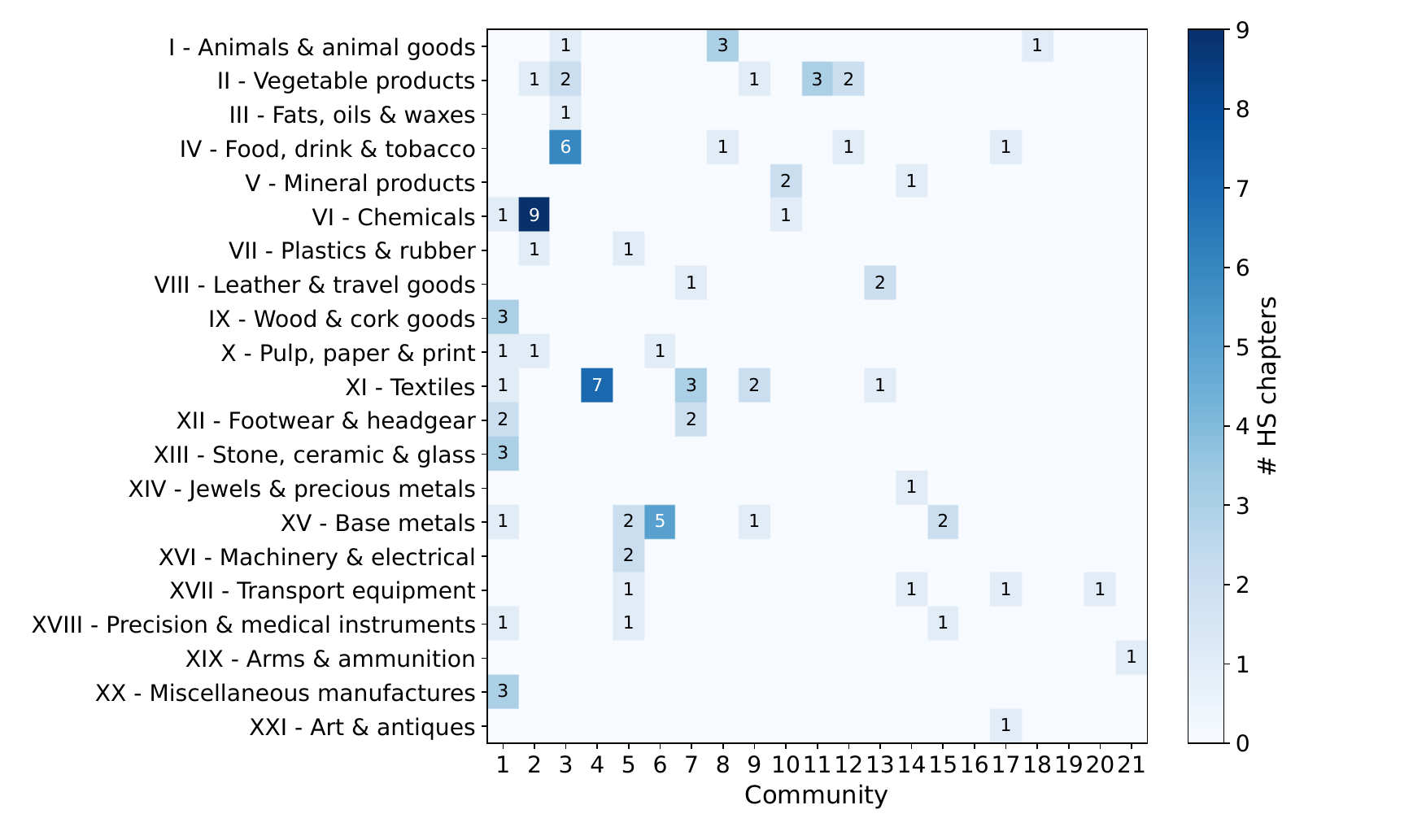}
\caption{\textbf{Official HS sections versus trade-network based product communities.} Heatmap comparing official HS sections with the communities obtained from export flow similarities in Fig.~\ref{fig:motivation_chapter_export}. Rows, indicating official HS sections, are labelled by Roman numerals and descriptive names, while columns correspond to communities and are reordered by decreasing number of product chapters they contain. Cell $(x, y)$ reports the number of HS product chapters belonging to official section $y$, that are assigned to community $x$ by the spectral clustering.}
\label{fig:cd3_contingency_export}
\end{figure}

The comparison between spectral clustering and Leiden reveals an export--import asymmetry. On the export side, spectral clustering achieves stronger agreement with the official taxonomy than Leiden, both at the chapter and section levels. On the import side, Leiden performs better or comparably in the fair-$k$ comparison. This suggests that export and import structures organise products according to different principles. Export-side similarity may reflect producer capabilities, industrial specialisation, and supply-chain position, whereas import-side similarity may reflect demand-side complementarities and common consumption or input bundles. The section--community matrix in Figure~\ref{fig:cd3_contingency_export} shows where the flow-based taxonomy preserves HS structure and where it reorganises it. Each row corresponds to an official HS section, while each column corresponds to one of the 21 trade-network communities obtained from export-flow similarities. Cell $(x,y)$ counts the number of HS chapters from official section $x$ assigned to community $y$. 
Some official HS sections are relatively coherent, whereas others are more fragmented across several communities. For example, Section~\textsc{VI} (``\texttt{Chemicals}'') has $9$ out of its $11$ chapters assigned to a single community, while Sections~\textsc{II}(``\texttt{Vegetable products}''), \textsc{XI}(``\texttt{Textiles}''), and \textsc{XV}(``\texttt{Base metals}'') are distributed across five communities. This selective alignment supports the idea that the HS taxonomy is neither irrelevant nor sufficient: it captures part of the organisation of trade flows, but the multilayer trade network also reveals additional cross-category structure. 

Complementary analyses are reported in the Supplementary Material. In particular, the analogous import-side section--community matrix shows that import-flow similarities reorganise HS sections differently from the export side, while the yearly divergence analysis indicates that the disagreement between trade-based communities and official HS sections is persistent over time rather than being driven by a few exceptional years.

Overall, the product-community results show that JS-Walk recovers a flow-based taxonomy that is complementary to the HS. The recovered communities partially overlap with the official classification but also reveal cross-HS groupings, fragmentation within broad chapters, and systematic export--import differences. The HS taxonomy remains essential as an administrative and descriptive system, but the multilayer network approach uncovers an additional organisation based on realised trade relationships.

\subsection{Influential actors of the worldwide trading networks}
\label{sec:economy_results}

\begin{figure}[!h]
    \centering
    \includegraphics[width=6.5in]{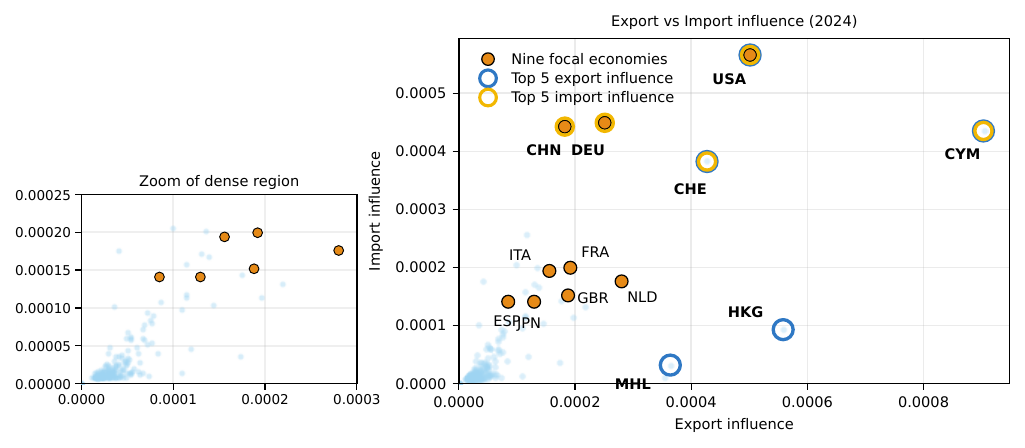}
    \caption{\textbf{Export- versus import-side influence of world economies in 2024.}
    Each point represents an economy's baseline network influence in 2024, with export-side influence on the horizontal axis and import-side influence on the vertical axis. The nine focal economies are highlighted and labelled. Blue open circles mark economies among the top five in export-side influence, while orange open circles mark economies among the top five in import-side influence. The inset enlarges the densely populated lower-left region, where most economies have relatively small influence on both sides. Economies in the upper-right are structurally influential on both sides, whereas differences between export- and import-side scores indicate asymmetry between export-side and import-side network roles.}
    \label{fig:economy_export_import_scatter}
\end{figure}

The final analysis shifts from products to economies, by asking which economies' removal would most perturb the centrality structure of the GTN. Our influence framework is designed to capture network-structural effects rather than trade volume alone: a large economy can be partly substitutable when its links sit on dense, redundant pathways, whereas a smaller economy can be influential if it occupies a structurally pivotal position in specific product layers. As defined in Section~\ref{sec:influence}, baseline influence combines the trade-share weight of each product layer with the leave-one-out PageRank perturbation caused by removing the economy from that layer. The shock-modulated variant further weights this contribution by the year-on-year JS-Walk change in the economy's own trade-partner profile.
Figure~\ref{fig:economy_export_import_scatter} first provides a cross-sectional comparison of baseline influence in the export and import sides in 2024. Economies in the upper-right corner are influential on both sides, while differences between export- and import-side scores indicate asymmetric network roles. The United States is influential on both sides, while China and Germany show particularly high import-side influence relative to their export-side influence. This confirms that the two perturbation effects are not interchangeable: some economies are more influential through export-flow structure, while others are more influential through import-origin structure.

\begin{figure}[!h]
    \centering
    \includegraphics[width=5.5in]{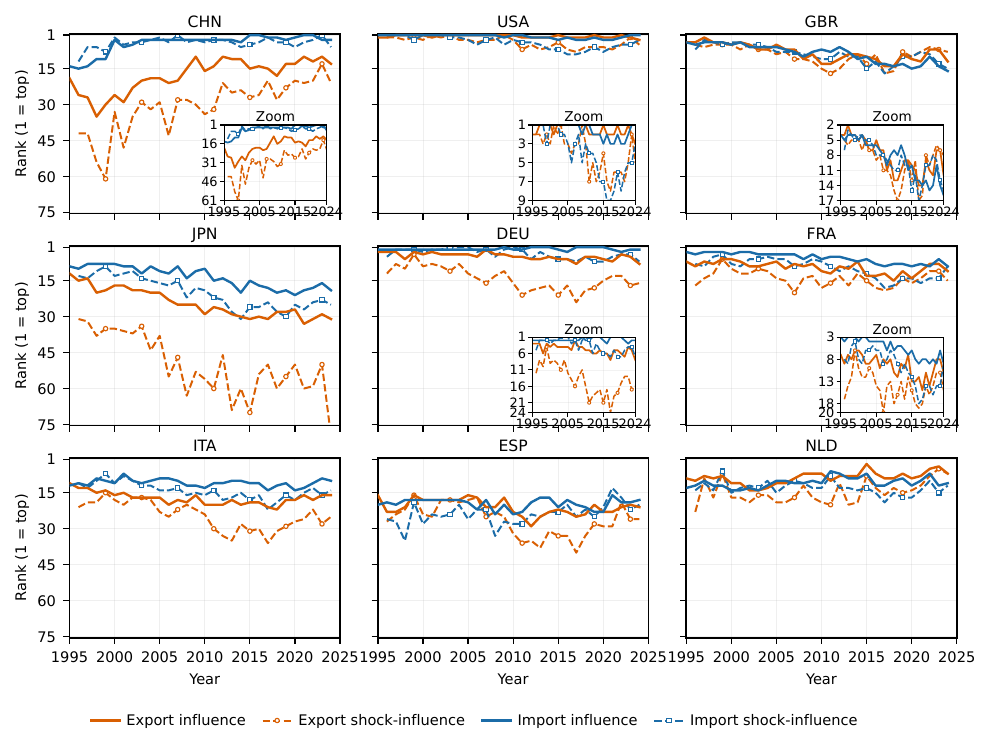}
    \caption{\textbf{Temporal changes in export- and import-side influence ranks of selected world economies.}
    Annual influence ranks are shown for nine selected economies---China, the United States, the United Kingdom, Japan, Germany, France, Italy, Spain, and the Netherlands---over 1995--2024. Orange curves represent export-side influence and blue curves import-side influence. Solid lines show baseline influence ranks, while dashed lines show shock-modulated influence ranks. Rank one denotes the most influential economy in a given year, so curves closer to the top indicate stronger network-removal impact. The baseline measure captures trade-share-weighted leave-one-out PageRank impact, whereas the shock-modulated measure additionally emphasises product layers in which the economy's export-destination or import-origin profile changes between consecutive years. Insets for China, the United States, United Kingdom, Germany, and France enlarge the upper-ranking region where the four trajectories overlap closely. The common main-panel scale allows direct comparison of influence dynamics across economies, while the insets reveal finer differences among closely ranked specifications.}
    \label{fig:economy_influence_all_view}
\end{figure}

We then examine how influence evolves over time for nine major trading economies, namely China, the United States, the United Kingdom, Japan, Germany, France, Italy, Spain, and the Netherlands. These are used as illustrative cases to compare different export- and import-side trajectories. Figure~\ref{fig:economy_influence_all_view} reports the  annual product-level influence ranks under four specifications: baseline and shock-modulated influence, respectively on the export and on the import side. 
Three patterns stand out. First, influence is persistent for some economies but not for all. The United States stays near the top on both sides throughout the period, indicating persistently high network-structural influence, whereas Germany, France, Italy, Japan, the United Kingdom, Spain, and the Netherlands follow more heterogeneous, side-dependent paths. China provides the clearest distinction between trade scale and network structure: its import-side rank rises close to the top, while its export-side rank remains markedly lower, consistent with an influence measure based on network removal rather than trade volume.

Second, the export- and import- sides rankings frequently differ for the same economy, confirming that the two orientations capture distinct forms of network-structural influence. Export-side influence reflects the structural effect of removing an economy from exporter--destination flow patterns, whereas import-side influence reflects its role in importer--origin profiles. Persistent gaps between an economy's orange and blue curves therefore indicate side-specific influence.

Third, the gap between solid and dashed lines captures the dynamic component. Because the ranks are computed relative to the full set of economies, an economy tends to hold or improve its position under shock modulation when its influential layers are also those in which its partner profile is changing, whereas it tends to fall in the ranking when its influence is concentrated in more stable pathways. The shock-modulated rank therefore complements rather than replaces the baseline rank: it indicates whether an economy's structural importance rests mainly on stable trade pathways or on layers undergoing economy-specific reorientation.

The Supplementary Material provides additional checks on these results. Raw-score counterparts show that the rank dynamics are not merely artefacts of relative ordering; chapter-level results preserve the main qualitative patterns; comparisons with the 
fitness index introduced in Ref.~\citep{tacchella_new_2012} show that network-removal influence and productive capability capture distinct structural properties. Overall, network-structural influence in the GTN is multidimensional---shaped by product composition, partner structure, export versus import orientation, and temporal change---and the most influential economies are not necessarily the largest traders but those whose participation sustains central pathways in the multilayer network.

\FloatBarrier

\section{Discussion and Conclusions}
\label{sec:discussion}
An important methodological contribution of this work is the introduction of JS-Walk, a bounded and parameter-free similarity measure for comparing directed and weighted networks through node-level random-walk transition profiles. Its node-resolved construction allows the same framework to be applied across temporal comparisons, product-level clustering, and economy-level structural analysis. In the Global Trade Network, this unified perspective reveals structures that are not apparent from aggregate trade volumes alone, showing that the GTN is not only a system of trade volumes, but a structured and evolving multilayer network.

In particular, our analyses highlighted four main properties of the Global Trade Network. First, the GTN is temporally persistent but not stationary. Consecutive annual networks remain, on average, structurally similar, yet comparisons with the initial network and with the long-run average reveal gradual drift over three decades. This means that aggregate trade structure evolves through cumulative small year-to-year changes rather than through frequent abrupt rewirings. Such a pattern is consistent with the persistence of trade relationships, production capabilities, infrastructure, and institutional ties, while still allowing long-run changes in global production and demand to reshape the network.

Second, lower-level trade heterogeneity is essential. The aggregate network provides a useful summary, but it masks large differences across products, chapters, and sections. The product-level analyses show that individual layers (at different aggregation levels) can depart substantially from the aggregate trajectory in their stability, export--import asymmetry, and long-run drift. A multilayer approach is therefore necessary: collapsing all products into a single network removes precisely the variation that is needed to understand how different parts of the trade system evolve.

Third, the flow-based product communities show that administrative similarity and trade-flow similarity are related but distinct. The HS taxonomy remains a natural reference because it organises products consistently across economies and years. However, the moderate ARI values and the section--community structure show that the official taxonomy does not fully determine the organisation of trade flows. Some HS sections remain relatively coherent, while others are split across several trade-based groups. Supplementary chapter-fragmentation and divergence analysis support the same conclusion: products can be close in the trade network even when they are distant in the HS hierarchy. This suggests that trade-flow similarity captures information about production systems, input--output relationships, demand patterns, and supply-chain organisation that is not fully represented by customs categories.
The export--import asymmetry appears across multiple analyses. In temporal similarity, import-origin profiles are more persistent than export-destination profiles at the aggregate level. In product communities, export- and import-side similarity structures lead to different patterns of agreement with the HS taxonomy, and the import-side similarity graph displays stronger modular structure. In economy influence, export-side and import-side influence scores and ranks differ, indicating that economies can have different structural roles depending on whether one focuses on destination markets or origin profiles. This asymmetry is substantively important. Export structure may reflect producer capabilities and competitive positioning, while import structure may reflect demand, sourcing behaviour, and dependence on supplier sets. Treating the GTN as a single directed graph without separating these perspectives would obscure this distinction.

Fourth, our study of economies' influence  further shows that network-structural importance cannot be reduced to trade volume. Large trade volumes do not automatically imply high perturbation influence, because network redundancy, substitutability, and layer composition matter. The leave-one-out PageRank impact captures how much the centrality landscape of other economies changes when a given economy is removed. The shock-modulated extension then identifies whether this influence is located in stable layers or in layers undergoing reorganisation. This distinction is useful for separating entrenched structural dependence from influence associated with recent change.

There are limitations in our analyses. The present analysis treats BACI trade values as observed weighted flows and does not explicitly model price changes, reporting uncertainty, or re-export effects. JS-Walk is designed for weighted networks, but it compares row-normalised transition profiles rather than raw trade volumes. It therefore captures how an economy distributes trade across partners, rather than whether a product layer is large or small in absolute value. This is desirable for structural comparison: two products may have very different total trade volumes but similar partner distributions, or similar volumes but very different network organisation. At the same time, JS-Walk is invariant to proportional rescaling of all outgoing flows from the same economy, so differences in absolute scale alone are not reflected in the similarity score. In the economy influence framework, absolute trade scale enters through the trade-share weights: a layer contributes more to an economy's aggregate influence when it represents a larger share of that economy's export or import portfolio. Separately, the leave-one-out PageRank impact measures how much the centrality structure of the remaining network changes when that economy is removed from the layer. The shock multiplier is a further dynamic component: it upweights this removal impact when the economy's own trade-partner profile in that layer changes strongly between consecutive years. The community-detection results also depend on the chosen clustering framework and on the decision to match HS chapter and section counts in the constrained reconstruction. We therefore interpret the recovered product communities as a complementary structural view rather than as a proposed replacement for the HS taxonomy.

Future work could extend the analysis in several directions. One direction could study shocks around specific historical events, such as financial crises, supply-chain disruptions, or major policy changes, by examining whether particular products or economies deviate sharply from their long-run similarity trajectories. A second direction could combine JS-Walk communities with product complexity and input--output information, which may help explain why certain products cluster together despite belonging to different HS categories. A third direction is to develop predictive models in which temporal JS-Walk distances or economy influence scores are used to anticipate future changes in trade structure.

\section*{Acknowledgment}

V.L. acknowledges support from NERC (NE/Y001184/1).
A.C and V.L.  
acknowledge  support from the European Union-NextGenerationEU,
GRINS project (Grant No. E63-C22-0021-20006).

\section*{Data Availability}
The BACI trade dataset is publicly available from CEPII
(\url{http://www.cepii.fr/CEPII/en/bdd_modele/bdd_modele_item.asp?id=37}).
Code and processed data will be made available upon publication at [repository URL].

\bibliographystyle{comnet}
\bibliography{references}

@article{hoang_reshaping_2023,
  author  = {Hoang, V. P. and Piccardi, Carlo and Tajoli, Lucia},
  title   = {Reshaping the Structure of the World Trade Network: A Pivotal Role for China?},
  journal = {Applied Network Science},
  volume  = {8},
  number  = {35},
  year    = {2023}
}

@article{battiston_multiplex_2026,
    author = {Battiston, Federico and Frasca, Mattia and G{\'o}mez-Garde{\~n}es, Jes{\'u}s and Min, Byungjoon and Radicchi, Filippo and Santoro, Andrea and Latora, Vito},
    title = {Dynamical processes and emergent behaviors in multiplex networks},
    journal = {arxiv preprint arXiv:2605.04199},
    year = {2026}
}

@article{garlaschelli_fitness_2004,
    author = {Garlaschelli, Diego and Loffredo, Maria I.},
    title = {Fitness-Dependent Topological Properties of the World Trade Web},
    volume = {93},
    pages = {188701},
    number = {18},
    journal ={Phys. Rev. Lett.},
    year = {2004}
}

@Article{fagiolo_worldtrade_2010,
author={Giorgio Fagiolo and Javier Reyes and Stefano Schiavo},
title={The evolution of the world trade web: a weighted-network analysis},
pages={479-514},
volume={20},
number={4},
journal={Journal of Evolutionary Economics},
year={2010}
}

@article{garlaschelli_structure_2005,
author = {Diego Garlaschelli and Maria I. Loffredo},
title = {Structure and evolution of the world trade network},
volume = {355},
number = {1},
pages = {138-144},
journal = {Physica A: Statistical Mechanics and its Applications},
year = {2005},
}

@article{fagiolo_worldtrade_2009,
  author = {Fagiolo, Giorgio and Reyes, Javier and Schiavo, Stefano},
  title = {World-trade web: Topological properties, dynamics, and evolution},
  volume = {79},
  pages = {036115},
  number = {3},
  journal = {Phys. Rev. E},
  year = {2009}
}

@article{debenedictis_world_2011,
  author  = {De Benedictis, Luca and Tajoli, Lucia},
  title   = {The World Trade Network},
  journal = {The World Economy},
  volume  = {34},
  number  = {8},
  pages   = {1417--1454},
  year    = {2011}
}

@techreport{debenedictis_network_2013,
  author      = {De Benedictis, Luca and Nenci, Silvia and Santoni, Gianluca and Tajoli, Lucia and Vicarelli, Claudio},
  title       = {Network Analysis of World Trade using the {BACI-CEPII} dataset},
  institution = {CEPII},
  type        = {Working Paper},
  number      = {2013-24},
  year        = {2013}
}

@article{barigozzi_multinetwork_2010,
  author  = {Barigozzi, Matteo and Fagiolo, Giorgio and Garlaschelli, Diego},
  title   = {Multinetwork of international trade: A commodity-specific analysis},
  journal = {Physical Review E},
  volume  = {81},
  number  = {4},
  pages   = {046104},
  year    = {2010},
  doi     = {10.1103/PhysRevE.81.046104}
}

@article{barigozzi_identifying_2011,
  author  = {Barigozzi, Matteo and Fagiolo, Giorgio and Mangioni, Giuseppe},
  title   = {Identifying the community structure of the international-trade multi-network},
  journal = {Physica A: Statistical Mechanics and its Applications},
  volume  = {390},
  number  = {11},
  pages   = {2051--2066},
  year    = {2011},
  doi     = {10.1016/j.physa.2011.02.004}
}

@article{piccardi_communities_2012,
  author  = {Piccardi, Carlo and Tajoli, Lucia},
  title   = {Existence and significance of communities in the World Trade Web},
  journal = {Physical Review E},
  volume  = {85},
  number  = {6},
  pages   = {066119},
  year    = {2012},
  doi     = {10.1103/PhysRevE.85.066119}
}

@article{ren_bridging_2020,
  author  = {Ren, Zhuo-Ming and Zeng, An and Zhang, Yi-Cheng},
  title   = {Bridging nestedness and economic complexity in multilayer world trade networks},
  journal = {Humanities and Social Sciences Communications},
  volume  = {7},
  number  = {1},
  pages   = {156},
  year    = {2020},
  doi     = {10.1057/s41599-020-00651-3}
}

@article{hidalgo_product_2007,
  author  = {Hidalgo, Cesar A. and Klinger, Bailey and Barab{\'a}si, Albert-L{\'a}szl{\'o} and Hausmann, Ricardo},
  title   = {The Product Space Conditions the Development of Nations},
  journal = {Science},
  volume  = {317},
  number  = {5837},
  pages   = {482--487},
  year    = {2007},
  doi     = {10.1126/science.1144581}
}

@article{hidalgo_building_2009,
  author  = {Hidalgo, Cesar A. and Hausmann, Ricardo},
  title   = {The Building Blocks of Economic Complexity},
  journal = {Proceedings of the National Academy of Sciences},
  volume  = {106},
  number  = {26},
  pages   = {10570--10575},
  year    = {2009},
  doi     = {10.1073/pnas.0900943106}
}

@article{kosztyan_trade_2024,
  author  = {Koszty{\'a}n, Zsolt Tibor and Kiss, D{\'e}nes and Feh{\'e}rv{\"o}lgyi, Be{\'a}ta},
  title   = {Trade network dynamics in a globalized environment and on the edge of crises},
  journal = {Journal of Cleaner Production},
  volume  = {465},
  pages   = {142699},
  year    = {2024},
  doi     = {10.1016/j.jclepro.2024.142699}
}

@article{tacchella_new_2012,
  author  = {Tacchella, Andrea and Cristelli, Matthieu and Caldarelli, Guido and Gabrielli, Andrea and Pietronero, Luciano},
  title   = {A New Metrics for Countries' Fitness and Products' Complexity},
  journal = {Scientific Reports},
  volume  = {2},
  number  = {723},
  year    = {2012}
}

@article{battiston_structural_2014,
  author  = {Battiston, Federico and Nicosia, Vincenzo and Latora, Vito},
  title   = {Structural Measures for Multiplex Networks},
  journal = {Physical Review E},
  volume  = {89},
  number  = {3},
  pages   = {032804},
  year    = {2014}
}

@techreport{CEPII:2010-23,
  author      = {Gaulier, Guillaume and Zignago, Soledad},
  title       = {{BACI}: International Trade Database at the Product-Level. The 1994--2007 Version},
  institution = {CEPII},
  type        = {Working Paper},
  number      = {2010-23},
  year        = {2010}
}

@article{oliver_harmonised_1987,
  author  = {Oliver, Peter and Yataganas, Xenophon},
  title   = {The Harmonised System of Customs Classification},
  journal = {Yearbook of European Law},
  volume  = {7},
  number  = {1},
  pages   = {113--129},
  year    = {1987}
}

@misc{wco_hs_faq,
  author       = {{World Customs Organization}},
  title        = {Frequently Asked Questions Related to the Harmonized System},
  year         = {2024},
  howpublished = {World Customs Organization}
}

@misc{wco_hs_convention,
  author       = {{World Customs Organization}},
  title        = {International Convention on the Harmonized Commodity Description and Coding System},
  year         = {1983},
  howpublished = {World Customs Organization}
}

@misc{wco_chapter77,
  author       = {{World Customs Organization}},
  title        = {Chapter 77: Reserved for Possible Future Use in the Harmonized System},
  year         = {2022},
  howpublished = {Harmonized System Nomenclature 2022 Edition}
}

@article{endres2003new,
  author  = {Endres, Dominik M. and Schindelin, Johannes E.},
  title   = {A New Metric for Probability Distributions},
  journal = {IEEE Transactions on Information Theory},
  volume  = {49},
  number  = {7},
  pages   = {1858--1860},
  year    = {2003}
}

@inproceedings{koutra_deltacon_2013,
  author    = {Koutra, Danai and Vogelstein, Joshua T. and Faloutsos, Christos},
  title     = {{DeltaCon}: A Principled Massive-Graph Similarity Function},
  booktitle = {Proceedings of the 2013 SIAM International Conference on Data Mining},
  pages     = {162--170},
  year      = {2013},
  publisher = {SIAM}
}

@article{piccardi_metrics_2023,
  author  = {Piccardi, Carlo},
  title   = {Metrics for Network Comparison Using Egonet Feature Distributions},
  journal = {Scientific Reports},
  volume  = {13},
  number  = {14657},
  year    = {2023}
}

@article{bagrow_information-theoretic_2019,
    author  = {Bagrow, J. and Bollt, E.},
  title   = {An information-theoretic, all-scales approach to comparing networks},
  journal = {Appl Netw Sci},
  volume  = {4},
  pages   = {45},
  year    = {2019}
}

@article{felippe_network_2024,
  author  = {Felippe, Helcio and Battiston, Federico and Kirkley, Alec},
  title   = {Network Mutual Information Measures for Graph Similarity},
  journal = {Communications Physics},
  volume  = {7},
  number  = {335},
  year    = {2024}
}

@article{berlingerio2012netsimile,
  author  = {Berlingerio, Michele and Koutra, Danai and Eliassi-Rad, Tina and Faloutsos, Christos},
  title   = {{NetSimile}: A Scalable Approach to Size-Independent Network Similarity},
  journal = {arXiv preprint arXiv:1209.2684},
  year    = {2012}
}

@book{hand_principles_2001,
  author    = {Hand, David J. and Mannila, Heikki and Smyth, Padhraic},
  title     = {Principles of Data Mining},
  publisher = {MIT Press},
  address   = {Cambridge, MA},
  year      = {2001}
}

@book{cover_elements_2006,
  author    = {Cover, Thomas M. and Thomas, Joy A.},
  title     = {Elements of Information Theory},
  edition   = {2},
  publisher = {Wiley-Interscience},
  address   = {Hoboken, NJ},
  year      = {2006}
}

@incollection{lovasz_random_1993,
  author  = {Lov{\'a}sz, L{\'a}szl{\'o}},
  title   = {Random Walks on Graphs: A Survey},
  booktitle = {Combinatorics, Paul Erd{\H{o}}s is Eighty},
  volume    = {2},
  series    = {Bolyai Society Mathematical Studies},
  pages     = {1--46},
  year      = {1993}
}

@book{good_permutation_2005,
  author    = {Good, Phillip I.},
  title     = {Permutation, Parametric, and Bootstrap Tests of Hypotheses},
  edition   = {3},
  publisher = {Springer},
  address   = {New York},
  year      = {2005}
}

@article{holme_temporal_2012,
  author  = {Holme, Petter and Saram{\"a}ki, Jari},
  title   = {Temporal Networks},
  journal = {Physics Reports},
  volume  = {519},
  number  = {3},
  pages   = {97--125},
  year    = {2012}
}

@article{traag2019leiden,
  author  = {Traag, V. A. and Waltman, L. and van Eck, N. J.},
  title   = {From Louvain to Leiden: Guaranteeing Well-Connected Communities},
  journal = {Scientific Reports},
  volume  = {9},
  number  = {5233},
  year    = {2019}
}

@article{hubert1985comparing,
  author  = {Hubert, Lawrence and Arabie, Phipps},
  title   = {Comparing Partitions},
  journal = {Journal of Classification},
  volume  = {2},
  number  = {1},
  pages   = {193--218},
  year    = {1985}
}

@techreport{page1999pagerank,
  author      = {Page, Lawrence and Brin, Sergey and Motwani, Rajeev and Winograd, Terry},
  title       = {The {PageRank} Citation Ranking: Bringing Order to the Web},
  institution = {Stanford InfoLab},
  number      = {1999-66},
  year        = {1999}
}

@article{gleich2015pagerank,
  author  = {Gleich, David F.},
  title   = {{PageRank} Beyond the Web},
  journal = {SIAM Review},
  volume  = {57},
  number  = {3},
  pages   = {321--363},
  year    = {2015}
}

@article{lin1991divergence,
  author  = {Lin, Jianhua},
  title   = {Divergence Measures Based on the Shannon Entropy},
  journal = {IEEE Transactions on Information Theory},
  volume  = {37},
  number  = {1},
  pages   = {145--151},
  year    = {1991}
}

@inproceedings{ng2002spectral,
title     = {On Spectral Clustering: Analysis and an Algorithm},
author    = {Ng, Andrew Y. and Jordan, Michael I. and Weiss, Yair},
booktitle = {Advances in Neural Information Processing Systems},
volume    = {14},
pages     = {849--856},
year      = {2002},
publisher = {MIT Press}
}

%








\end{document}


\title{Supplementary Material for\\
Multilayer Analysis of the Global Trade Network}
\author{Chenyang Li, Leonardo Brogi, Andrea Civilini, Piero Mazzarisi, Nicola Perra, Vito Latora}
\maketitle

This Supplementary Material provides additional visualisations and robustness checks supporting the three empirical analyses reported in the main text. Section~\ref{app:network_temporal} complements the temporal analysis by reporting aggregation-level network-flow examples, year-label permutation null models, and additional summaries of short-run persistence across HS chapters and sections. Section~\ref{app:product_community_supp} provides further diagnostics for the flow-based reconstruction of product communities, including import-side results, chapter fragmentation, and yearly divergence from the official HS taxonomy. Section~\ref{app:economy_influence_supp} reports additional economy-influence results, including raw-score counterparts, shock-modulated comparisons, external comparisons with economic fitness, and robustness to coarser HS chapter aggregation.

\section{Supplementary Network Visualisations and Temporal Similarity}
\label{app:network_temporal}

The main text shows that the aggregate Global Trade Network is highly persistent in the short run but drifts gradually over the full sample period. This section provides supplementary evidence for that claim. We first show additional network-flow examples at chapter and section levels, then report a year-label permutation null model, and finally examine how temporal persistence varies across HS chapters and sections.

\subsection{Network-flow examples across aggregation levels}
\label{app:network_flow_examples}

These visual examples extend the product-level flow diagrams shown in the main text to coarser HS aggregation levels. They are intended as qualitative illustrations only; all quantitative results in the paper use the full weighted adjacency matrices rather than the thresholded networks displayed here.

\begin{figure}[!h]
    \centering
    \includegraphics[width=5.5in]{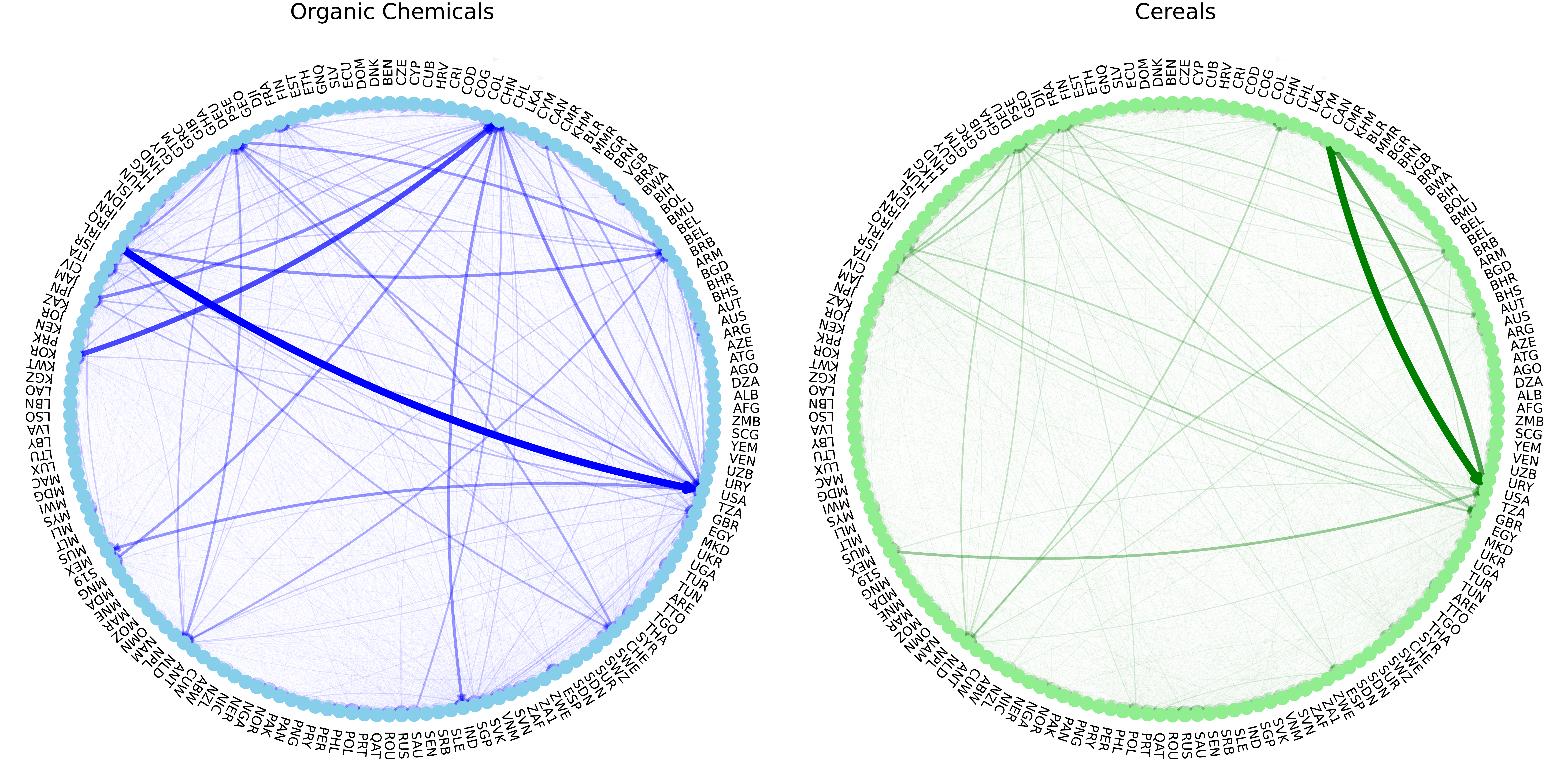}
    \caption{\textbf{Comparison of directed network-flow diagrams for two HS chapters, Organic Chemicals (HS~29) and Cereals (HS~10), in 2010.}
    Nodes correspond to economies and directed links to export relationships aggregated across all products within the chapter; edge width encodes total value~$w_{ij}$. For visual clarity, only the top 15\% of edges by weight are shown. The two chapters illustrate how, even at a coarser aggregation level, the same set of economies can sustain markedly different trade structures, with Organic Chemicals exhibiting a dense, multi-hub backbone and Cereals a sparser configuration anchored on a few dominant exporter--importer pairs. The qualitative contrast persists across thresholds.}
    \label{fig:app_layer_examples_chapters}
\end{figure}

\begin{figure}[!h]
    \centering
    \includegraphics[width=5.5in]{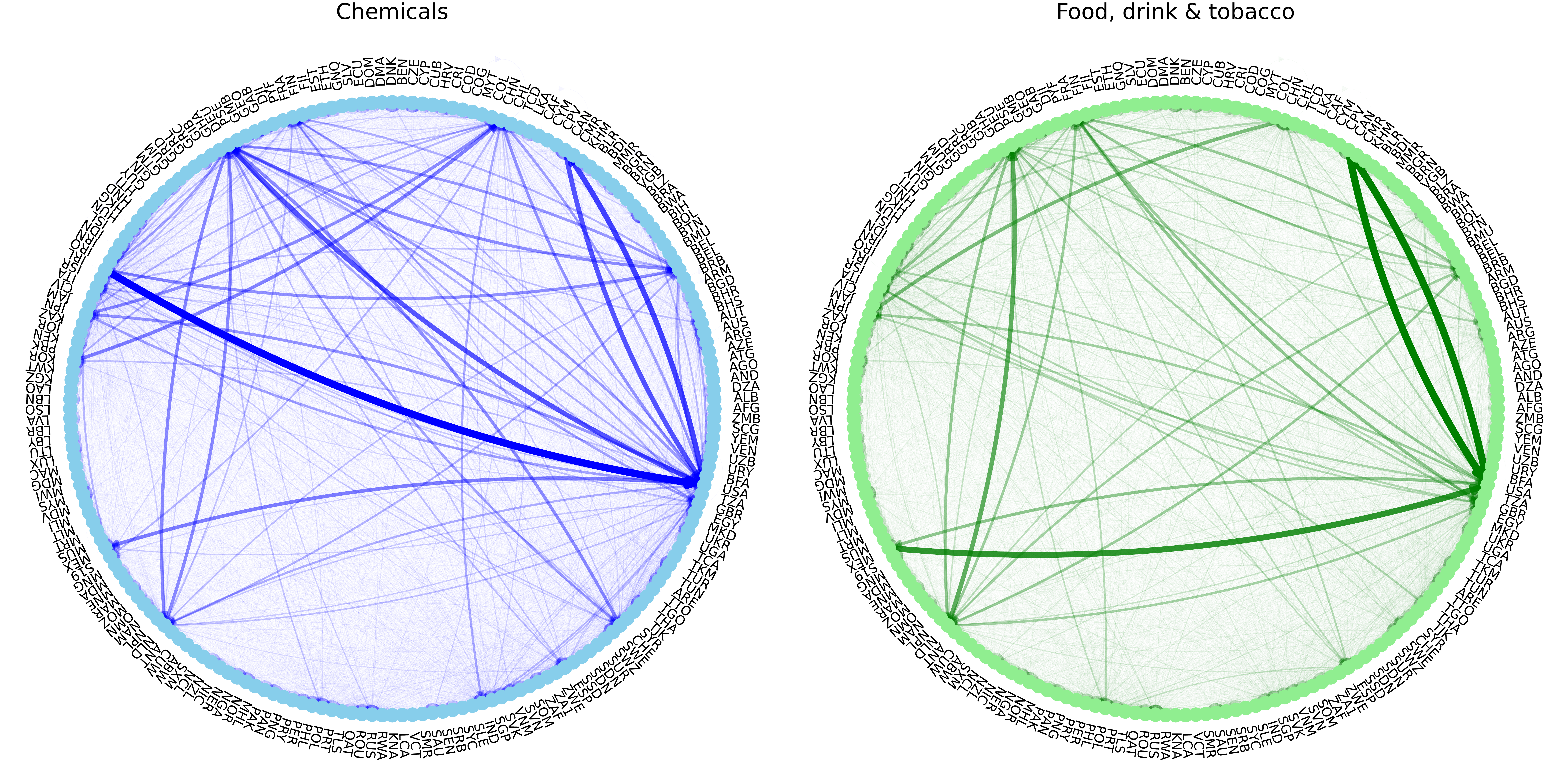}
    \caption{\textbf{Comparison of directed network-flow diagrams for two HS sections, Chemicals (Section~VI) and Food, drink \& tobacco (Section~IV), in 2010.}
    Nodes correspond to economies and directed links to export relationships aggregated across all products within the section; edge width encodes total value~$w_{ij}$. For visual clarity, only the top 15\% of edges by weight are shown. At the section level both networks engage essentially the full set of economies, yet their backbones differ in density and in the identity of the dominant flows. This confirms that the structural heterogeneity observed at the product and chapter levels is not erased by aggregation. The qualitative contrast persists across thresholds.}
    \label{fig:app_layer_examples_sections}
\end{figure}

\FloatBarrier

\subsection{Aggregate and disaggregated temporal similarity}
\label{app:temporal_similarity_supp}

We next examine whether the temporal patterns reported in the main text depend on the chronological ordering of the observed annual networks, and whether aggregation hides heterogeneity across HS sections and chapters.

\begin{figure}[!h]
\centering
\includegraphics[width=5.5in]{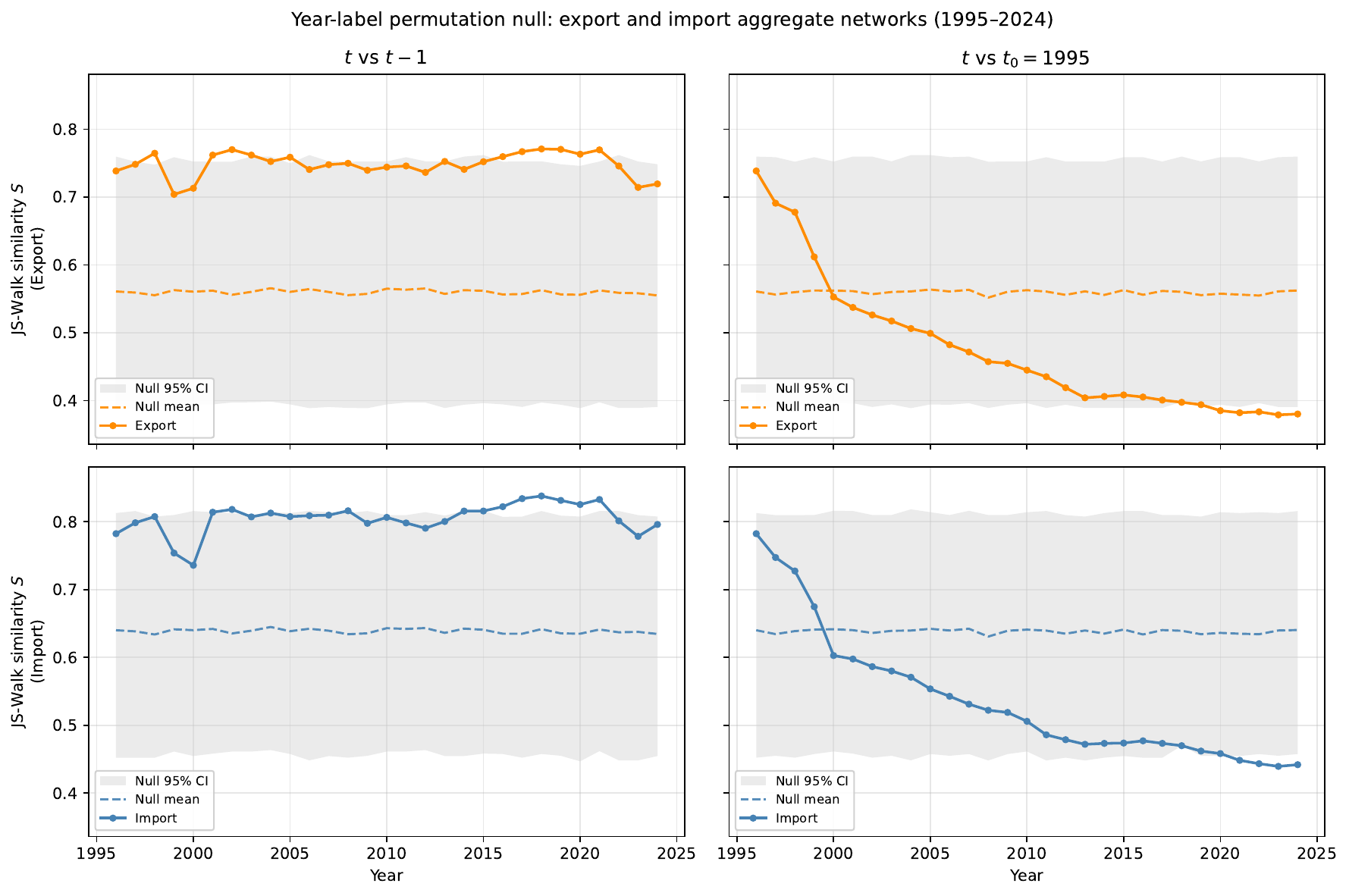}
\caption{\textbf{Year-label permutation null for aggregate export and import networks, 1995--2024.}
Rows distinguish the export side (\emph{top}) and import side (\emph{bottom}); columns report the year-on-year comparison (\emph{left}, $t$ vs.\ $t-1$) and the fixed-initial-year comparison (\emph{right}, $t$ vs.\ $t_0=1995$). Solid lines show the observed JS-Walk similarity, dashed lines show the null mean, and shaded bands show the 95\% confidence interval from 1{,}000 random reshufflings of the annual network sequence. The null model preserves the observed annual networks but destroys their chronological ordering. The year-on-year observed curves lie above the null mean, indicating stronger short-run persistence than expected under random temporal ordering. The fixed-initial-year panels show a systematic decline that is not reproduced by the nearly flat null mean, supporting the interpretation of cumulative long-run drift in the aggregate GTN.}
\label{fig:app_temp_null_aggregate}
\end{figure}

\begin{figure}[!h]
    \centering
    \includegraphics[width=5.5in]{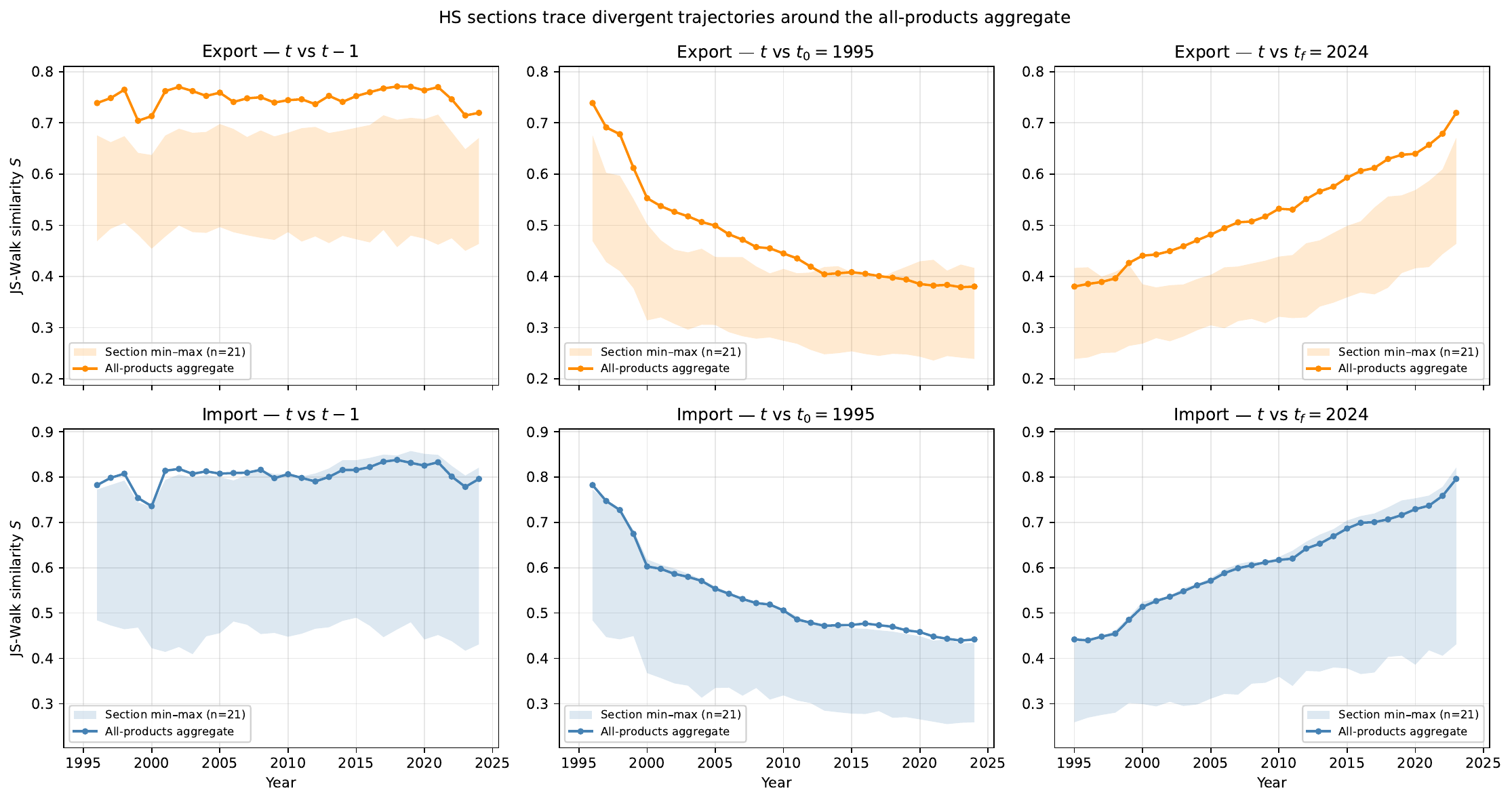}
    \caption{\textbf{HS sections trace divergent trajectories around the all-products aggregate.}
    Each panel reports JS-Walk similarity $S$ for the all-products aggregate network (solid line) together with the per-section min--max envelope across the 21 HS sections (shaded band). Rows distinguish the export (\emph{top}) and import (\emph{bottom}) sides. Columns report three temporal references: year-on-year comparison ($t$ vs.\ $t-1$), comparison with the initial year ($t$ vs.\ $t_0=1995$), and comparison with the final year ($t$ vs.\ $t_f=2024$). The aggregate trajectory exhibits smaller relative variation than the section-level envelope, indicating that aggregation averages out section-specific dynamics. The width of the envelope, especially under the fixed-initial-year and fixed-final-year references, documents substantial cross-section heterogeneity in long-run structural change.}
    \label{fig:app_temp_aggregate_vs_section}
\end{figure}

\begin{figure}[!h]
    \centering
    \includegraphics[width=5.5in]{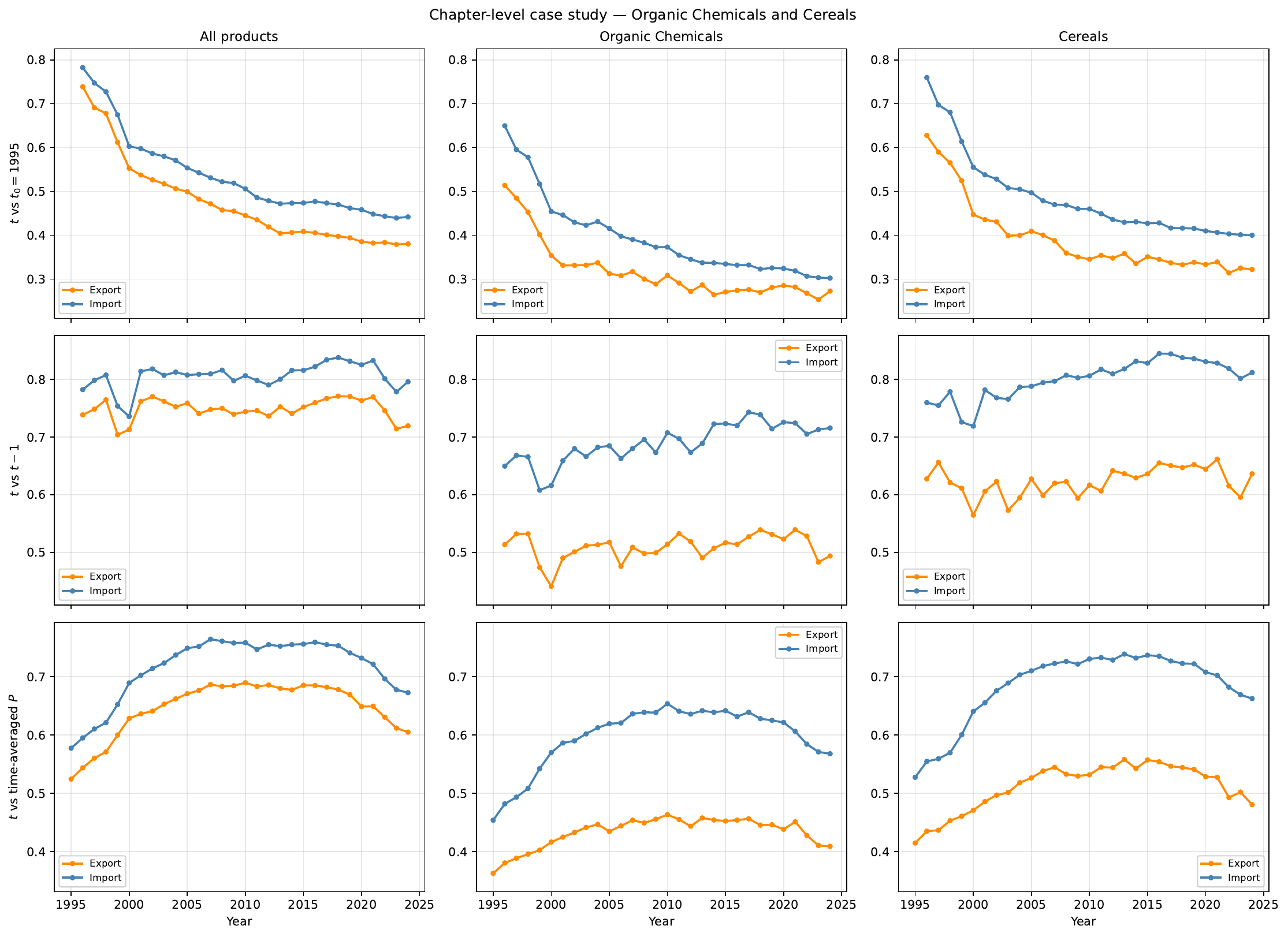}
    \caption{\textbf{Chapter-level case study: Organic Chemicals (HS~29) and Cereals (HS~10).}
    Columns indicate the all-products aggregate (\emph{left}), Organic Chemicals (\emph{centre}), and Cereals (\emph{right}); rows report the three temporal references $t$ vs.\ $t_0=1995$, $t$ vs.\ $t-1$, and $t$ vs.\ the time-averaged transition matrix $\bar{P}$. Export and import sides are shown jointly in each panel. The two chapters share the long-run drift seen at the aggregate level but display chapter-specific differences in year-on-year stability and in the magnitude of the export--import gap, supporting the interpretation of chapter-level layers as structurally distinct from the aggregate.}
    \label{fig:app_temp_case_chapter}
\end{figure}

\begin{figure}[!h]
    \centering
    \includegraphics[width=5.5in]{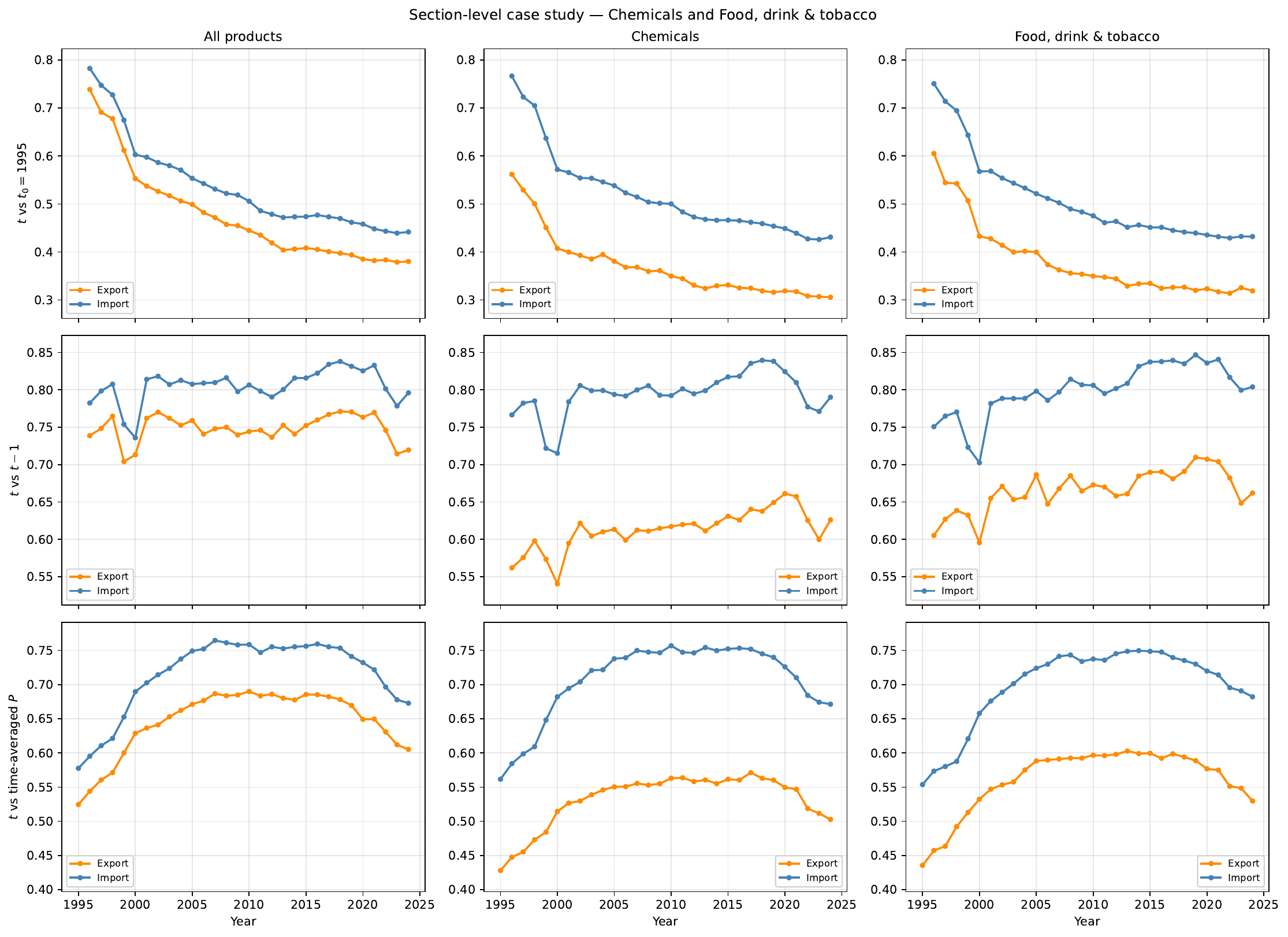}
    \caption{\textbf{Section-level case study: Chemicals (HS section~VI) and Food, drink \& tobacco (HS section~IV).}
    Columns indicate the all-products aggregate (\emph{left}), Chemicals (\emph{centre}), and Food, drink \& tobacco (\emph{right}); rows report the three temporal references $t$ vs.\ $t_0=1995$, $t$ vs.\ $t-1$, and $t$ vs.\ the time-averaged transition matrix $\bar{P}$. Export and import sides are shown jointly in each panel. Both sections reproduce the aggregate long-run drift while differing in the persistent export--import gap and in the smoothness of their year-on-year similarity, complementing the chapter-level case study in Figure~\ref{fig:app_temp_case_chapter}.}
    \label{fig:app_temp_case_section}
\end{figure}

\begin{figure}[!h]
    \centering
    \includegraphics[width=6in]{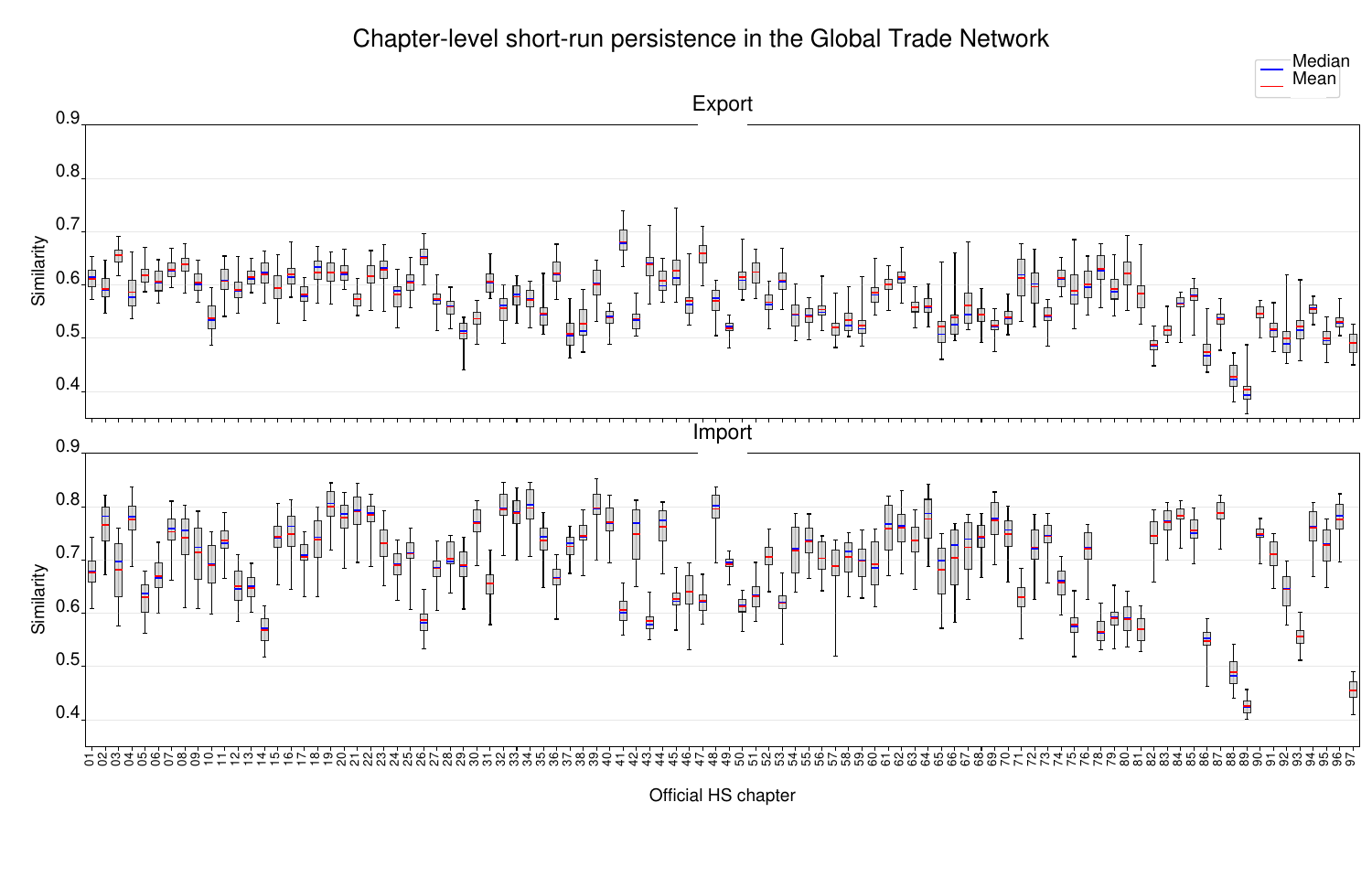}
    \caption{\textbf{Chapter-level summary of short-run persistence.}
    For each official HS chapter, products are first aggregated into a chapter-level trade layer. The figure summarises the year-on-year JS-Walk similarity $S^{\alpha,r}_{\mathrm{yoy}}(t)=S(P^{\alpha,r}_{t},P^{\alpha,r}_{t-1})$ across the $T-1$ consecutive-year comparisons over 1995--2024. Whiskers indicate the minimum and maximum similarity, the grey box indicates the interquartile range, the blue line marks the median, and the red line marks the mean. Differences across chapters show that the aggregate GTN masks substantial heterogeneity in short-run persistence across lower-level trade layers.}
    \label{fig:app_chapter_yoy_persistence}
\end{figure}

\FloatBarrier

\section{Flow-Based Product Communities: Supplementary Results}
\label{app:product_community_supp}

The main text reports that product communities recovered from trade-flow similarity only partially align with the official HS taxonomy. This section provides complementary diagnostics showing where the reconstructed communities agree with HS groupings, where official chapters fragment across trade-based groups, and whether these patterns differ between export- and import-side similarity.

\subsection{Import-side chapter similarity and section contingency}
\label{app:community_import_similarity}

The main text presents the export-side chapter similarity network and export-side section contingency matrix. Here we report the corresponding import-side results, where rows of the transition matrices represent importing economies and columns represent import origins.


\begin{figure}[!h]
\centering
\includegraphics[width=4.5in]{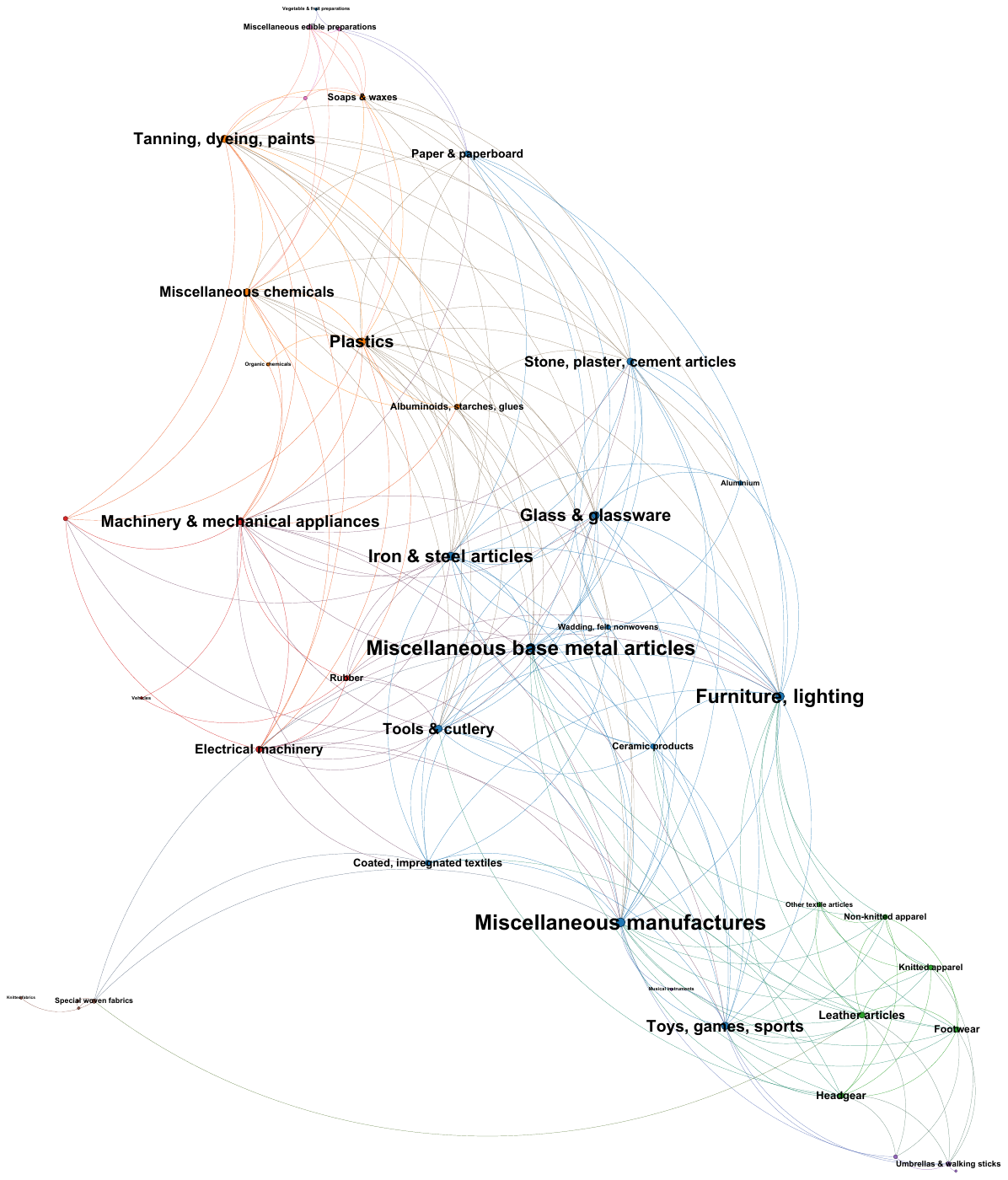}
\caption{%
  \textbf{Chapter similarity network on the import side.}
  The figure is the import-side counterpart of the export-side chapter similarity network reported in the main text. Each node is one of the 96 HS chapters active in 2010, sized by total chapter trade volume and coloured by spectral community ($k=21$). Edges show the top 5\% of pairwise JS-Walk similarities computed on import-side row-normalised adjacency matrices, where rows are importers and columns are exporters. The import-side network is sparser at the chapter level under the same threshold but exhibits a stronger modular signature, indicating that consumer-side trade structures cluster more tightly than producer-side ones. Layout: ForceAtlas2 in Gephi.%
}
\label{fig:app_motivation_chapter_import}
\end{figure}

\begin{figure}[!h]
\centering
\includegraphics[width=5.5in]{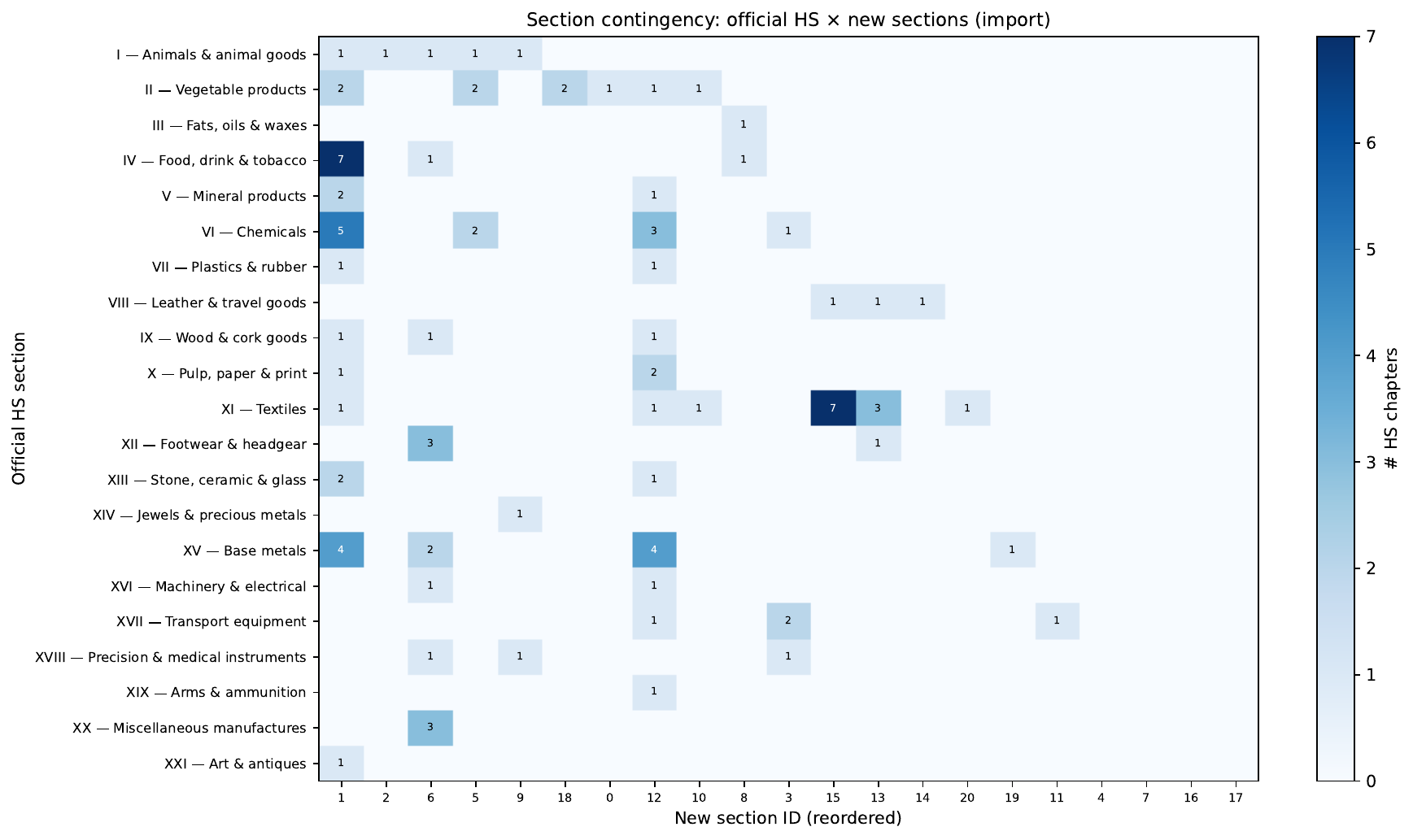}
\caption{
  \textbf{Import-side section contingency between official HS sections and trade-based new sections.}
  The heatmap compares official HS sections with the new sections obtained from import-side trade-flow similarity. Cell $(i,j)$ counts the number of HS chapters that belong to official section $i$ and are assigned to new section $j$ by spectral clustering ($k=21$) on the import-side JS-Walk similarity. Columns are reordered to ease visual grouping. Import-side contingency is more concentrated than the export-side counterpart: Section~IV \emph{Food, drink \& tobacco} maps 7 of its 9 chapters into one new section; Section~VI \emph{Chemicals} concentrates 5 chapters in one new section; and Section~XI \emph{Textiles} concentrates 7 chapters in one new section. This stronger concentration is consistent with the higher modularity of the import similarity graph.}
\label{fig:app_cd3_contingency_import}
\end{figure}

\FloatBarrier

\subsection{Chapter fragmentation and divergence from official HS sections}
\label{app:community_fragmentation_divergence}

The following diagnostics move beyond global ARI scores by asking which parts of the HS hierarchy are reorganised by trade-flow similarity. Chapter fragmentation measures how products from the same official HS chapter are distributed across reconstructed new chapters, while yearly divergence measures the stability of the alignment between new sections and official HS sections over time.

\begin{figure}[!h]
\centering
\includegraphics[width=5.5in]{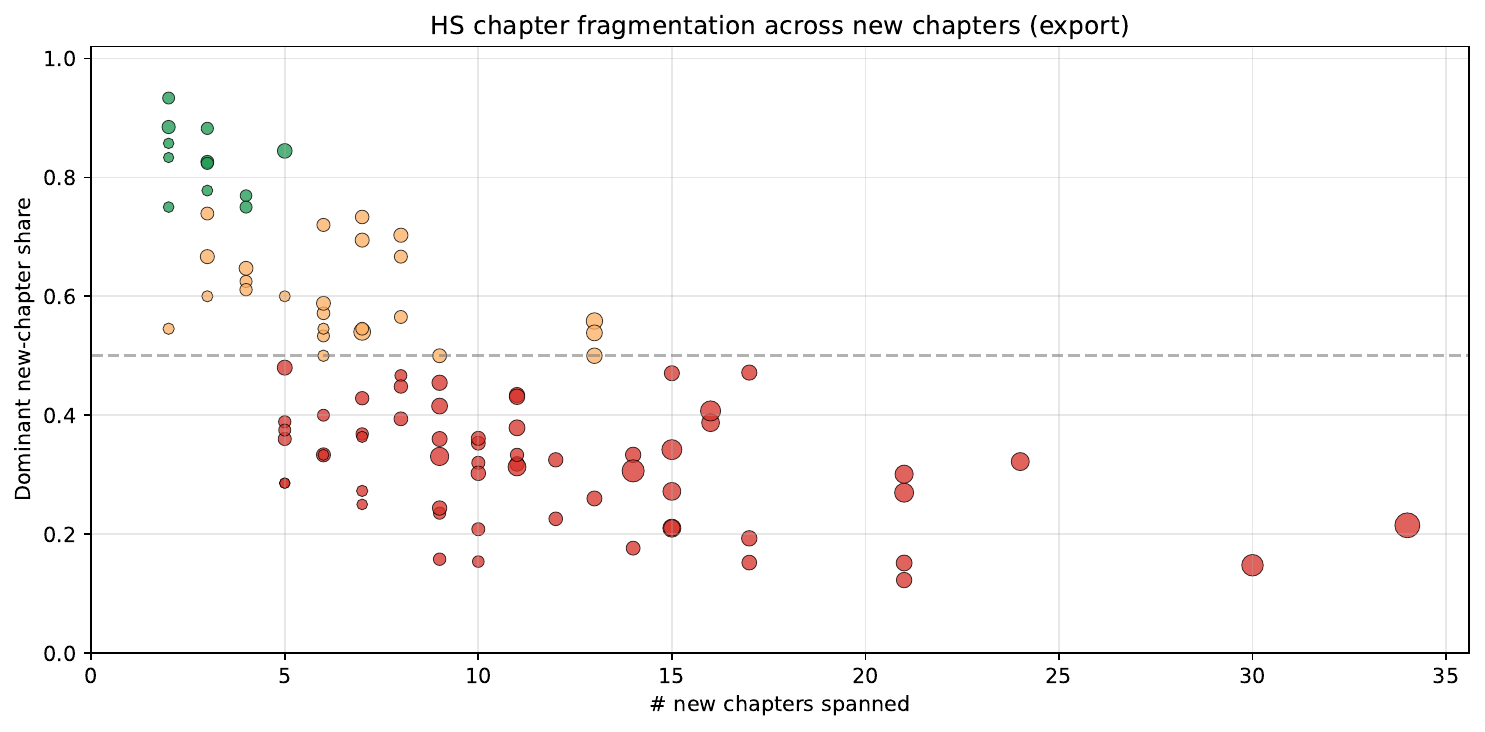}
\caption{
  \textbf{Export-side chapter fragmentation across trade-based new chapters.}
  Each marker is one of the 96 official HS chapters active in the data. The $x$-axis shows the number of distinct new chapters its products span under spectral clustering ($k=96$, matching the number of active HS chapters) on the export-side JS-Walk similarity; the $y$-axis shows the share of that chapter's products contained in its single dominant new chapter; marker size scales with the chapter's product count. The dashed line at $y=0.5$ separates relatively coherent chapters, whose dominant reconstructed chapter contains at least half of the official chapter's products, from fragmented chapters. The L-shaped pattern indicates that small chapters often remain concentrated, while large heterogeneous chapters split into many trade-based groups. The two most fragmented chapters are HS~84 \emph{Machinery \& mechanical appliances} and HS~85 \emph{Electrical machinery}.}
\label{fig:app_cd4_fragmentation_export}
\end{figure}

\begin{figure}[!h]
\centering
\includegraphics[width=5.5in]{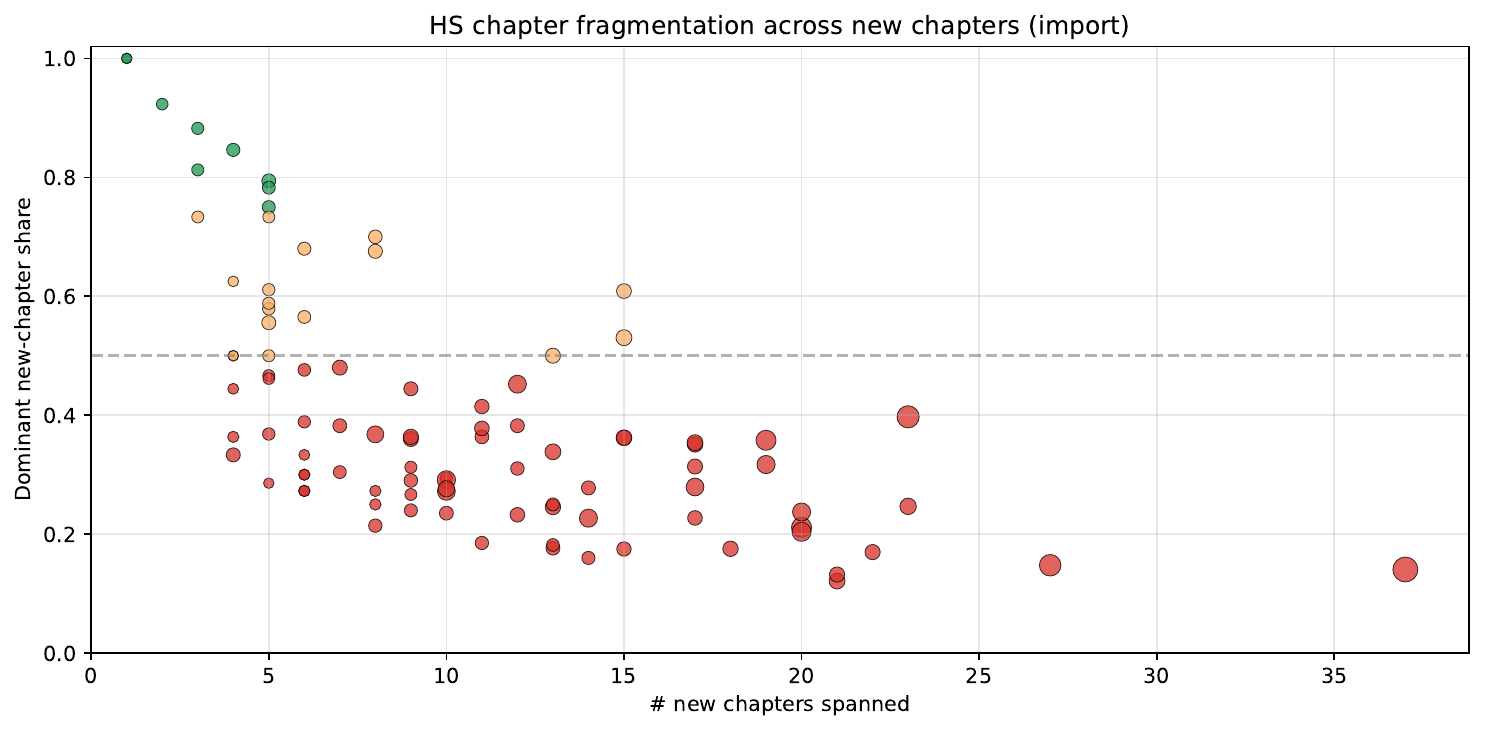}
\caption{
  \textbf{Import-side chapter fragmentation across trade-based new chapters.}
  This figure repeats the fragmentation analysis in Figure~\ref{fig:app_cd4_fragmentation_export} using import-side JS-Walk similarity. The qualitative L-shaped pattern is preserved, with HS~84 \emph{Machinery \& mechanical appliances} and HS~85 \emph{Electrical machinery} again located at the highly fragmented extreme. Import-side fragmentation is slightly more pronounced than on the export side, particularly among chapters spanning more than 25 reconstructed new chapters with dominant shares below 0.2.}
\label{fig:app_cd4_fragmentation_import}
\end{figure}

\begin{figure}[!h]
\centering
\includegraphics[width=5.5in]{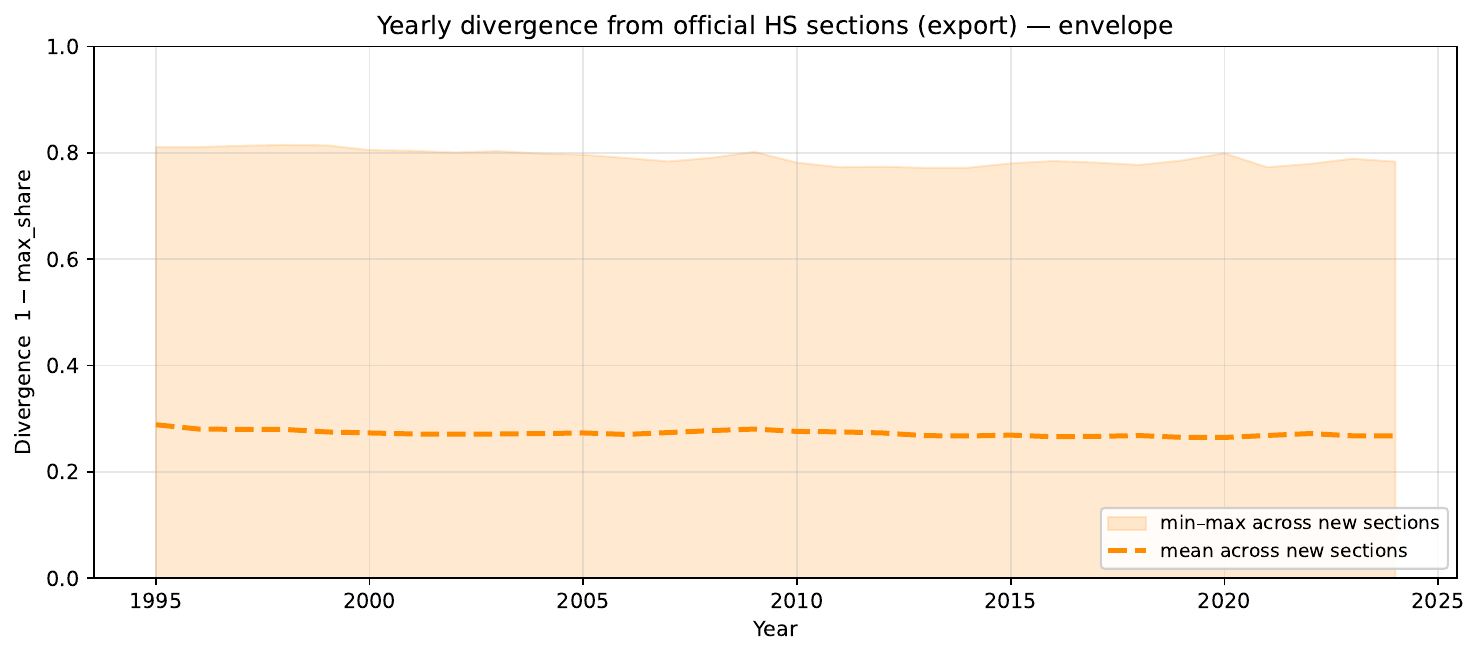}
\caption{
\textbf{Yearly divergence from official HS sections on the export side.}
For each trade-based new section, divergence is defined as
$\Delta_s(t)=1-\max_o V_{s,o}(t)/V_s(t)$, where $V_s(t)$ is the total trade value in new section $s$ in year $t$ and $V_{s,o}(t)$ is the part assigned to official HS section $o$. The shaded band reports the min--max range across the 21 new sections, and the dashed line reports the mean. The near-flat trajectory indicates that the disagreement between flow-based and official HS groupings is persistent rather than driven by a few exceptional years.}
\label{fig:app_cd1a_envelope_export}
\end{figure}

\begin{figure}[!h]
\centering
\includegraphics[width=5.5in]{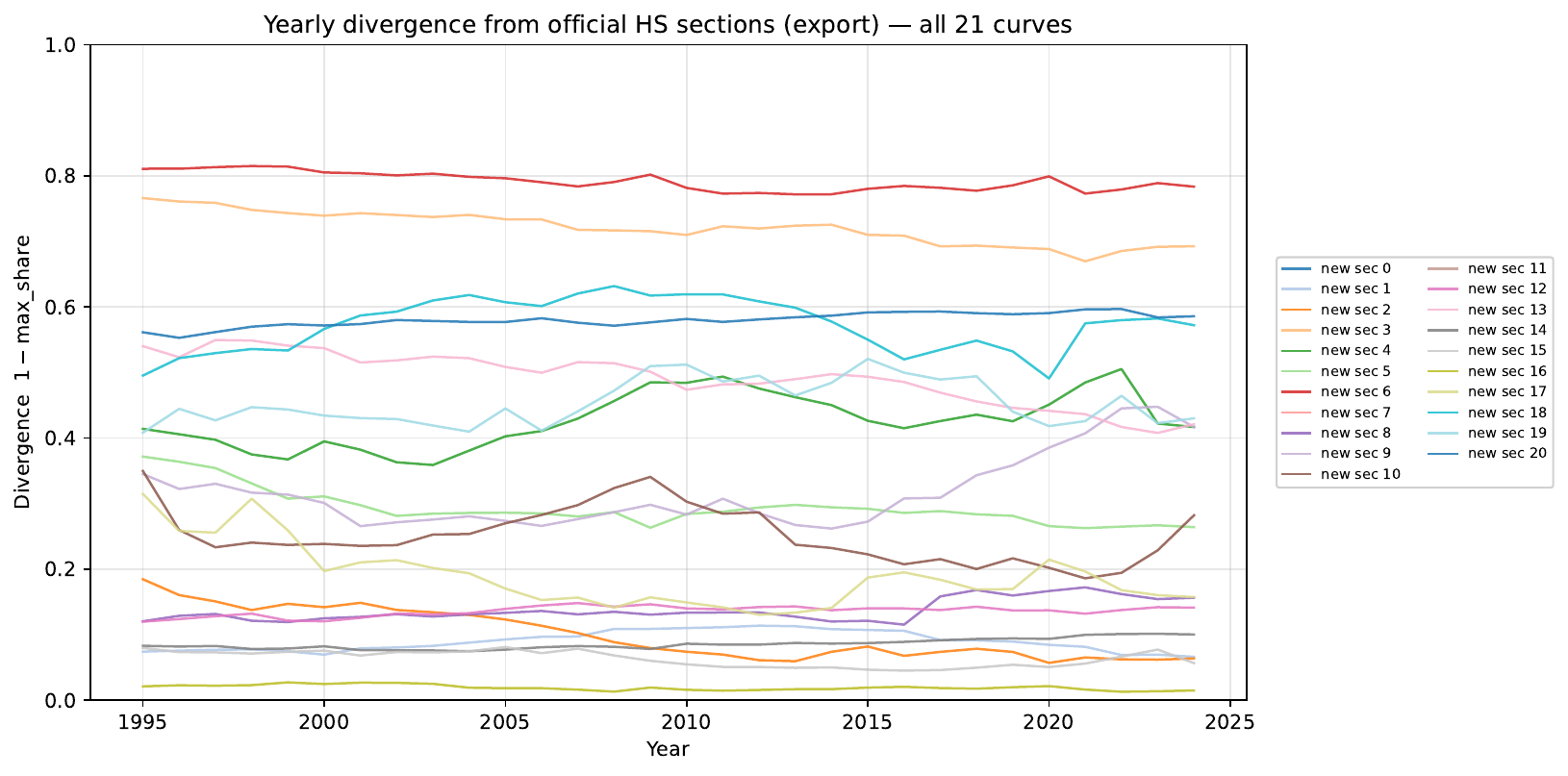}
\caption{
  \textbf{Per-section export-side divergence trajectories underlying Figure~\ref{fig:app_cd1a_envelope_export}.}
  Each of the 21 new sections is traced individually over 1995--2024. The min--max band in Figure~\ref{fig:app_cd1a_envelope_export} is resolved here into individual trajectories: some new sections remain strongly mixed across HS sections, while others are almost entirely aligned with a single official HS section. Most curves are stable over time, consistent with structurally fixed alignment patterns rather than transient year-specific deviations.}
\label{fig:app_cd1b_curves_export}
\end{figure}

\begin{figure}[!h]
\centering
\subfloat[Envelope across the 21 new sections.]{%
  \includegraphics[width=0.95\textwidth]{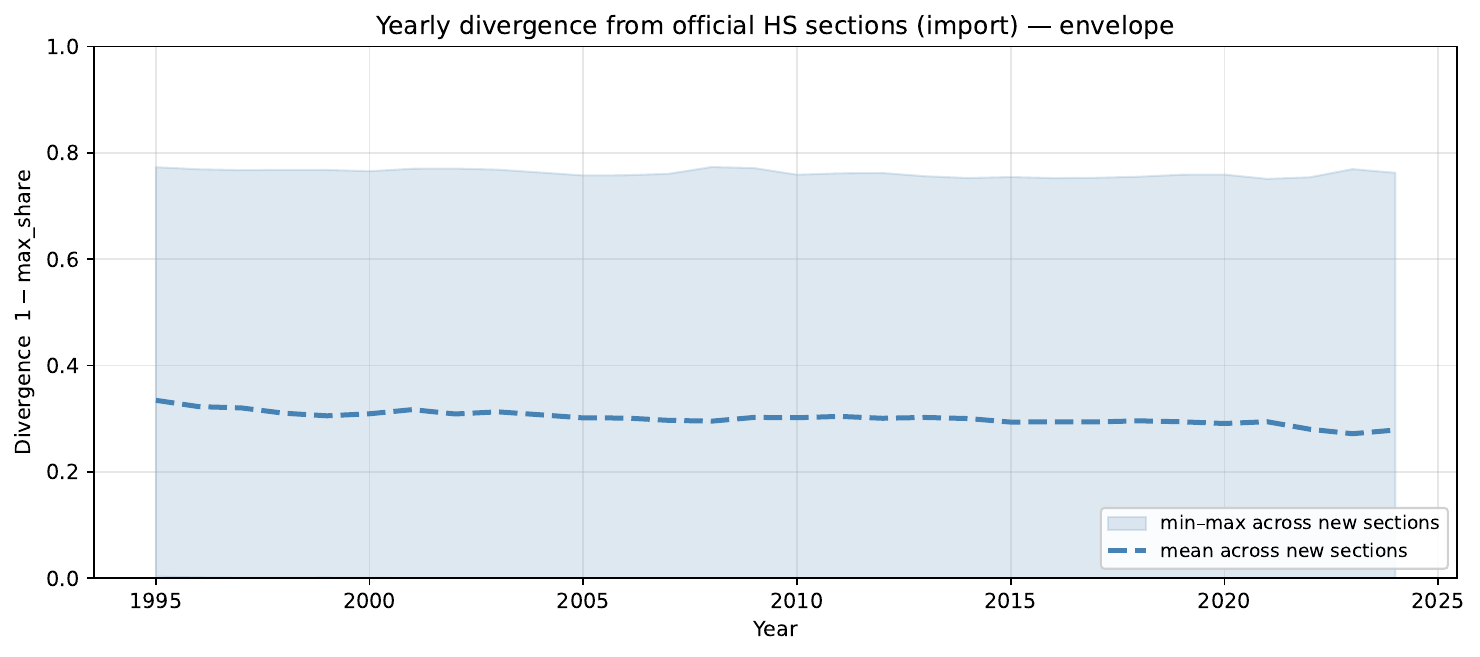}%
  \label{fig:app_cd1a_envelope_import}
}

\vspace{6pt}

\subfloat[Per-section trajectories.]{%
  \includegraphics[width=0.95\textwidth]{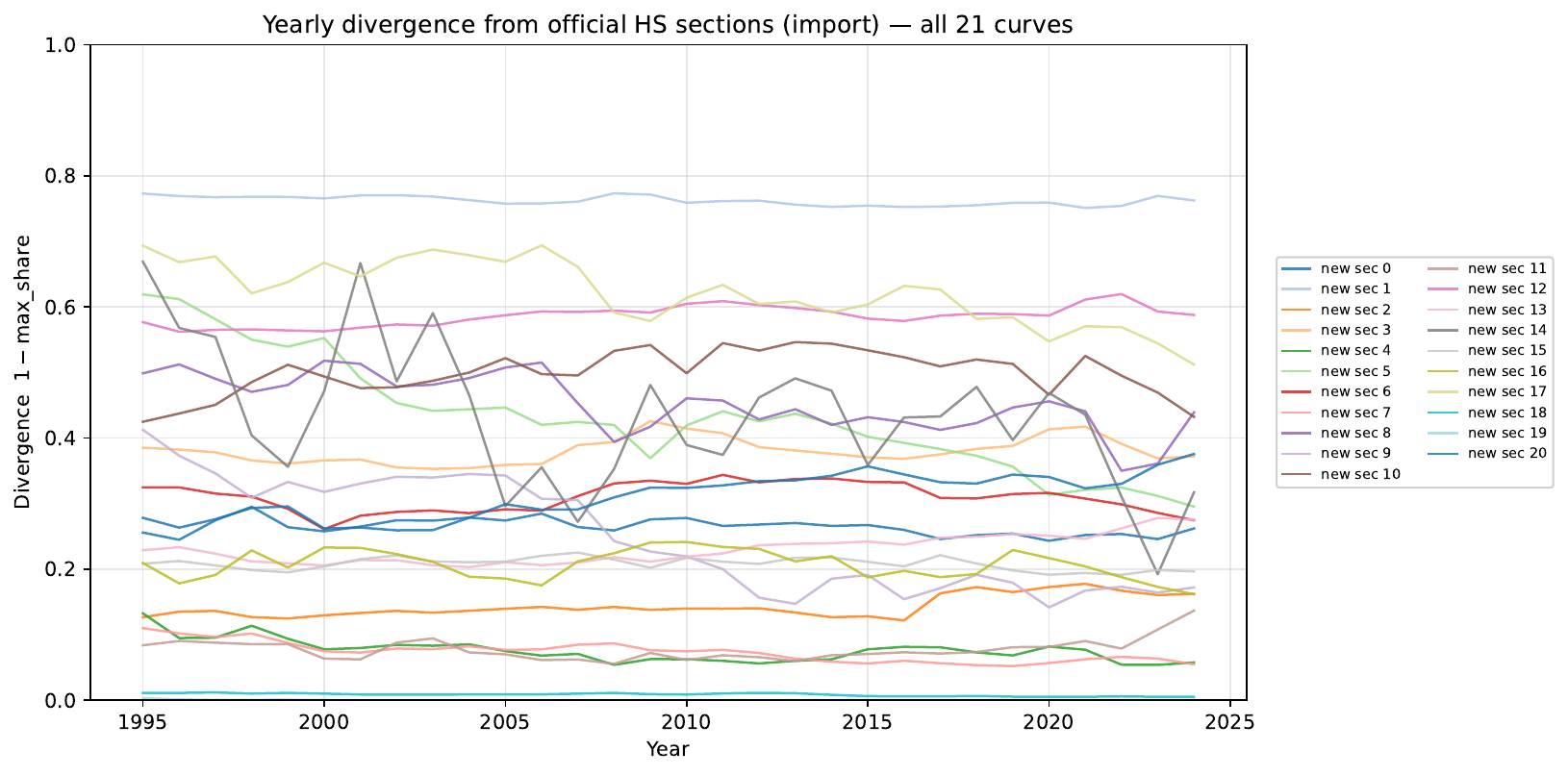}%
  \label{fig:app_cd1b_curves_import}
}

\caption{
  \textbf{Yearly divergence from official HS sections on the import side.}
  This figure repeats the divergence analysis of Figures~\ref{fig:app_cd1a_envelope_export} and~\ref{fig:app_cd1b_curves_export} for import-side JS-Walk similarity. Mean divergence is slightly higher on the import side than on the export side, while the envelope shows comparable long-term stability over 1995--2024. The result supports the interpretation that disagreement between flow-based and official HS groupings is persistent in both trade orientations.}
\label{fig:app_cd1_import}
\end{figure}

\FloatBarrier

\section{Economy Influence Robustness and Supplementary Results}
\label{app:economy_influence_supp}

The main text presents economy influence primarily through annual ranks and export--import comparisons for selected major economies. This section reports additional checks supporting those results. Raw-score counterparts verify that the rank dynamics are not only artefacts of relative ordering. Shock-modulated scatter plots examine whether export-side and import-side asymmetries persist when influence is weighted by year-on-year restructuring. Comparisons with the Tacchella--Pietronero fitness index provide an external reference for productive capability, while chapter-level results test robustness to coarser product aggregation.

\subsection{Raw-score counterparts of the rank-based influence results}
\label{app:economy_raw_scores}

\begin{figure}[!h]
    \centering
    \includegraphics[width=5.5in]{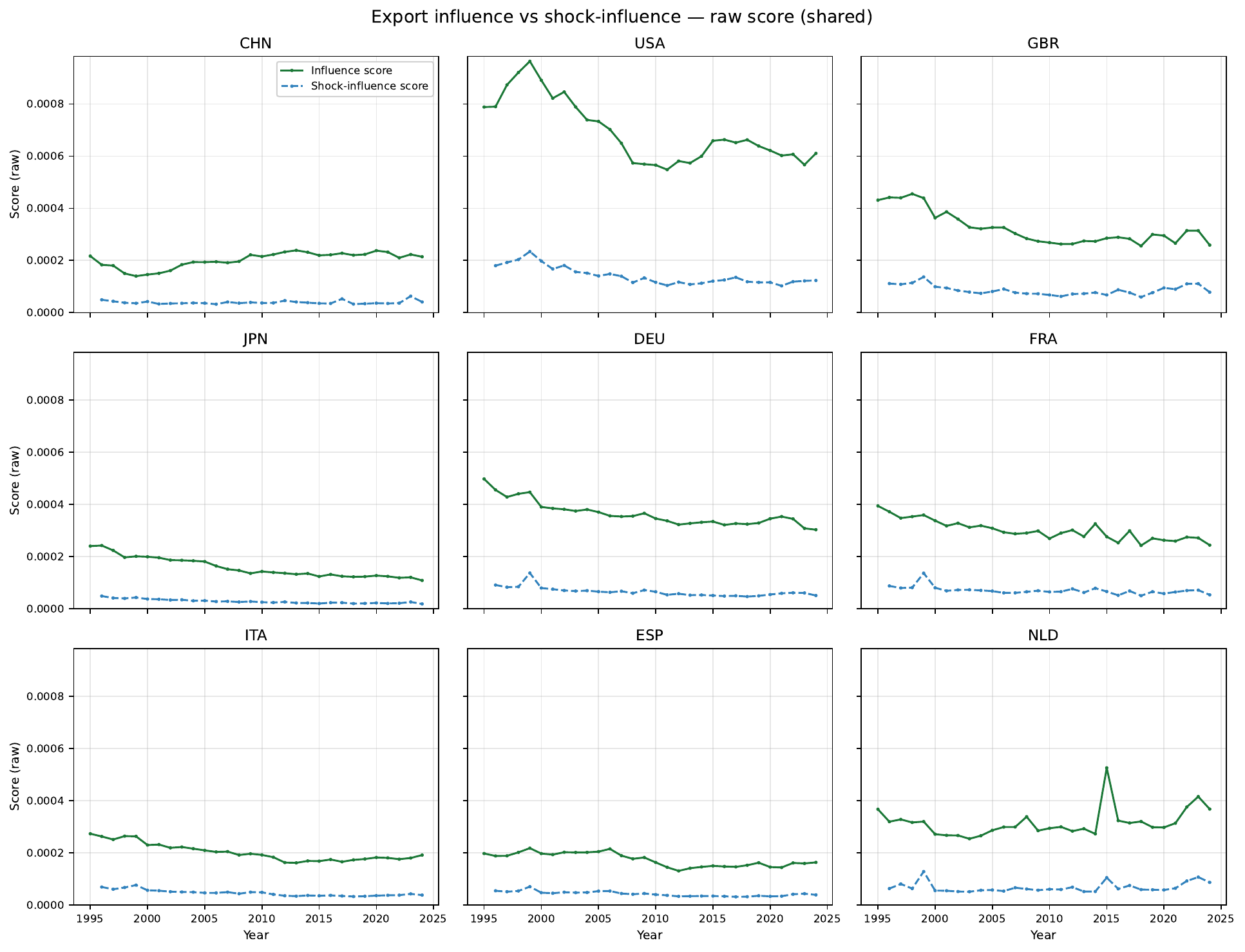}
    \caption{\textbf{Raw baseline and shock-modulated export-side influence scores, 1995--2024.}
    The solid line reports the annual product-level baseline export-side influence score, while the dashed line reports the corresponding shock-modulated score. The shock-modulated measure weights each layer-level PageRank perturbation by the node-level JS-Walk distance between consecutive years, so gaps between the two curves indicate whether an economy's export-side network influence is concentrated in structurally changing or relatively stable product layers.}
    \label{fig:app_product_export_infl_vs_shock_score_raw}
\end{figure}

\begin{figure}[!h]
    \centering
    \includegraphics[width=5.5in]{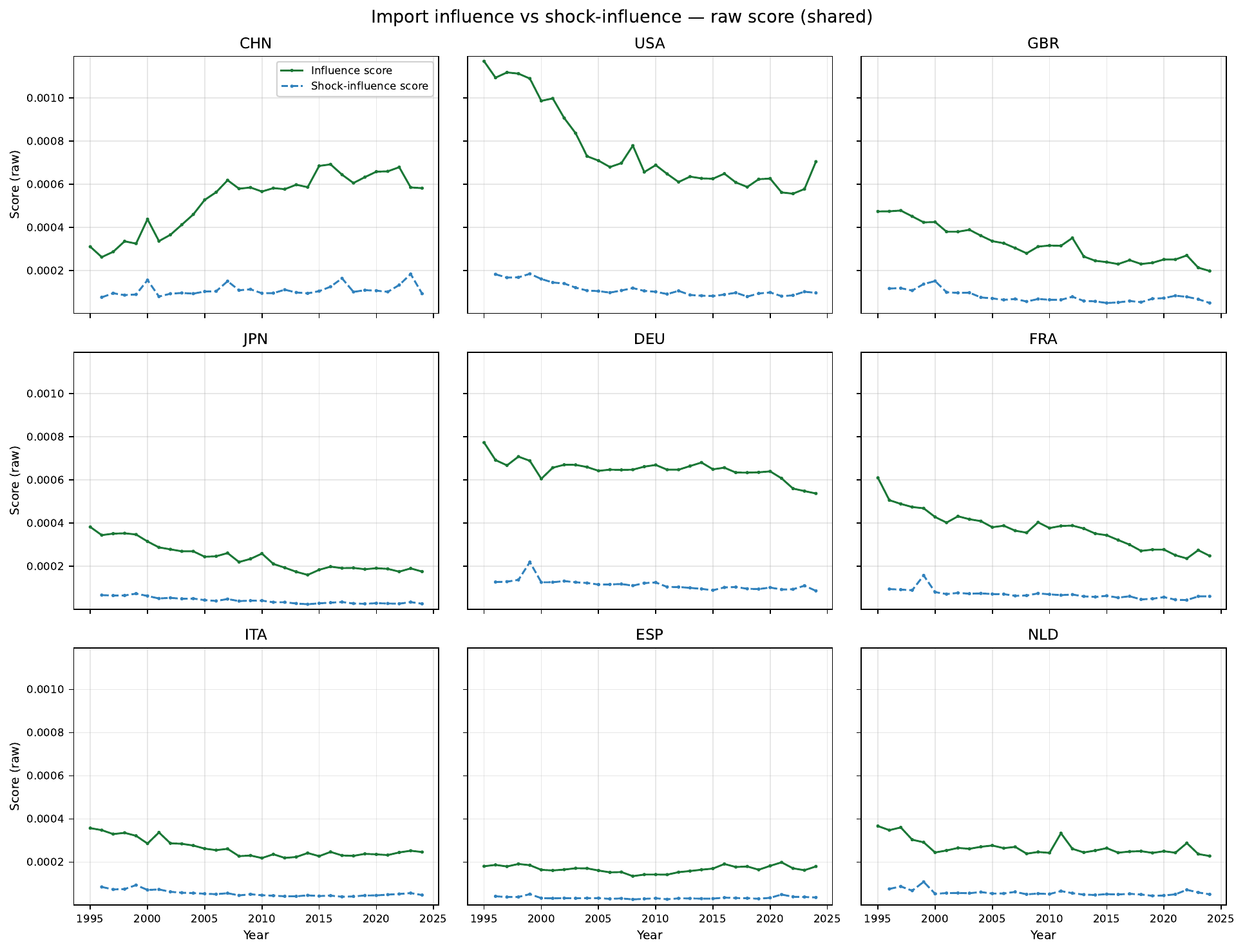}
    \caption{\textbf{Raw baseline and shock-modulated import-side influence scores, 1995--2024.}
    The solid line reports the annual product-level baseline import-side influence score, while the dashed line reports the corresponding shock-modulated score. Shock modulation identifies the component of import-side network influence associated with changes in an economy's import-origin profile across consecutive years. The shared scale allows direct comparison of the magnitude of stable and restructuring import-side influence across the focus economies.}
    \label{fig:app_product_import_infl_vs_shock_score_raw}
\end{figure}

Figures~\ref{fig:app_product_export_infl_vs_shock_score_raw} and~\ref{fig:app_product_import_infl_vs_shock_score_raw} report the raw-score counterparts of the rank-based influence trajectories in the main text. They show whether the observed rank differences correspond to differences in score magnitude and how much of each economy's influence is associated with structurally changing product layers.

\FloatBarrier

\subsection{Shock-modulated export--import asymmetry}
\label{app:economy_shock_scatter}

We next compare shock-modulated export-side and import-side influence in the final year of the sample. This complements the baseline export--import scatter in the main text by focusing on the part of influence associated with recent changes in trade-partner profiles.

\begin{figure}[!h]
    \centering
    \includegraphics[width=5.5in]{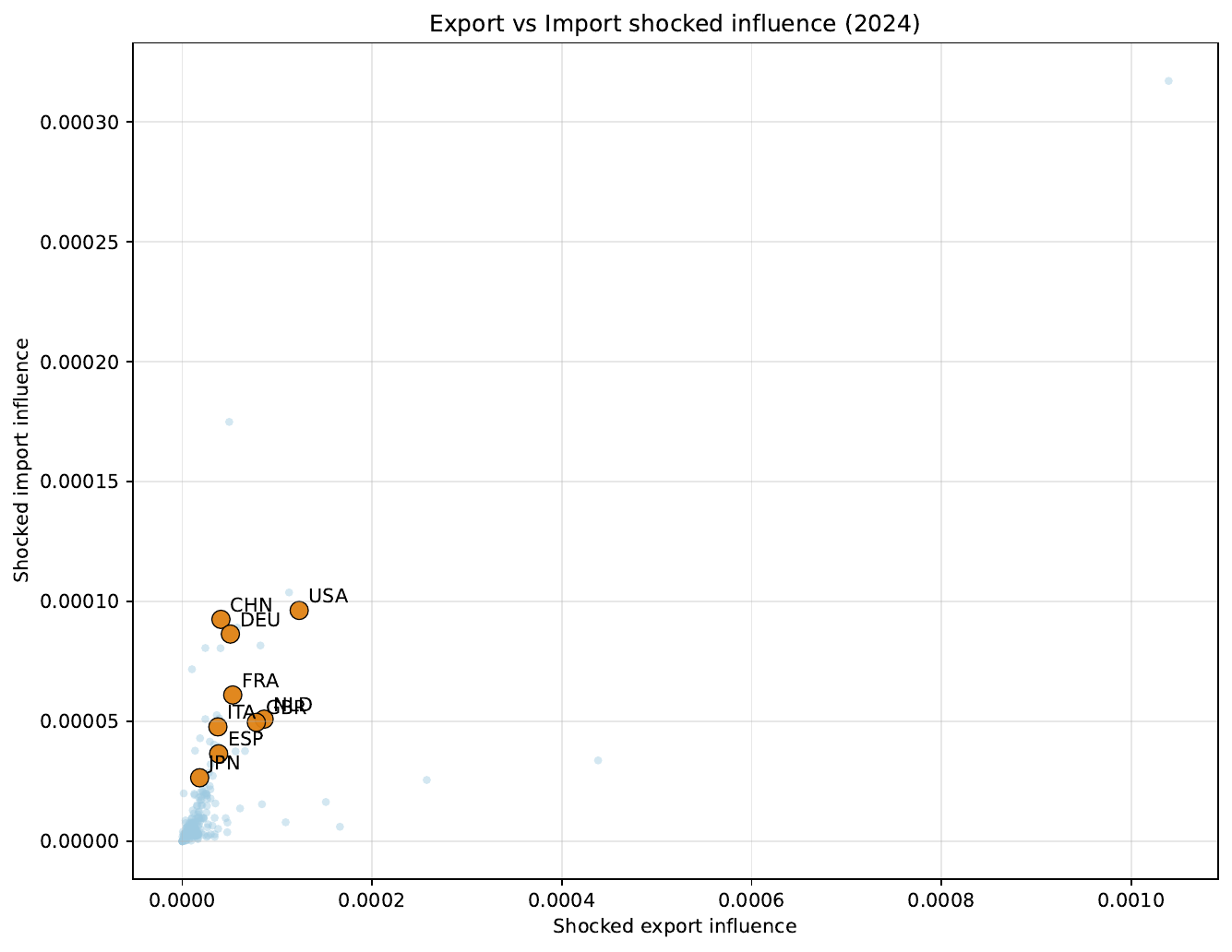}
    \caption{\textbf{Shock-modulated export-side versus import-side influence in 2024.}
    Each point represents an economy's product-level shock-modulated network influence in 2024, with the horizontal axis showing shocked export-side influence and the vertical axis showing shocked import-side influence. The nine focus economies are highlighted and labelled. Economies far from the origin combine high network-removal impact with substantial year-on-year restructuring of their trade-partner profiles, while deviations from the diagonal indicate asymmetry between shock-sensitive export-side and import-side roles.}
    \label{fig:app_product_export_import_shock_scatter}
\end{figure}

\FloatBarrier

\subsection{Comparison with economic fitness}
\label{app:economy_fitness}

The following figures compare network-removal influence with the Tacchella--Pietronero fitness index computed from the yearly RCA matrix. This comparison is not intended as a benchmark for the proposed influence measure; rather, it shows that network perturbation and productive capability capture different structural dimensions of the trade system.

\begin{figure}[!h]
    \centering
    \includegraphics[width=5.5in]{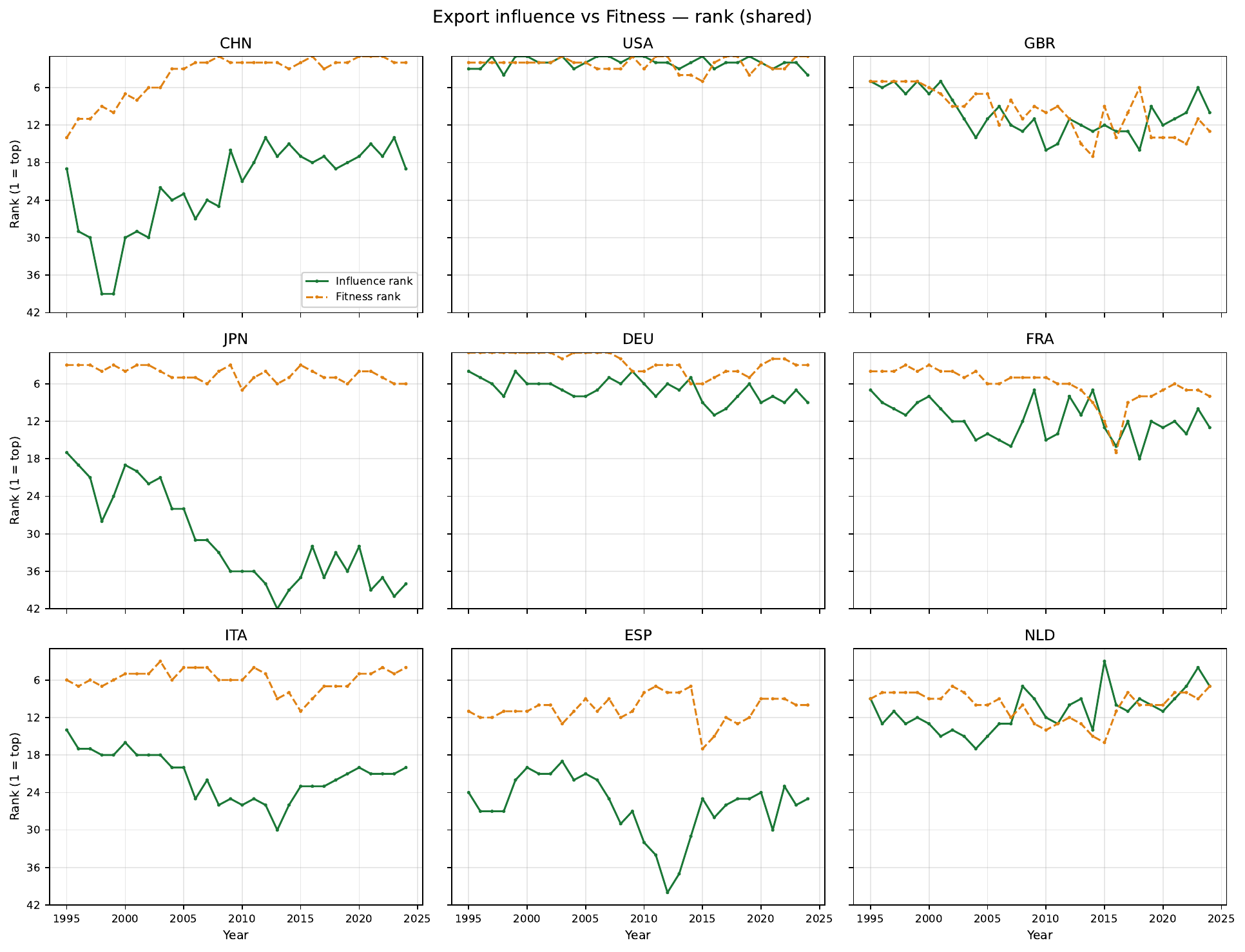}
    \caption{\textbf{Export-side influence rank versus economic fitness rank, 1995--2024.}
    The green line reports the annual product-level baseline export-side influence rank, while the orange line reports the corresponding Tacchella--Pietronero fitness rank computed from the yearly RCA matrix. Rank one denotes the highest-ranked economy in a given year. The comparison shows how network-removal influence differs from productive-capability rankings: an economy can be central in the perturbation structure of export flows without having the same position in the fitness hierarchy, and vice versa.}
    \label{fig:app_product_export_infl_vs_fitness_rank}
\end{figure}

\begin{figure}[!h]
    \centering
    \includegraphics[width=5.5in]{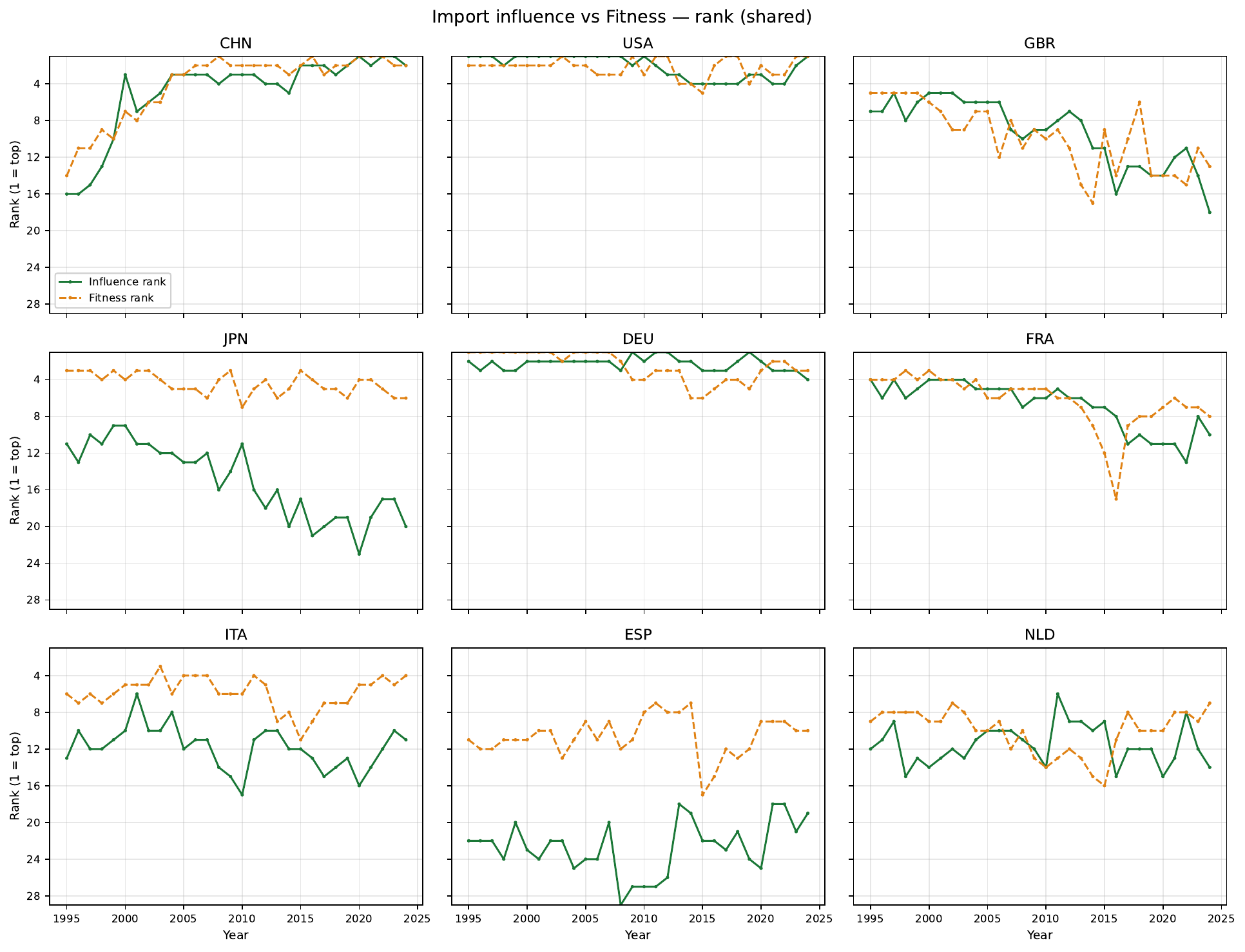}
    \caption{\textbf{Import-side influence rank versus economic fitness rank, 1995--2024.}
    The green line reports the annual product-level baseline import-side influence rank, while the orange line reports the corresponding Tacchella--Pietronero fitness rank. The comparison assesses whether economies that strongly perturb the import-origin structure of the GTN also occupy high positions in the productive-capability ranking. Differences between the two trajectories indicate that import-side network influence and economic fitness capture distinct structural dimensions.}
    \label{fig:app_product_import_infl_vs_fitness_rank}
\end{figure}

\FloatBarrier

\subsection{Full-sample rankings and special reporting economies}
\label{app:economy_top_rankings}

The next two figures summarise economies with persistently high baseline influence over the full sample period. Some entries correspond to BACI reporting areas or historical/special economy codes rather than current sovereign states; these are retained because the influence calculation is performed on the full BACI economy set used throughout the analysis.

\begin{figure}[!h]
    \centering
    \includegraphics[width=5.5in]{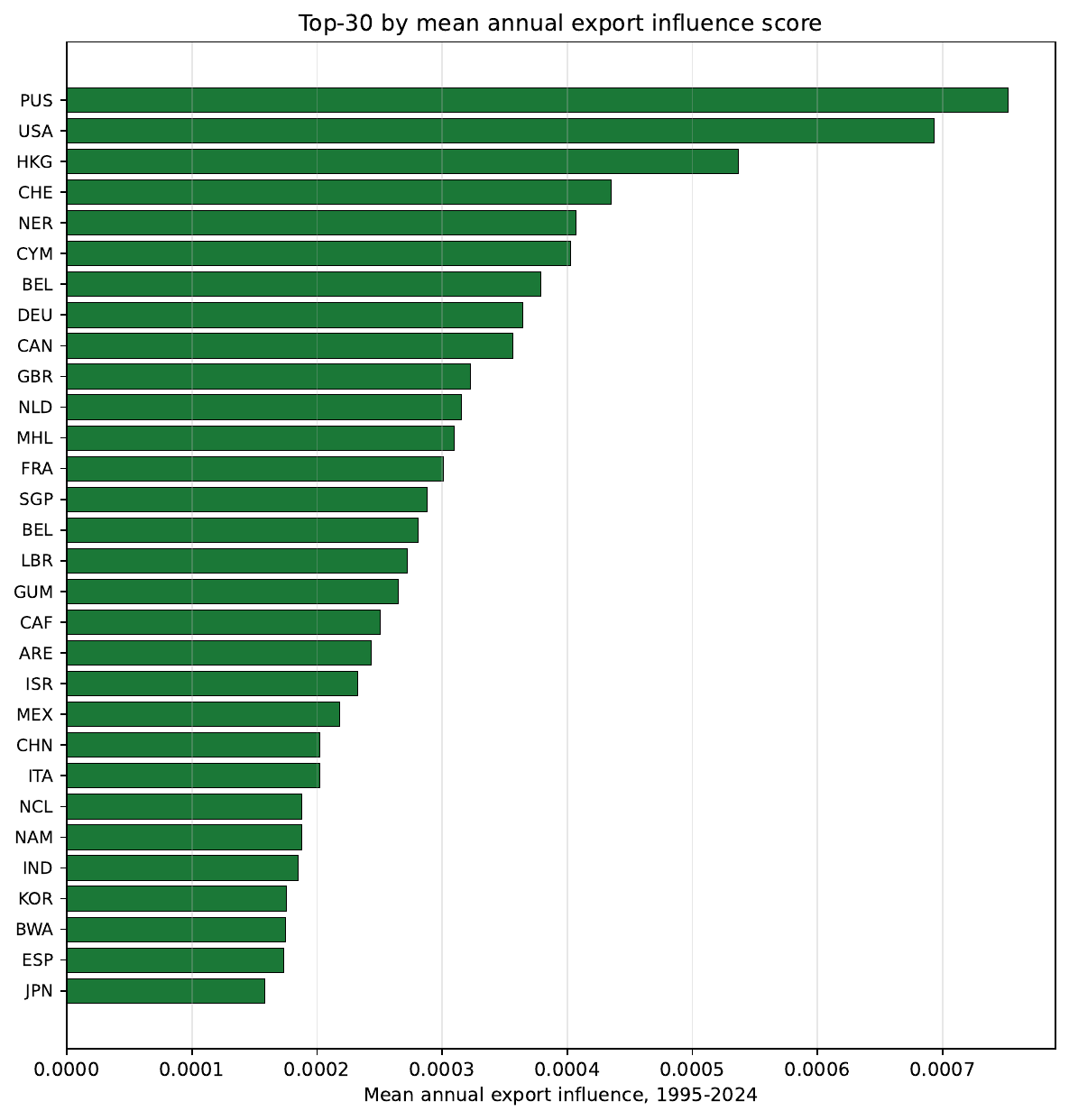}
    \caption{\textbf{Top 30 economies by mean annual baseline export-side influence, 1995--2024.}
    Bars report the average annual product-level baseline export-side influence score over the full sample period. Averaging across years smooths short-run fluctuations and identifies economies with persistently high export-side network-removal impact. The ranking complements the 2024 cross-section by distinguishing long-run influence from end-of-sample influence. Some labels correspond to BACI reporting areas or historical/special economy codes retained in the full economy set.}
    \label{fig:app_product_avgscore_export_infl}
\end{figure}

\begin{figure}[!h]
    \centering
    \includegraphics[width=5.5in]{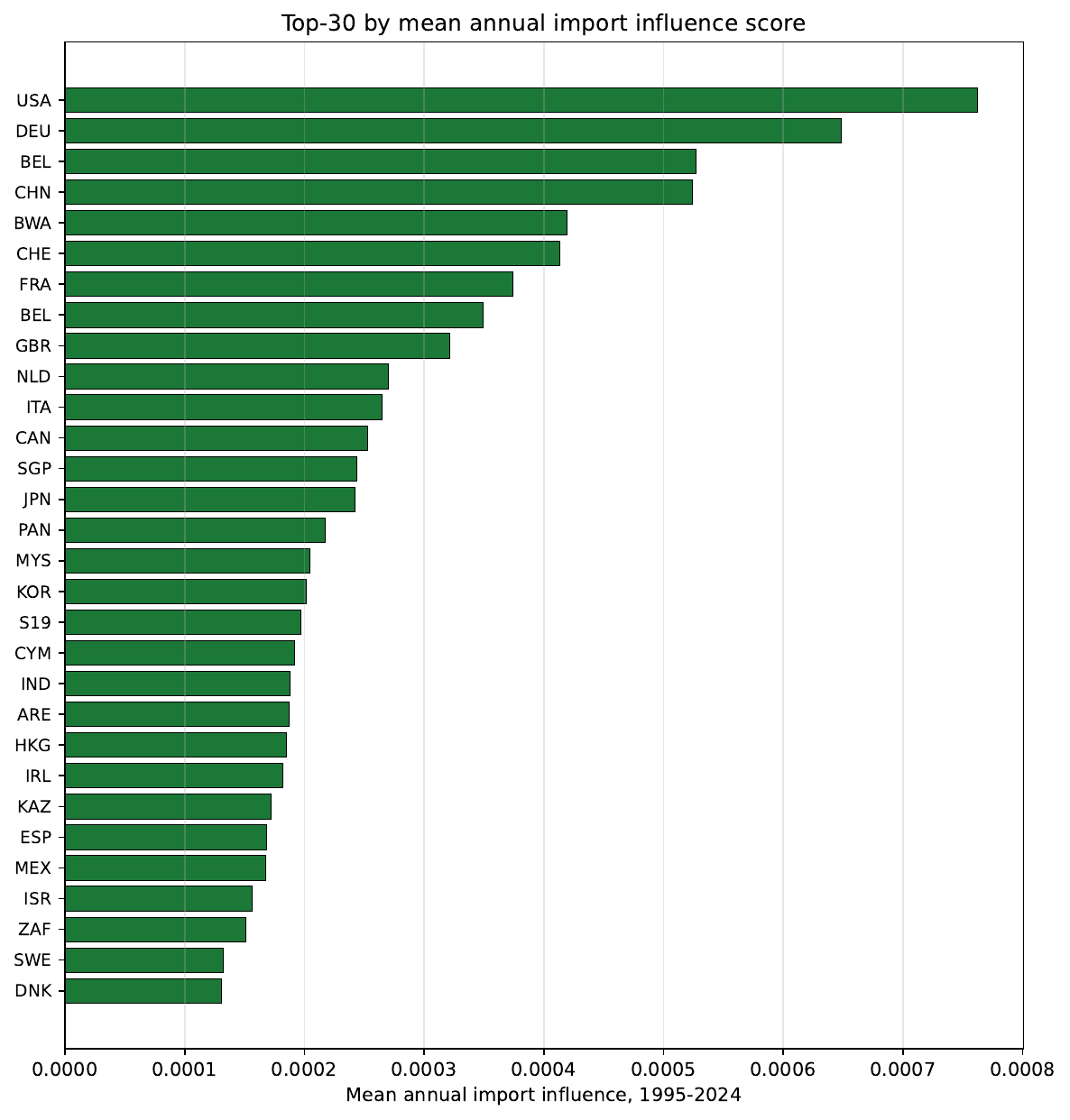}
    \caption{\textbf{Top 30 economies by mean annual baseline import-side influence, 1995--2024.}
    Bars report the average annual product-level baseline import-side influence score over the full sample period. The ranking identifies economies that persistently perturb the import-origin structure of the GTN when removed from product layers. Comparing this panel with the 2024 import ranking separates long-run import-side influence from recent cross-sectional influence. Some labels correspond to BACI reporting areas or historical/special economy codes retained in the full economy set.}
    \label{fig:app_product_avgscore_import_infl}
\end{figure}

\FloatBarrier

\subsection{Chapter-level robustness}
\label{app:economy_chapter_robustness}

Finally, we repeat the influence analysis after aggregating products to HS chapters. These results test whether the main export--import asymmetries are robust to a coarser representation of product layers.

\begin{figure}[!h]
    \centering
    \includegraphics[width=5.5in]{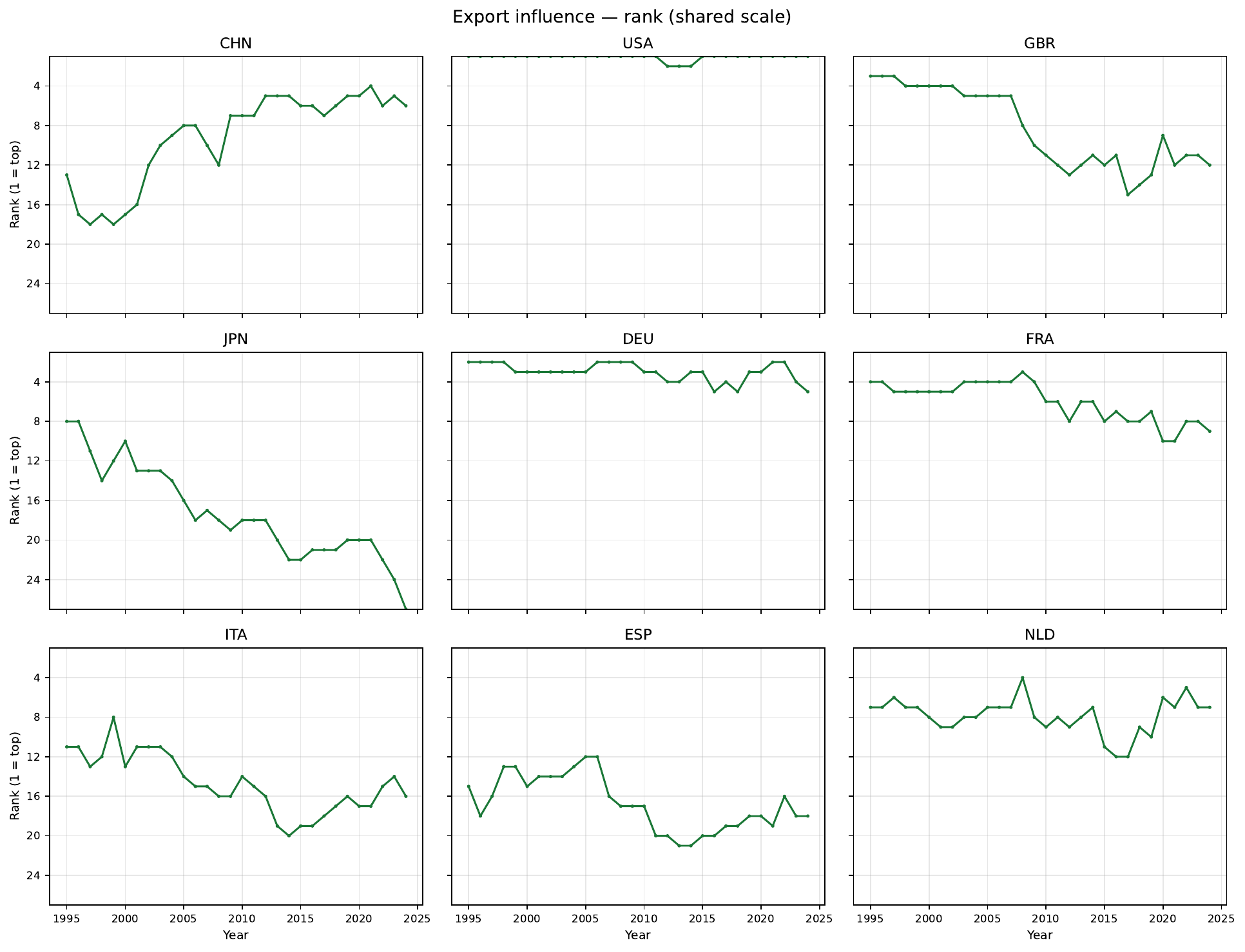}
    \caption{\textbf{Chapter-level export-side influence ranks for selected major economies, 1995--2024.}
    This figure repeats the baseline export-side influence ranking after aggregating products to HS chapters. Rank one denotes the highest influence in a given year. The comparison with the product-level results tests whether the main export-side rankings are robust to coarser product aggregation.}
    \label{fig:app_chapter_export_infl_rank}
\end{figure}

\begin{figure}[!h]
    \centering
    \includegraphics[width=5.5in]{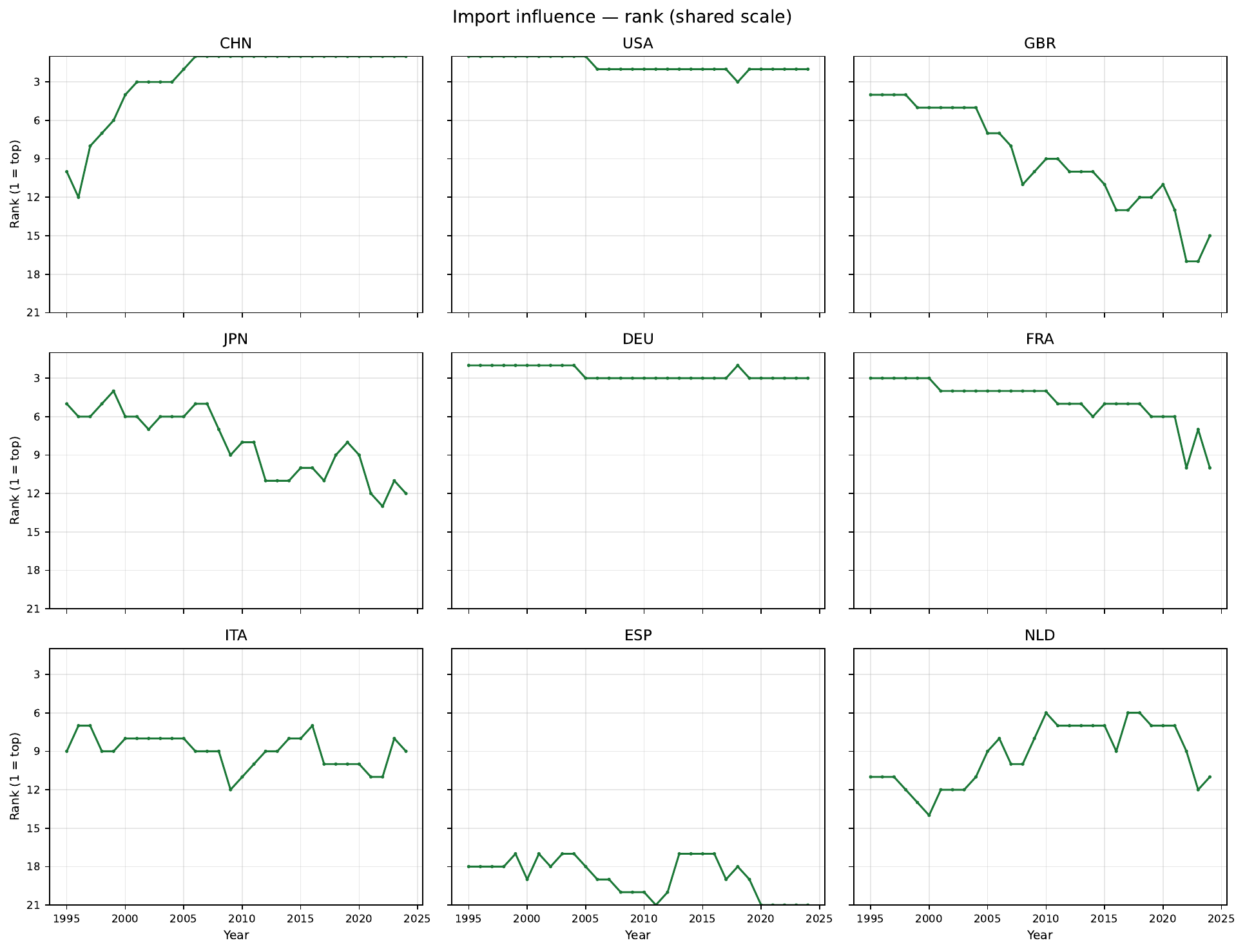}
    \caption{\textbf{Chapter-level import-side influence ranks for selected major economies, 1995--2024.}
    This figure repeats the baseline import-side influence ranking after aggregating products to HS chapters. Influence is computed on the transposed importer--exporter graph and rank one denotes the highest import-side perturbation influence in a given year. Similar temporal ordering across product and chapter levels supports the robustness of the import-side results to aggregation.}
    \label{fig:app_chapter_import_infl_rank}
\end{figure}

\begin{figure}[!h]
    \centering
    \includegraphics[width=5.5in]{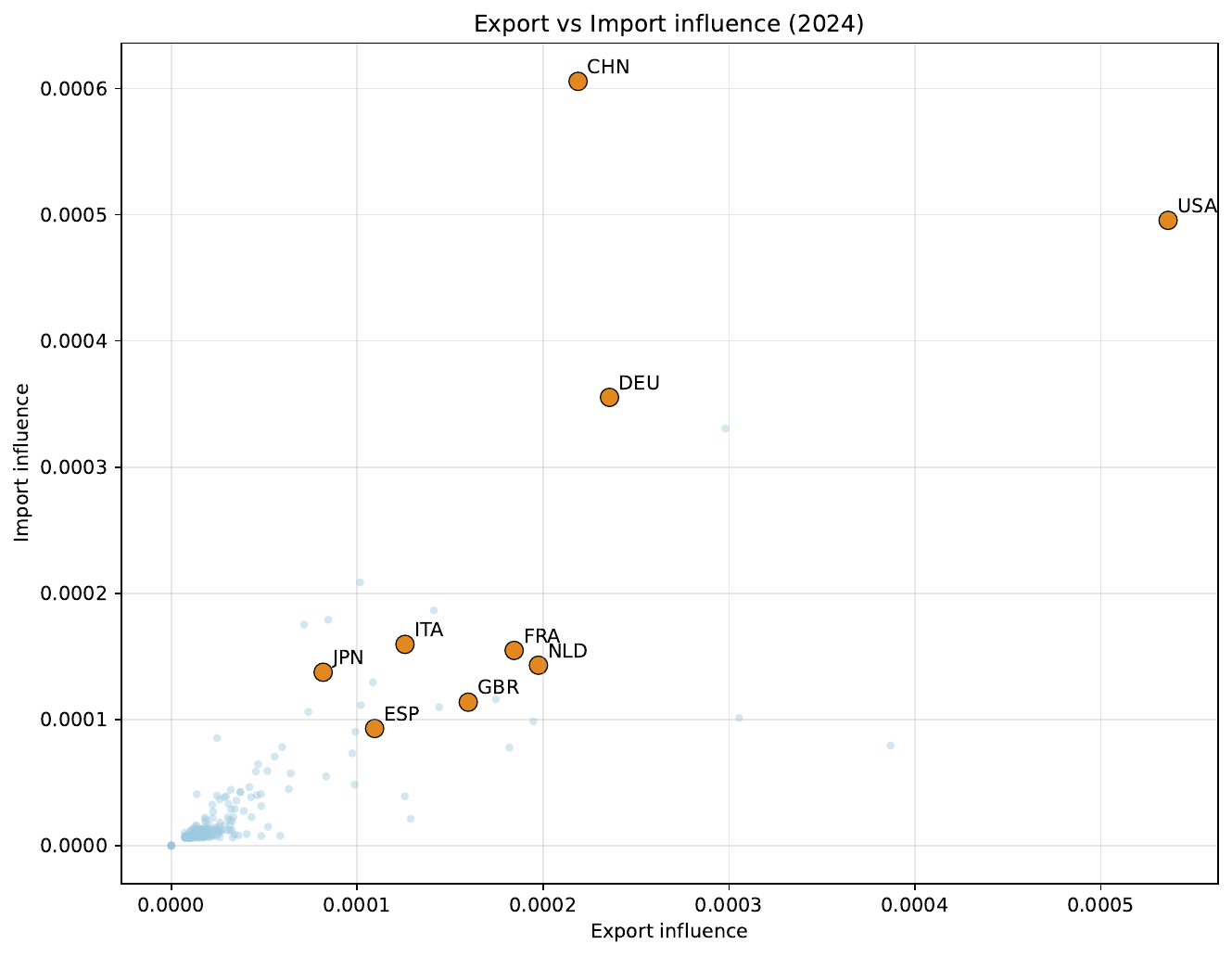}
    \caption{\textbf{Chapter-level export-side versus import-side influence in 2024.}
    Each point represents an economy's baseline chapter-level network influence in 2024, with export-side influence on the horizontal axis and import-side influence on the vertical axis. The nine focus economies are highlighted and labelled. The panel provides a chapter-level robustness check for the export--import asymmetry reported at the product level.}
    \label{fig:app_chapter_export_import_scatter}
\end{figure}

\FloatBarrier

\subsection{Alternative role-direction sensitivity}
\label{app:economy_role_direction}

As an additional sensitivity check, we evaluate the alternative role-direction specification described in the methods. In this variant, export-side and import-side portfolio weights are retained, but the PageRank perturbation step is evaluated using the alternative graph orientation. This test assesses whether the export--import asymmetry observed in the main specification is preserved when influence is interpreted through supplier-side and demand-side roles.

\begin{figure}[!h]
    \centering
    \includegraphics[width=5.5in]{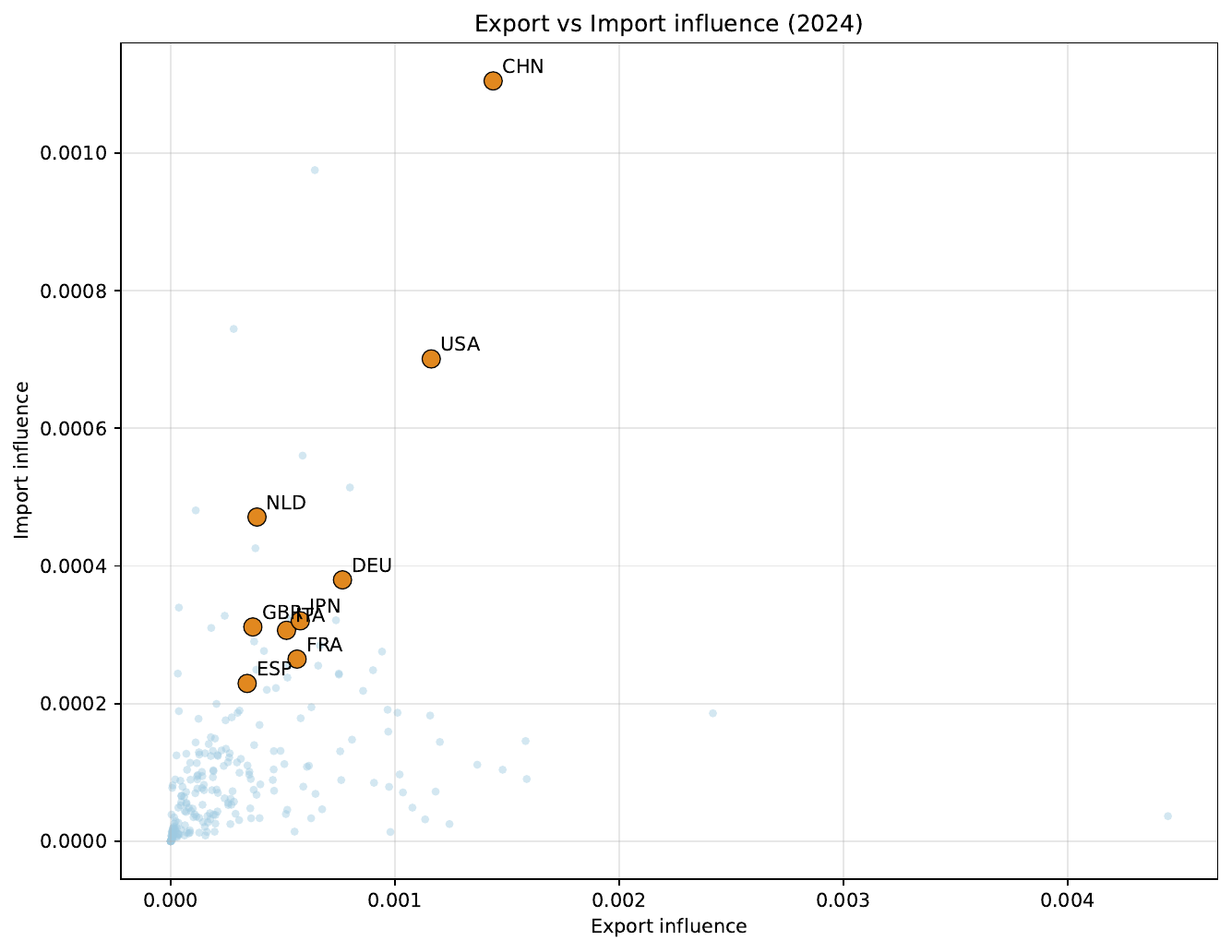}
    \caption{\textbf{Export-side versus import-side influence under the alternative role-direction specification in 2024.}
    Each point represents an economy's 2024 influence score under the role-based sensitivity specification, with the horizontal axis showing the export-side variant and the vertical axis showing the import-side variant. The nine focus economies are highlighted and labelled. The scatter assesses whether the export--import asymmetry observed in the main specification is preserved when PageRank perturbations are evaluated using the alternative role-direction orientation.}
    \label{fig:app_role_export_import_scatter}
\end{figure}

\FloatBarrier